\documentclass[prb,amssymb,amsmath,superscriptaddress,twocolumn]{revtex4-2}
\usepackage[english]{babel}
\usepackage[utf8]{inputenc}
\usepackage{graphicx}
\graphicspath{{./images/}}
\usepackage{url}
\usepackage{hyperref}
\usepackage{color}
\usepackage{xcolor}
\usepackage{cancel}
\usepackage{enumerate}
\usepackage{enumitem}
\usepackage{pifont}
\usepackage{mathtools}
\usepackage{bbold}
\usepackage{lipsum}
\usepackage{tikz}
\usepackage{tikz-3dplot}
\usetikzlibrary{positioning}
\usetikzlibrary{backgrounds}
\usetikzlibrary{arrows.meta}
\usepackage[caption=false]{subfig}
\usepackage{bm}
\usepackage[normalem]{ulem}

\usepackage[defaultcolor=blue]{changes}
\definechangesauthor[name=Yantao, color=cyan]{y}
\definechangesauthor[name=Babak, color=orange]{b}
\definechangesauthor[name=Adam, color=purple]{a}
\definechangesauthor[name=Herb, color=red]{h}

\begin{document}
\title{Renormalized Magic Angles in Asymmetric Twisted Graphene Multilayers}

\author{Adam Eaton}
\affiliation{Department of Physics, Indiana University, Bloomington, Indiana 47405, USA}

\author{Yantao Li}
\affiliation{Department of Physics, Indiana University, Bloomington, Indiana 47405, USA}

\author{H. A. Fertig}
\affiliation{Department of Physics, Indiana University, Bloomington, Indiana 47405, USA}
\affiliation{Quantum Science and Engineering Center, Indiana University, Bloomington, Indiana 47405, USA}

\author{Babak Seradjeh}
\affiliation{Department of Physics, Indiana University, Bloomington, Indiana 47405, USA}
\affiliation{Quantum Science and Engineering Center, Indiana University, Bloomington, Indiana 47405, USA}
\affiliation{IU Center for Spacetime Symmetries, Indiana University, Bloomington, Indiana 47405, USA}

\date{\today} 

\begin{abstract}
Stacked graphene multilayers with a small relative twist angle between each of the layers have been found to host flat bands at a series of ``magic'' angles. We consider the effect that Dirac point asymmetry between the layers, and in particular different Fermi velocities in each layer, may have on this phenomenon.  Such asymmetry may be introduced by unequal Fermi velocity renormalizations through Coulomb interactions with a dielectric substrate. 
It also arises in an approximate way in tetralayer systems, in which the outer twist angles are large enough that there is a dominant moir{\'e} peridocity from the stacking of the inner two layers.  We find in such models that the flat band phenomenon persists in spite of this asymmetry, and that the magic angles acquire a degree of tunability through either controlling the screening in the bilayer system or the twist angles of the outer layers in the tetralyer system. Notably, we find in our models that the quantitative values of the magic angles are increased.
\end{abstract}

\maketitle

\section{Introduction}
In recent years, the discovery that electronic properties of twisted stacked graphene multilayers can be controlled by the twist angle, which modulates the interlayer tunneling between the graphene layers, has led to the burgeoning field of ``twistronics"~\cite{Santos_2007,Morrell_2010,Andrei_2020, Carr_2017, Ren_2020}. A remarkable discovery~\cite{BM} in the single-particle physics of these systems is that they host flat bands at certain ``magic'' angles, as first shown for the simplest case of twisted bilayer graphene (TBG).  The flatness of these bands suggests that when interactions are included, they should host correlated electron states. And indeed, with improving sample preparation techniques, such states have been observed, most prominently Mott insulating states and superconductivity~\cite{Cao_2018a,Cao_2018b}.
Such exotic correlated electron states are not unique to TBG~\cite{Yankowitz_2019,Wong_2020}, but are also present in other twistronic systems including those involving hexagonal boron nitride~\cite{hBN1, hBN2, hBN3, Yang_2020, Andelkovic2020, Wang2019}, twisted tungsten selenide 
and other transition metal dichalcogenides~\cite{Wang_2020, Zhang_2019, Li_2021, Naik_2018, Naik_2020, Zhan_2020}, twisted double bilayer graphene~\cite{Haddadi, Cheroblu, Burg2019, Koshino2019, He_2021,Zhang_2021, Cao_2020, Fang_2016, Culchac_2020, Choi_2019,Liu_2020,Lee_2019, Adak_2020}, twisted trilayer graphene~\cite{Kaxiras,Park_2021,Hao_2021, Suarez_2013, Chen_2016, Zuo_2018, Ma_2021, Xu_2021, Li_2019}, as well as other systems of stacked twisted graphene multilayers~\cite{Liu_2020, Hierarchy, Denner2020, Tritsaris_2020, Gupta2020}.  They are even present in systems that do not possess a moir\'e potential~\cite{Kerelskye_2021, 2021, zhou2021isospin, delabarrera2021cascade, seiler2021quantum} and may also arise in other twisted graphene structures without flat bands due to low-energy van Hove singularities and Lifshitz transitions ~\cite{Li_2022}.

Theoretical understanding of this system has greatly benefited from the introduction of the Bistritzer-MacDonald (BM) model~\cite{BM}, in which the graphene sheets are individually treated in the long wavelength limit as Dirac point Hamiltonians, while the interlayer tunneling is treated in a spatially periodic model, represented in a small moir\'e Brillouin zone (mBZ). Intriguingly, although there has been important progress~\cite{Tarnopolsky_2019, Balents, Morrell_2010, Nam_2017, Zou_2018, Yuan_2018, Lin_2018, Zhang_2019, Kang_2018, Koshino2019, Rademaker_2018, Po_2019, Qiao_2018, Carr_2019, Guinea_2019}, a full explanation for the band flatness within the mBZ at magic twist angles remains elusive.   One question that this naturally raises is the role of symmetry in producing flat bands in such models.  In addition to translational and discrete rotational symmetries, a mirror symmetry operation maps the $K_M$ and $K_M'$ points of the mBZ onto one another~\cite{Po_2018}.  Indeed the eigenstates of the BM model at the $K_M$ and $K'_M$ points largely reside in one layer or the other.  Moreover, the energy dispersions in their vicinities are essentially identical, i.e., they have the same Fermi velocities. The symmetry of these Dirac points can be broken with a perpendicular electric field~\cite{Po_2018}, in which case the flatness of the low-energy bands at the magic angles are not expected to survive.  However, the symmetry of the Dirac points in a mBZ may be broken in more subtle ways, and whether the flat band phenomenon survives the lifting of this symmetry in general is, to our knowledge, not known.

In this work we explore this question by investigating models in which the symmetry between the Dirac points of the layers that are tunnel-coupled has been broken, in effect through {\it different} Fermi velocities at the two coupled Dirac points.
We consider two concrete situations where this can occur.  The first involves a dielectric screening substrate applied on only one side of the TBG system.  In general, Coulomb interactions renormalize the Fermi velocity at the Dirac points of a graphene layer~\cite{Gonzalez_1994}, through the effects of high momentum states on those at low-momentum.  Because the two graphene layers are at different distances from the substrate, screening sets in at different length scales for each of them and leads to different Fermi velocities at low energies for the coupled Dirac points.
We estimate this effect and show that it can be considerable for high dielectric substrates, such as SrTiO$_3$~\cite{Veyrat_2020}.

A second such model involves twisted tetralayer graphene with three independent twist angles $\theta_{12}$ (top pair), $\theta_{23}$ (middle pair), and $\theta_{34}$ (bottom pair).
By considering situations where the $\theta_{12}$ and $\theta_{34}$ are not too small, we approximate the four-layer system as two coupled systems comprised of the top and bottom pairs of layers supporting Dirac points, which are themselves in turn tunnel-coupled with effective twist angle $\theta_{23}$ between them.  Having three independent twist angles is useful because it allows engineering of the relevant properties of the system.  In our treatment, one finds that in addition to renormalized Fermi velocities at the Dirac points of the top and bottom pairs of layers, there are also changes in the precise form of the tunneling between the two coupled systems.

\begin{figure}[t]
    \includegraphics[width=\linewidth]{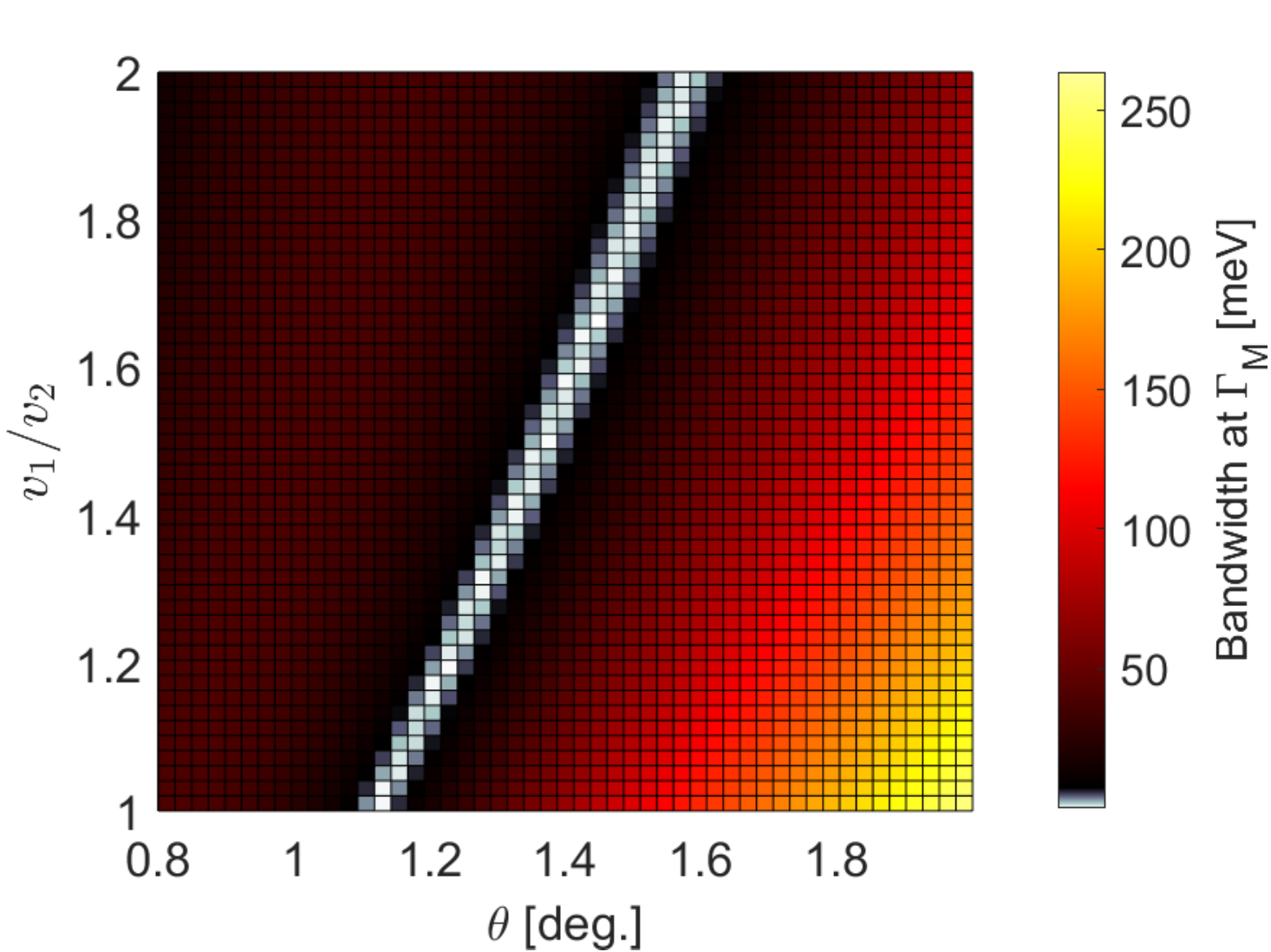}
    \caption{The bandwidth of the asymmetric TBG at the $\Gamma_M$ point of the moir\'e Brillouin zone as a function of both the twist angle $\theta$ and the Fermi velocity asymmetry $v_{1}/v_{2}$. Locations where the bandwidth is less than 5 meV are shown in white. Here the tunneling amplitude $w = 110$ meV and $v_{2} = 0.88 \times 10^6$ m/s is the Fermi velocity of bare monolayer graphene.}
    \label{fig:lambda vs theta}
\end{figure}

Our main result is that magic angles at which flat bands arise do indeed survive symmetry breaking between Dirac points even when it is relatively strong.
Figure \ref{fig:lambda vs theta} illustrates a typical result for the asymmetric bilayer system, in which one sees that engineering the Fermi velocity ratio allows for controlling the value of the magic angle.
The locations of the magic angle can be predicted to quite a good approximation by perturbation theory~\cite{BM}, which results in
the condition
${\hbar k_{\theta} \sqrt{v_{1} v_{2}}} / {w} = \sqrt{3}$,
where $v_{1}$ and $v_{2}$ are the Fermi velocities associated with the Dirac points in the two coupled layers, $w$ is the tunneling strength between layers, and $k_{\theta} = 2k_D\sin(\theta/2)$ is the separation between twisted Dirac points as determined by the twist angle $\theta$ and $k_D$, the separation between the $K$ and $K'$ points of a single graphene sheet.  Qualitatively similar results are obtained for the tetralayer system when the twist angles for the outer layers are not too small, and again the values of magic angles can be accounted for by a perturbation theory analysis.  While our basic approach does not include the effects of incommensuration  arising at most sets of twist angles in this system, an estimate of these using degenerate perturbation theory suggests that they do not eliminate the basic flat band phenomenon.

The rest of this article is organized as follows.
In Sec.~\ref{sec:asymTBG} we provide an analysis of twisted bilayer graphene with unequal Fermi velocities in the layers and describe how such asymmetry can emerge for a bilayer system with different dielectric screening in each layer.
In Sec.~\ref{sec:4TBG} we focus on an effective realization of this model in a graphene tetralayer in which the outer twist angles are unequal and not too small. We model this system by treating the effects of twisting in the outer layers via ${\it k \cdot p}$ perturbation theory, which essentially renormalizes the Dirac point velocities, and then numerically solve for the spectrum in an effective bilayer BM model. We then provide a perturbative analysis for magic angles in this system and compare them with numerical results for representative sets of angles.
We conclude in Sec.~\ref{sec:sum} with a summary and discussion. We present a study of effects incommensuration between outer and inner twist angles in Appendices. Appendix~\ref{app:wave} provides some results that motivate our treatment of the tetralayer system as an effective bilayer system, in particular showing conditions under which incommensuration effects should be very small. Appendix~\ref{app:incomm} provides a degenerate perturbation theory estimate of the effects of scattering by incommensurate wavevectors from the outer two twisted layers in our idealization of the tetralayer as an effective bilayer system.

\section{Asymmetric Twisted Bilayer Graphene}\label{sec:asymTBG}
We consider ansymmetic twisted bilayer system with unequal Fermi velocities, described by the continuum Hamiltonian
\begin{equation}
H_\text{ATBG} =
\begin{bmatrix}
    h_1 & T \\
    T^\dagger & h_2
\end{bmatrix},
\label{HATBG}
\end{equation}
where $h_l =  \hbar v_{l} {\boldsymbol \sigma} \cdot \left[-i\boldsymbol{\nabla} + (-1)^l {\bf q}_0/2\right]$ is the Hamiltonian of layer $l=1,2$, with $\boldsymbol{\sigma} = (\sigma_x,\sigma_y)$ the vector of Pauli matrices and $\boldsymbol{\nabla} = (\partial_x, \partial_y)$,
and the tunneling
$T = w\sum_{j = 0}^2 \exp(-i {\bf Q}_j \cdot {\bf r}) T_j$, with ${\bf Q}_j = {\bf q}_j - {\bf q}_0$, ${\bf q}_0 = k_\theta(0,-1)$, ${\bf q}_1=k_{\theta}(-\frac{\sqrt{3}}{2},\frac{1}{2})$, and ${\bf q}_2=k_{\theta}(\frac{\sqrt{3}}{2},\frac{1}{2})$~\cite{Balents}. Note that we have ignored the effect of the small rotation angle $\theta$ on Pauli matrices in each layer and assumed the Dirac points that are tunnel-coupled by $T$ reside in the same valley of their host graphene sheets, and we are only describing the low-energy bands from those valleys. The tunneling matrices
$T_j$ are given by~\cite{Balents}
\begin{equation}    \label{eq:Tj}
    T_0 = \begin{bmatrix}
         u & 1 \\
         1 & u \\
    \end{bmatrix},
    \quad
    T_1 = \begin{bmatrix}
         u & e^{2\pi i/3} \\
         e^{-2\pi i/3} & u \\
    \end{bmatrix},
    \quad
    T_2 = T_1^*.
\end{equation}
In the situations of interest to us, $v_{1}\neq v_{2}$, and $u\neq 1$ allows for different tunneling amplitudes between atoms on the same sublattice and those on different sublattices, which represents a simple model of lattice relaxation in the layers~\cite{Nam_2017,Koshino_2018,Carr_2019}. Except where otherwise indicated, in our numerical results we take the tunneling amplitude to be $w = 110$ meV, and the effective ratio of tunneling between sites on the same sublattice and opposite sublattice $u = 0.8$.

\subsection{Perturbative Estimate of Magic Angle}
\label{sec:per}
The effect of the tunneling $T$ on the low-energy dispersion in each layer can be computed perturbatively as corrections to poles of the resolvent operator, $G(E) = (E - H_\text{ATBG})^{-1}$. It is useful to employ the projector $P_l$ onto the space of each layer to define an energy-dependent effective Hamiltonian $h^\text{eff}_l(E)$ in layer $l$,
\begin{equation}
E-h^\text{eff}_l(E) \equiv \left[P_l {G}(E) P_l \right]^{-1},
\end{equation}
which then yields
\begin{subequations}
\begin{align}
h^\text{eff}_1(E)
	&= h_1 + T(E - h_2)^{-1}T^\dagger,\\
h^\text{eff}_2(E)
	&= h_2 + T^\dagger(E - h_1)^{-1}T.
\end{align}
\end{subequations}

We next evaluate the matrix elements of $h_l^\text{eff}$ in the plane-wave basis $|{\bf k}_l\rangle$, where ${\bf k}_l$ is measured from the Dirac point of layer $l$. Then $T_j$ only connects states with wavevectors that differ by ${\bf q}_j$, so that
\begin{equation}\label{eq:heff1}
\langle {\bf k}_1| h_1^\text{eff} - h_1 | {\bf k}_1\rangle = w^2 \sum_j \frac{T_j \left[ E + \hbar v_{2} \boldsymbol\sigma \cdot ({\bf k}_1 - {\bf q}_j) \right] T_j}{E^2 - (\hbar v_{2}|{\bf k}_1 - {\bf q}_j|)^2}.
\end{equation}
A similar expression for $\langle {\bf k}_2| h_2^\text{eff} - h_2 | {\bf k}_2\rangle$ is obtained by replacing $v_{2}$ with $v_{1}$ and ${\bf q}_j$ with $-{\bf q}_j$. Since we are interested in solutions $E \sim |{\bf k}_1|$, we expand
\begin{equation}
\frac1{E^2 - (\hbar v_{2}|{\bf k}_1 - {\bf q}_j|)^2} = -\frac{1 + 2{\bf k}_1\cdot{\bf q}_j}{(\hbar v_{2} k_\theta)^2} + \mathcal{O}(|{\bf k}_1|^2).
\end{equation}
Finally, using the identities
\begin{subequations}
\begin{align}
&\sum_j T_j^2
	= 3(1+u^2), \\
&\sum_j T_j \boldsymbol\sigma T_j
	= 3u^2 \boldsymbol\sigma,\\
&\sum_j T_j (\boldsymbol\sigma \cdot {\bf q}_j) T_j
	= 0,	 \\
&\sum_j T_j (\boldsymbol\sigma \cdot {\bf q}_j) {\bf q}_j T_j
	= \frac32(u^2-1) \boldsymbol\sigma,		
\end{align}
\end{subequations}
the matrix elements up to $\mathcal{O}(|{\bf k}|^2,w^4)$ simplify to
\begin{equation}
\langle {\bf k}_l| h_l^\text{eff} - h_l | {\bf k}_l\rangle \approx - 3  \alpha_{\bar l}^2\left[ (1+u^2) E + \hbar v_{\bar l} \boldsymbol\sigma\cdot{\bf k}_l \right], 
\end{equation}
where $\alpha_l = w/(\hbar k_{\theta} v_{l})$ and we have denoted opposite layers by $l \neq \bar l$.

\begin{figure}
\includegraphics[width=\linewidth]{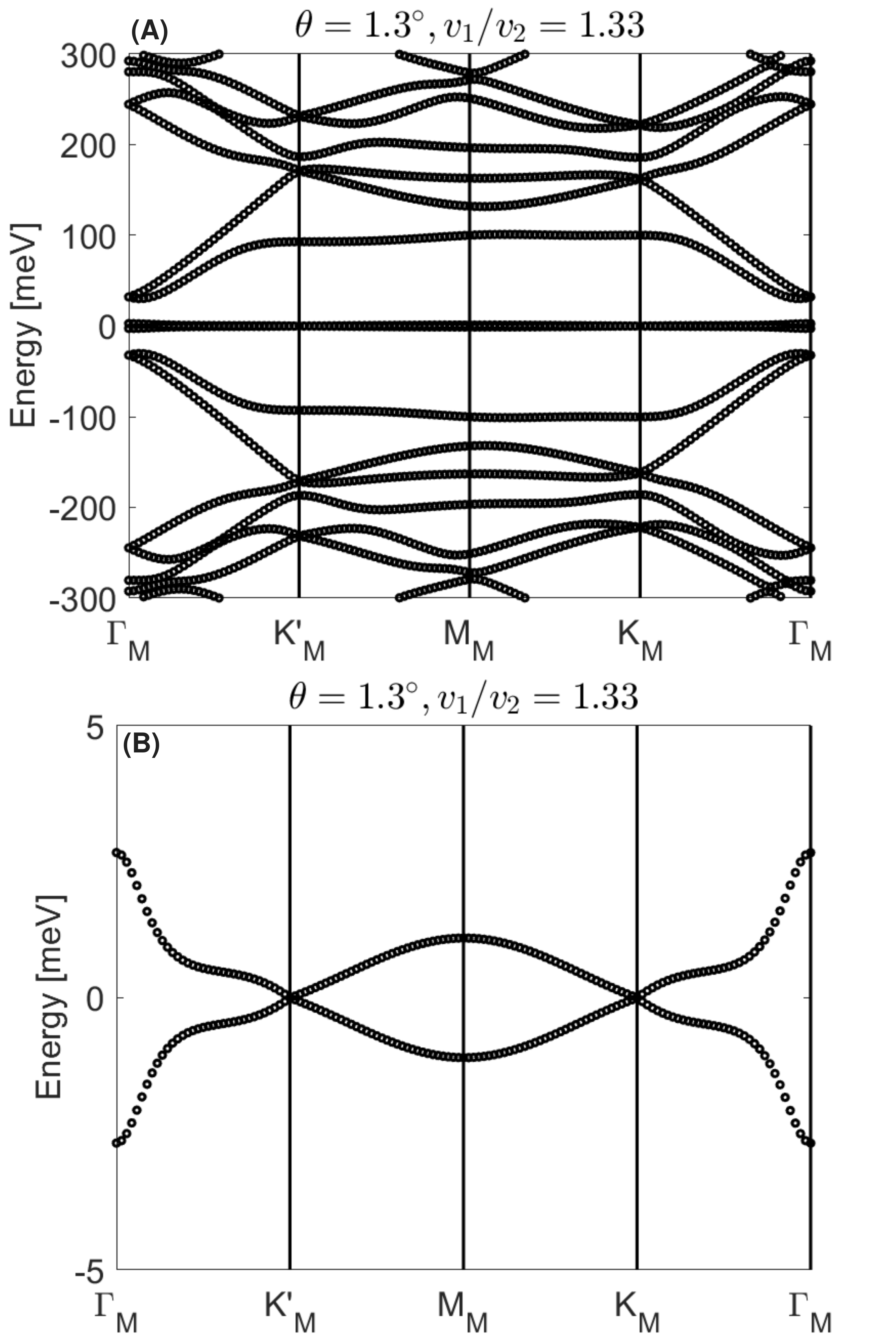}
\caption{(a) Band spectrum for asymmetric TBG with Fermi velocity asymmetry $v_1/v_2 = 1.33$ for twist angle $\theta=1.3^\circ$, which yields nearly flat bands. (b) Detail of low energy band spectrum.}
\end{figure}

Solving for the eigenvalue $E$ of $h_l^\text{eff}(E)$ self-consistently, we find $E = \pm \hbar v'_{Fl} |{\bf k}_l|$ with a renormalized Fermi velocity,
\begin{equation}
v'_{l} = \frac{ v_{l} - 3 \alpha_{\bar l}^2 v_{\bar l}}{1+ 3(1+u^2) \alpha_{\bar l}^2}.
\end{equation}
Thus the renormalized Fermi velocities both vanish when
\begin{equation}
\overline\alpha = \frac1{\sqrt 3}, \quad \overline\alpha \equiv \frac{w}{\hbar k_\theta \sqrt{v_{1}v_{2}}}.
\end{equation}
Thus, within this perturbative analysis,
the ``magic'' angle persists in the presence of Fermi velocity asymmetry between the twisted layers and is set by their geometric mean.

We note here that by defining $\beta_l = 1+3(1+u^2)\alpha_l^2$, we can write (in the plane-wave basis)
\begin{equation}
h_l^\text{eff} \approx (1-\beta_{\bar l}) E + \beta_{\bar l} \hbar v'_{l} \boldsymbol\sigma \cdot {\bf k},
\end{equation}
and the projected resolvent operator takes the form
\begin{equation}
(E-h_l^\text{eff})^{-1} \approx \frac{1}{\beta_{\bar l}}
\left[E - \hbar v'_{l} \boldsymbol\sigma \cdot {\bf k}\right]^{-1}.
\end{equation}
The poles of this operator occur at $E = \pm \hbar v'_l |{\bf k}_l|$ with a residue $1/\beta_{\bar l}$. The square-root of this residue, $1/\sqrt{\beta_{\bar l}}$, signifies the renormalization of the wavefunction amplitude due to projection to layer $l$.

\begin{figure}[t]
        \includegraphics[width=\linewidth]{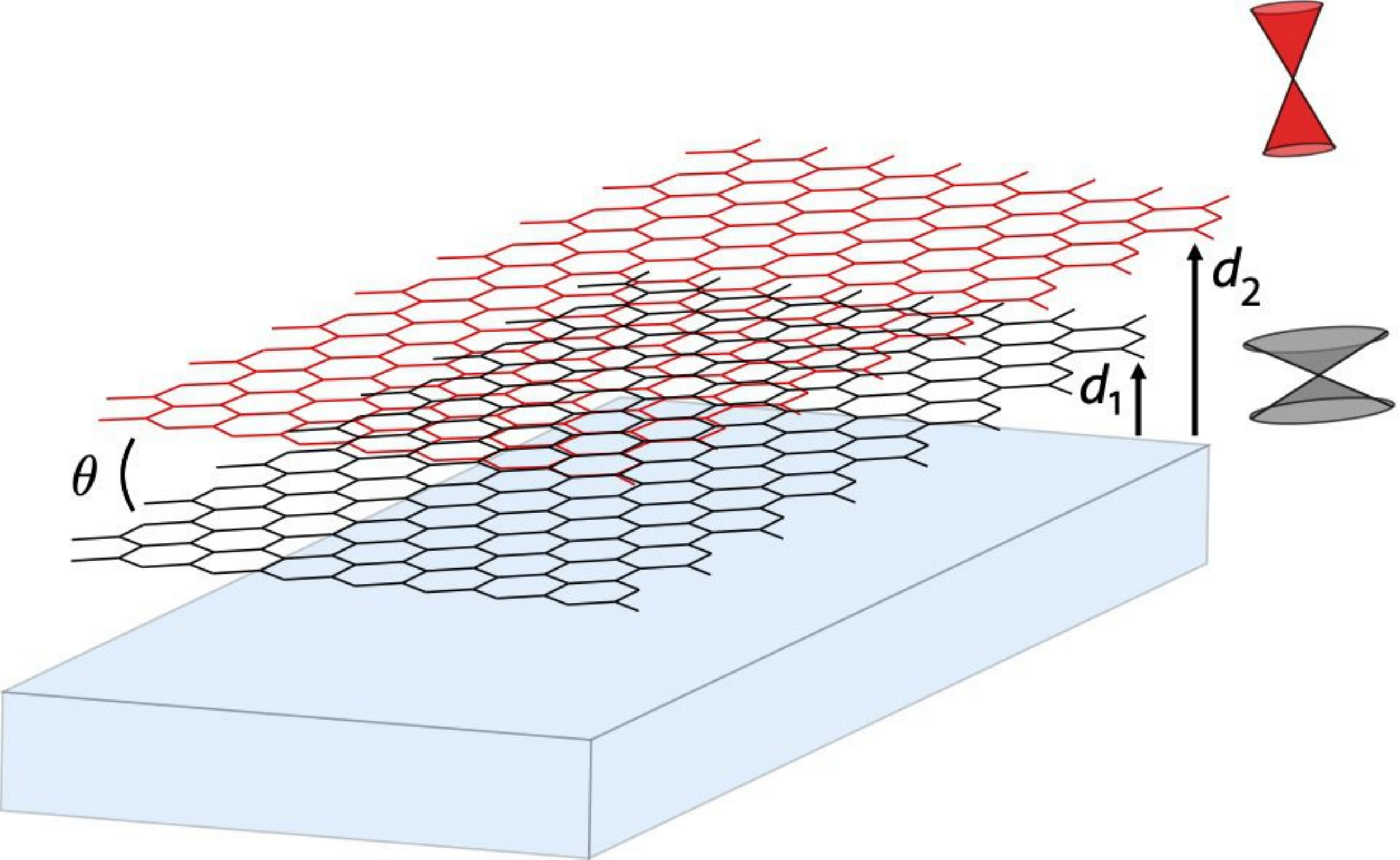}
    \caption{Sketch of the asymmetric TBG system composed of two layers of twisted graphene with a dielectric applied beneath the bottom layer. Dirac cones are shown representing the different Fermi velocities in the two layers.}
    \label{fig:ATBGsketch}
\end{figure}

\subsection{Realization by Asymmetric Dielectric Screening}
We now briefly discuss a mechanism through which different Fermi velocities could be generated for the two layers of a TBG system by exploiting the renormalization of the Fermi velocity via Coulomb interactions~\cite{Gonzalez_1994}. In particular we focus on a situation in which a dielectric layer is present only on one side of the TBG system, as sketched in Fig.~\ref{fig:ATBGsketch}, with $d_1$ and $d_2$ denoting the distances between the dielectric and each of the graphene sheets. For concreteness we take $d_1 < d_2$.  We expect for such geometries $d_2 \approx 2d_1$.

For wavevectors ${\bf k}$ with $|{\bf k}| \gg 2\pi/d_1 \equiv \Lambda_1$ the dielectric will have little effect, while for $k \ll 2\pi/d_2 \equiv \Lambda_1$, dielectric screening is essentially the same for both layers.  We model the difference in dielectric screening between the layers by an effective potential that applies only to the layer closer to the dielectric, of the form
\begin{equation}
\delta V(|{\bf k}|) =
\begin{cases}
\left(\frac1\kappa - \frac1{\kappa_0} \right)\frac{2\pi e^2}{ |{\bf k}|}, & \Lambda_2  < |{\bf k}| < \Lambda_1, \\
 0, &\text{otherwise},
\end{cases}
\end{equation}
where $\kappa_0$ is a dielectric constant due to the intrinsic screening of graphene applying to both layers, and $\kappa$ is the dielectric constant applied to the layer closer to the dielectric.

\begin{figure}[t]
        \includegraphics[width=\linewidth]{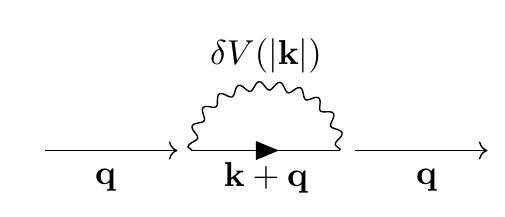}
    \caption{Feynman diagram for the Asymmetric TBG self-energy. Because of the presence of the dielectric, an effective potential difference $\delta V({\bf k})$ contributes to corrections to the propagator in one of the layers. Here the solid line with the arrow is the bare propagator $G_0({\bf k}+{\bf q},i\omega)$.}
    \label{fig:feynman}
\end{figure}

Because $\delta V$ has a cutoff on the low momentum side, we can estimate its effect perturbatively through an exchange self-energy correction, $\Sigma$, to the Matsubara Green's function,
\begin{equation}
G^{-1}({\bf k},i\omega) = G_0^{-1}({\bf k},i\omega) - \Sigma({\bf k}),
\label{eq:dyson}
\end{equation}
which to the lowest order and in $\delta V$ in the zero-temperature limit (see Fig.~\ref{fig:feynman}) has the form~\cite{Tang_2018}
\begin{equation}\label{eq:SigmaForm}
    \Sigma({\bf q}) = \int_{-\infty}^{\infty} \frac{d\hbar\omega}{2\pi}\int\frac{d^2 {\bf k}}{(2\pi)^2} \delta V(|{\bf k}|) G_0({\bf k}+{\bf q},i\omega).
\end{equation}
Here, 
\begin{equation}
G_0(\mathbf{k}, i\omega) 
= -\frac1\hbar\frac{i\omega+ v_F{\bf k} \cdot {\boldsymbol\sigma}}{\omega^2 + v_F^2|\mathbf{k}|^2},
\end{equation}
is the unperturbed Green's function, where we have set the chemical potential to zero so that we work near the charge-neutrality point, $v_F$ is the bare Fermi velocity, and $\boldsymbol{\sigma}=(\sigma_x,\sigma_y)$ are Pauli matrices.

Since we are interested in the renormalization of Fermi velocity, we consider small values of $|{\bf q}|$ in Eq.~\eqref{eq:SigmaForm} while the form of $\delta V$ guarantees that $|{\bf q}| \ll |{\bf k}|$ for non-vanishing values of the integrand. Then, integrating over $\omega$ first and expanding, up to $\mathcal{O}(q^2)$, $1/|{\bf k}+{\bf q}| \approx 1/|{\bf k}| - {\bf k}\cdot{\bf q}/|{\bf k}|^2$, we have
\begin{subequations}
\begin{align}
\Sigma(\mathbf{q}) 
	&= -\frac12\int\frac{d^2 {\bf k}}{(2\pi)^2} \delta V(|{\bf k}|)\frac{({\bf k}+{\bf q}) \cdot {\boldsymbol\sigma}}{|{\bf k}+{\bf q}|} \\
	&\approx -\frac{{\bf q}\cdot\boldsymbol{\sigma}}{8\pi} \int \delta V(k) dk \\
	&= -\left[\frac{e^2}4\left(\frac{1}{\kappa} - \frac{1}{\kappa_0} \right) \ln\frac{\Lambda_2}{\Lambda_1}\right] {\bf q}\cdot\boldsymbol{\sigma}.
\end{align}
\end{subequations}
Using Eq.~\eqref{eq:dyson} one sees that the Green's function retains its non-interacting form, albeit with
a renormalized velocity.  Since this renormalization applies only to the layer closer to the dielectric substrate, the ratio of effective Fermi velocities for the two layers becomes
\begin{equation}
\frac{v_{1}}{v_{2}}=1 + \frac{(\kappa_0-\kappa)e^2}{4\kappa\kappa_0 \hbar v_F}\ln \frac{\Lambda_2}{\Lambda_1}.
\end{equation}
With $\Lambda_2/\Lambda_1 \approx 2$, a very large value of $\kappa$ (as would be appropriate for example to SrTiO$_3$~\cite{Veyrat_2020}), and a background dielectric constant of $\kappa_0=4$, one finds $v_{2}/v_{1} \sim 1.1$.
The Fermi velocities of the two layers of TBG can thus be made different by $\sim 10\%$ due to such one-sided dielectric screening.
Finally, note that in these perturbative corrections we are including a contribution that makes the two Fermi velocities different, but do not include higher order logarithmic corrections due to Coulomb interactions, which cause the Fermi velocities to acquire some momentum dependence ~\cite{Kotov_12}.

\begin{figure}[t]
            \includegraphics[width=\linewidth]{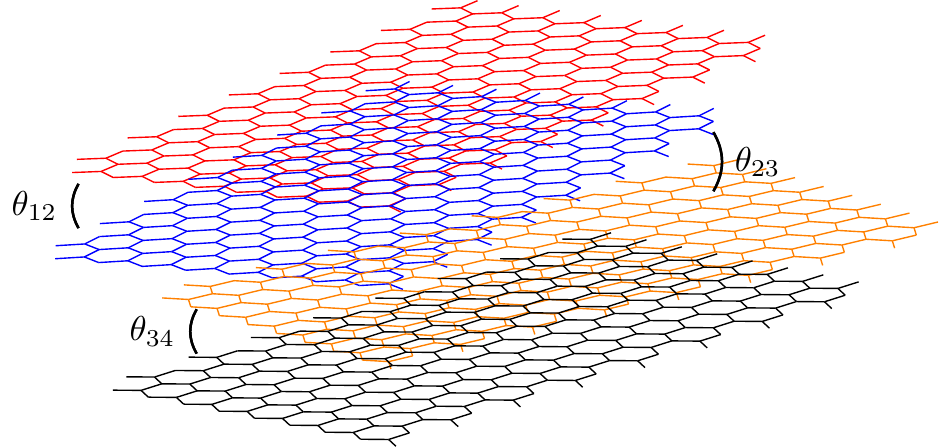}
    \caption{Sketch of the tetralayer graphene system. The angles between the two layers in the top and bottom bilayer are $\theta_{12}$ and $\theta_{34}$, respectively, and the angle between the two bilayers is $\theta_{23}$.}
    \label{fig:4TTGsketch}
\end{figure}

\section{Asymmetric Twisted Tetralayer}\label{sec:4TBG}
A second platform which approximately realizes the asymmetric Dirac point models we consider is a graphene tetralayer with three independent twist angles in which the outer two are not too small, as sketched in Fig.~\ref{fig:4TTGsketch}. The idea is that at such relatively large twist angles, the main effect of the outer twists is to renormalize the Fermi velocities of the inner layers, which in turn would implement the asymmetric twisted bilayer discussed above at smaller inner twist angles. The renormalized Fermi velocity asymmetry found increases when $\theta_{12}$ and $\theta_{34}$ are significantly different (while neither one is too small). For example, taking $\theta_{12} = 2.5^\circ$ and $\theta_{34} = 10^\circ$, we find $v'_{3}/v'_{2} \approx 1.57$.

 We shall model this system systematically below and provide perturbative estimates as well as numerical results for its spectra.

\subsection{Hamiltonian}\label{sec:4LHamiltonian}
The Hamiltonian for the system as a whole may be written as
\begin{equation} \label{fullH}
H =
\begin{bmatrix}
	H_\text{TBG}(\theta_{12}) & T_{23}({\bf r}) \\
	T_{23}^{\dagger}({\bf r}) & H_\text{TBG}(\theta_{34}) \\
\end{bmatrix},
\end{equation}
where $H_\text{TBG}(\theta_{ij})$ is the Hamiltonian~\cite{BM} for the bilayer $ij$ with twist angle $\theta_{ij}$,
\begin{equation}
H_\text{TBG}(\theta) =
\begin{bmatrix}
	h_+ & T({\bf r}) \\
	T^\dagger({\bf r}) &  h_- \\
\end{bmatrix},
\end{equation}
with $h_\pm = \hbar v \boldsymbol{\sigma}_{\pm\theta/2} \cdot \left(-i\boldsymbol{\nabla}\mp{\bf q}_0/2 \right)$ the Hamiltonian in each layer, $\boldsymbol{\sigma}_{\theta/2} = e^{-i\theta\sigma_z/4} \boldsymbol{\sigma} e^{i\theta\sigma_z/4}$,
$T({\bf r}) = w\sum_{j = 0}^2 \exp(-i {\bf Q}_j \cdot {\bf r}) T_j$ as before, and $T_{23}$ implements tunneling between the two bilayers by coupling layers 2 and 3.

Solving for the spectrum of the full Hamiltonian $H$ in general is very challenging, in particular because for an arbitrary set of twist angles the system is not spatially periodic.  For our purposes we are interested in parameter regimes in which there is approximate spatial periodicity, and in which the twist angles $\theta_{12}$ and $\theta_{34}$ are exploited to create Dirac points with different velocities, which can be coupled together to form an approximate moir{\'e} lattice. We note that in principle there are deviations from perfect discrete translational symmetry because, for general twist angles, tunneling may be accompanied by scattering by many different discrete wavevectors.  Ref.~\onlinecite{BM} demonstrated that for a single twisted graphene bilayer, the scattering involved is dominated by just two wavevectors and their linear combinations, so that the resulting bands fall in a two-dimensional Brillouin zone.  In the four-layer systems we consider, the outer two layers have relatively large twist angles compared to their neighbors, so that their single particle states near zero energy are well-approximated by a single plane wave.  This allows us to adopt the BM strategy for tunneling between the two middle layers.  We discuss in more detail below the justification for this, and in Appendix B estimate the effect of retaining plane wave states not included in our basic approach. Indeed, we find their effect to be quite small provided the outer twist angles are not too small.
%
%

In general, the Hamiltonians $H_\text{TBG}(\theta_{12})$ and $H_\text{TBG}(\theta_{34})$ in Eq.~\eqref{fullH} host Dirac points associated with each of their valleys, and the two degenerate states of those Dirac points reside mostly in one of the two members of the bilayer. Out of the four Dirac points hosted by (a single valley) of the two bilayers, we focus on those with the most weight in layers 2 and 3, respectively, and
model the diagonal components of Eq.~\eqref{fullH} using a ${k} \cdot {p}$ approximation.  Note that the remaining two Dirac points are remote in wavevector from low energy states in the opposite bilayer, and so are largely decoupled from states of the two Dirac points we retain.
This yields a simple linearly dispersing mode near each Dirac point with some Fermi velocity,
as well as wavefunctions associated with eigenstates.  We can then use these dispersive states to create a model for tunneling between the bilayers, as we now explain.

\subsection{Interbilayer Tunneling} \label{subsection: Interbilayer Tunneling}
To formulate the interbilayer tunneling, in analogy with Ref. \onlinecite{BM} we begin by calculating the matrix element $\langle {\bf k} \mu | H | {\bf k}' \mu' \rangle$ where ${\bf k}$ is the wavevector for an electron state and $\mu$ and $\mu'$ are indices labeling positive and negative energy states of a Dirac cone in bilayer 12 and 34, respectively. To compute these matrix elements we need wavefunctions for the states in the uncoupled bilayers, which in the BM model take the approximate form
\begin{equation}
\psi^{(12),\mu}_{\bf k} ({\bf r}) \propto \sum_{{\bf g}} e^{i{\bf g} \cdot {\bf r}}
    \begin{bmatrix}
    a^\mu_1({\bf g}) \\
    b^\mu_1({\bf g})e^{i({\bf g}+{\bf k})\cdot{\boldsymbol \tau}_1} \\
    a^\mu_2({\bf g})e^{-i({\bf g}+{\bf k})\cdot{\boldsymbol \tau}_2} \\
    b^\mu_2({\bf g}) \\
    \end{bmatrix}
    e^{i {\bf k} \cdot {\bf r}},
\label{eq:wavefunctions}
\end{equation}
for the 12 bilayer, and similarly for the 34 bilayer.
In this expression, $a_j^{\mu}$ denotes an amplitude on the $A$ sublattice of sheet $j = 1,2$, and $b_j^{\mu}$ is the corresponding amplitude on the $B$ sublattice.  The wavevectors ${\bf g}$ depend on the twist angle and are in general different for the 12 and 34 bilayers.  The values of $a_j^{\mu}({\bf g})$ and $b_j^{\mu}({\bf g})$ are determined by numerically solving the BM model for the isolated twisted bilayer.  Finally the vectors ${\boldsymbol \tau}_{1}$ and ${\boldsymbol \tau}_{2}$ denote the separations of different sublattice atoms within a unit cell of sheet $j$ of the bilayer. Note that for the $12$ and $34$ bilayers, our tunneling matrices are those in Ref.~\onlinecite{Balents},
which correspond to choices of ${\boldsymbol \tau}_{1}$ and ${\boldsymbol \tau}_{2}$ that lead to AA stacking in the zero twist angle limit.

Within this model, the tunneling matrix element takes the form
\begin{align}
\langle {\bf k} \mu | H | {\bf k}' \mu' \rangle
	&= \frac{\Omega_c}{\Omega}\sum_{{\bf R},{\bf R}'} t({\bf R} - {\bf R}') \sum_{{\bf g}, {\bf g}'} f^{\mu\mu'} ({\bf g}, {\bf g}',{\bf k},{\bf k}') \nonumber \\
	&\quad\times  e^{-i({\bf k}+{\bf g}) \cdot {\bf R}}
     e^{i({\bf k}'+{\bf g}') \cdot {\bf R}'}
     \label{matrix element}
\end{align}
where ${\bf R},{\bf R}'$ are Bravais lattice sites for sheets 2 and 3, respectively, $\Omega_c$ is a primitive unit cell area for the graphene Bravais lattice, and  $\Omega$ is the system area.  A tunneling amplitude $t({\bf R}-{\bf R}')$ has been introduced, which is assumed to depend only on the lateral separation ${\bf R}-{\bf R}'$ between points in different sheets~\cite{BM}, and
\begin{widetext}
\begin{align}
    f^{\mu\mu'} ({\bf g}, {\bf g}',{\bf k},{\bf k}') = 
    \left[a^\mu_1({\bf g}),
    b^\mu_1({\bf g})e^{i({\bf g}+{\bf k})\cdot{\boldsymbol \tau}_1},
    a^\mu_2({\bf g})e^{-i({\bf g}+{\bf k})\cdot{\boldsymbol \tau}_2} ,
    b^\mu_2({\bf g}) \right]^*
     M
    \begin{bmatrix}
    a^{\mu'}_3({\bf g}') \\
    b^{\mu'}_3({\bf g}')e^{i({\bf g}'+{\bf k}')\cdot{\boldsymbol \tau}_3}  \\
    a^{\mu'}_4({\bf g}')e^{-i({\bf g}'+{\bf k}') \cdot{\boldsymbol \tau}_4}  \\
    b^{\mu'}_4({\bf g}')
    \end{bmatrix}.
\end{align}
\end{widetext}
Here ${\bf g}$ and ${\bf g}'$ are reciprocal lattice vectors for the 12 and 34 bilayers, respectively, and $M$ is a $4\times4$ matrix that describes the tunneling between bilayers 12 and 34.
Following Ref. \onlinecite{BM}, we express the amplitude $t$ in terms of its Fourier transform, and after summing over the lattice sites one finds
\begin{align}
\langle {\bf k} \mu | H | {\bf k}' \mu' \rangle
	= \sum_{{\bf G}, {\bf G}'}\sum_{{\bf g}, {\bf g}'}\ & t({\bf k}+{\bf g}+{\bf G})  f^{\mu\mu'} ({\bf g}, {\bf g}',{\bf k},{\bf k}') \nonumber \\
	&\quad \times   \delta_{{\bf k}+{\bf g}+{\bf G}, {\bf k}' + {\bf g}'+{\bf G}'}.
\end{align}
Here the vectors ${\bf G}$ and ${\bf G}'$ correspond to the reciprocal lattice vectors of the two inner graphene sheets, 2 and 3, respectively.

We next adopt two simplifications which limit the values of twist angles for which our analysis gives a reasonable approximation.  Firstly, we note that when the angles $\theta_{12}$ and $\theta_{34}$ are not too small, the overlaps $f^{\mu \mu'} ({\bf g}, {\bf g}',{\bf k},{\bf k}')$ are sharply peaked at ${\bf g}={\bf g}'=0$.   This is discussed in more detail in Appendix~\ref{app:wave}.  Exploiting this feature allows one to set ${\bf g}={\bf g}'=0$.  This is a crucial simplification because retaining further values of ${\bf g},{\bf g}'$ spoils the spatial periodicity of the system, rendering it a quasicrystal~\cite{Kaxiras}.

\begin{figure}[t]
            \includegraphics[width=\linewidth]{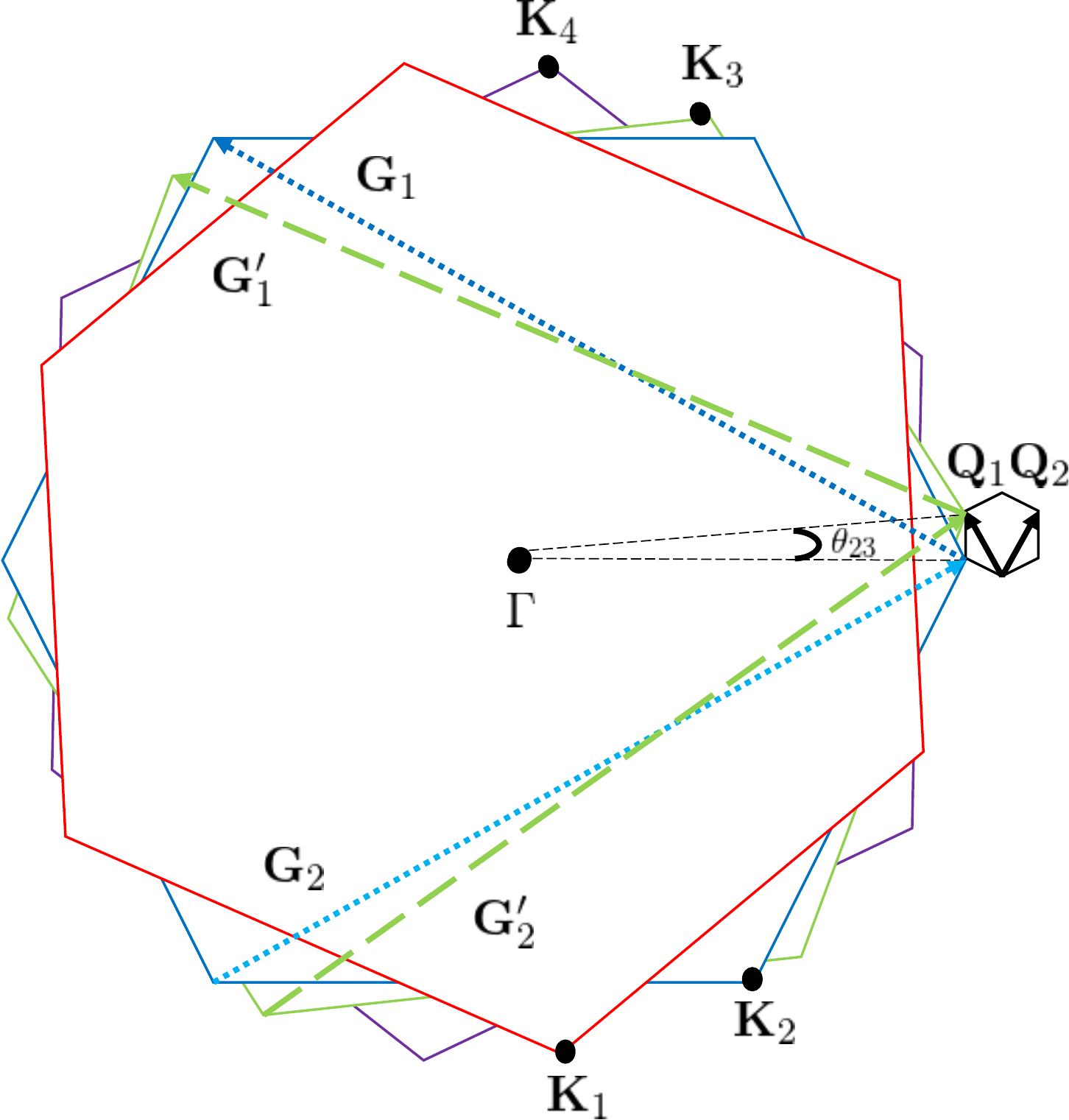}
    \caption{Geometry of reciprocal lattice vectors relevant to tunneling matrix elements in the tetralayer graphene system. The angle between the two bilayers  ($\theta_{23}$) is enlarged for clarity; for the small angles $\theta_{23}$ we consider in this work, $|\mathbf{G}_i| \gg |\mathbf{Q}_j|$ for all $i,j$. }
    \label{fig:geometry}
\end{figure}

The second simplification is commonly made for the BM model. In the expected situation where the distance between layers is larger than the graphene lattice constant, $t({\bf q})$ vanishes very rapidly for $|{\bf q}|$ larger than the inverse of the spacing between the two sheets.  Moreover, we are interested in the bands near zero energy, for which the values of ${\bf k}$, ${\bf k}'$ lie in the vicinity of Dirac points of the 2 and 3 layers.  We thus focus on values of ${\bf k}+{\bf G} = {\bf k}'+{\bf G}'$
in the vicinity of ${\bf K}_2$, the $K$ point of layer 2, which (assuming small $\theta_{23}$) are also near ${\bf K}_3$, a $K$ point of layer 3. On the scale of the Brillouin zone of a single graphene sheet, the set of wavevectors coupled together by $\langle {\bf k} \mu | H | {\bf k}' \mu' \rangle$ in the low energy bands are very close together, so we ignore the small wavevector variations in $t({\bf k}+{\bf G})$, and retain only values of ${\bf G}$ such that $\bf k+{\bf G}$ is near a Dirac point for the two inner layers, and for which ${\bf K_2}+{\bf G}$ has the smallest possible value.  There are three such choices for ${\bf G}$, and for all of them $t({\bf K_2}+{\bf G})$  has the same value $t$; other choices of ${\bf G}$ yield values for $t({\bf k}+{\bf G})$ which are negligibly small.  Thus in our reciprocal lattice sum we retain only ${\bf G}={\bf G}_{0,1,2}$, with ${\bf G}_0 = 0, {\bf G}_1 = k_D(-\frac32,\frac{\sqrt3}2), {\bf G}_2= k_D(\frac32,\frac{\sqrt3}2) $. In other words, we take $t({\bf k}+{\bf G}) \approx t({\bf G})$.
Furthermore, because the reciprocal lattice vectors of a single sheet are very large compared to the scale of a small-angle twisted bilayer mBZ, for each ${\bf G}_j, j=0,1,2$, we retain only a single ${\bf G}'_j={\bf G}_j+{\bf Q}_j$: the other combinations 
couple together states with very large single particle energy differences, which will have little effect on the bands near zero energy.  
{A sketch of the geometry with the relevant wavevectors is shown in Fig.~\ref{fig:geometry}}.

With this reasoning, the tunneling matrix element we adopt takes the form
\begin{equation}
\langle{\bf k}\mu|H|{\bf k}'\mu'\rangle
	= \frac{t}{\Omega} \sum_{j} f^{\mu\mu'}(0,0,{\bf K}_2+{\bf G}_j,{\bf K}_3+{\bf G}_j^{\prime}) \delta_{{\bf k}-{\bf k}',{\bf Q}_j}.
\end{equation}
The resulting system is now formally very similar to the BM model.

Finally we must choose a concrete form for the matrix $M$ entering the $f^{\mu\mu'}$ factors. To do this, we first note that tunneling between remote sheets is much smaller in amplitude that that between neighboring sheets, so we retain non-zero matrix elements only for the $2 \times 2$ block the connects sheets 2 and 3.  A natural choice is then $M_{23} = \mathbb{1}+\sigma_x$, since there is no distinction between atoms of the two sublattices in graphene beyond their locations in the unit cell, which are explicitly taken into account in the wavefunctions, Eq.~\eqref{eq:wavefunctions}. With this choice, we arrive at our model for tunneling between the bilayers,
\begin{equation}
\langle {\bf k} \mu | H | {\bf k}' \mu' \rangle
	= \dfrac{t}{\Omega} \sum_j f^{\mu\mu'}_j \delta_{{\bf k}-{\bf k}', {\bf Q}_j},
\label{MatrixElement}
\end{equation}
where
\begin{subequations}\label{eqn: fj}
\begin{align}
f^{\mu\mu'}_{0}&=[a_2^{\mu}(0)+b_2^{\mu}(0)]^*[a_3^{\mu'}(0)+b_3^{\mu'}(0)],\\
f^{\mu\mu'}_{1}&=[a_2^{\mu}(0)e^{-i\phi}+b_2^{\mu}(0)]^*[a_3^{\mu'}(0)+b_3^{\mu'}(0)e^{i\phi}],\\
f^{\mu\mu'}_{2}&=[a_2^{\mu}(0)e^{i\phi}+b_2^{\mu}(0)]^*[a_3^{\mu'}(0)+b_3^{\mu'}(0)e^{-i\phi}],
\end{align}\end{subequations}
with $\phi=2\pi/3$. The constants $a_2^{\mu}(0), b_2^{\mu}(0), \cdots$ are found by numerically obtaining the bilayer wavefunction by diagonalizing the Bistritzer-MacDonald model Hamiltonian~\cite{BM} for the individual 12 and 34 bilayers at their Dirac points.

Thus, in terms of the matrices $f_j$ defined in Eqs.~\eqref{eqn: fj}, we have
\begin{equation}
    T_{23}(\mathbf{r}) = w \sum_{j = 0}^3 f_j \exp(-i\mathbf{Q}_j \cdot \mathbf{r}).
    \label{Tmumu}
\end{equation}
Note that in the limit where layers 2 and 3 are coupled to one another but not to layers 1 and 4, the matrices $f_{j}$ become precisely the same as the tunneling matrices $T_{j+1}$ in Ref. \onlinecite{BM},
which differ slightly from what was used for the (12) and (34) bilayers as described above \cite{Balents}.
This corresponds to adopting values of ${\boldsymbol \tau}_{i,j}$, the displacements of the two atoms in sheets $i$ and $j$ that are tunnel coupled, which differ in the two cases: in the zero twist angle limit, the 12 and 34 displacements correspond to AA stacking, while in the 23 case they correspond to AB stacking.  However for non-zero twist angles, the local alignment varies among all possibilities, so that other possible choices for untwisted layer alignment should not qualitatively change our results.

\subsection{Perturbation Theory}\label{section:PT}
In this section
we use a low-energy perturbation theory in the interlayer tunneling to estimate the Fermi velocity at a Dirac point of the mBZ in our tetralayer model, and look for situations in which it vanishes, as an indicator for the flat bands~\cite{BM}.
For the tetralayer system,
our starting Hamiltonian has the form
\begin{equation}    \label{Full Hamiltonian}
H =
\begin{bmatrix}
    h_{1} & T_{12} & 0 & 0 \\
    T_{12}^\dagger & h_2 & T_{23}  & 0 \\
    0 & T_{23}^\dagger & h_3 & T_{34}  \\
    0 & 0 & T_{34}^\dagger & h_4  \\
\end{bmatrix}
\end{equation}
where $h_l$ are the Hamiltonians for layer $l$ and $T_{ll'}$ are the tunneling matrices between layers $l$ and $l'$.
Projecting the resolvent operator into the subspace of layers 2 and 3, we can write energy-dependent effective Hamiltonians in the vicinity of the Dirac points at ${\bf K}_2$ and ${\bf K}_3$ respectively in the forms
\begin{align}
h_2^\text{eff}(E)
	&= \tilde h_2(E) + T_{23}\left[ E - \tilde h_3(E) \right]^{-1} T_{23}^\dagger, \\
h_3^\text{eff}(E)
	&= \tilde h_3(E) + T_{23}^\dagger \left[ E - \tilde h_2(E) \right]^{-1} T_{23},
\end{align}
where, $\tilde h_2 = h_2 + T_{12}^\dagger g_1 T_{12}$ and $\tilde h_3 = h_3 + T_{34}g_4T_{34}^\dagger$.

Following the derivation in Sec.~\ref{sec:per} we may write
\begin{align}
\tilde h_{2}(E) &\approx (1-\beta_1)E + \beta_1 h'_{2} \\
\tilde h_{3}(E) &\approx (1-\beta_4) E + \beta_4 h'_{3},
\end{align}
where $h'_{2} = \hbar v'_{2} \boldsymbol\sigma \cdot {\bf k}_2$ for ${\bf k}_2$ measured from the ${\bf K}_2$ and $h'_{3} = \hbar v'_{3} \boldsymbol\sigma \cdot {\bf k}_3$ for ${\bf k}_3$ measured from the ${\bf K}_3$. Here, $\beta_l \equiv [1+3(1+u^2)\alpha_l^2]$ and the renormalized Fermi velocities in layers 2 and 3 are $v'_{2} = (1-3\alpha_1^2) v / \beta_1$, $v'_{3} = (1-3\alpha_4^2) v / \beta_4$.
This yields
\begin{align}
h_2^\text{eff}(E)
	&\approx (1-\beta_1) E + \beta_1 h_2' + \beta_4^{-1} T_{23} g'_3(E) T_{23}^\dagger, \\
h_3^\text{eff}(E)
	&\approx (1-\beta_4) E + \beta_4 h_3' + \beta_1^{-1} T_{23}^\dagger g'_2(E) T_{23},
\end{align}
where $g'_2(E) = \big(E - h'_2 \big)^{-1}$ and $g'_3(E) = \big(E - h'_3 \big)^{-1}$.

The analysis may be straightforwardly generalized to examine situations in which the tunneling amplitude between layers 2 and 3 is different that between the other layers.  Assuming $T_{23}$ has the same form as the tunneling in the BM model with a multiplicative factor $z$ and solving for the eigenvalue $E$ self-consistently, we find
\begin{align}
v^\text{eff}_2
	&= \frac{\beta_1\beta_4 v'_2 - 3z^2 (\alpha'_3)^2 v'_3}{z\beta_1\beta_4 + 3(1+u^2)(\alpha'_3)^2}, \\
v^\text{eff}_3
	&= \frac{\beta_1\beta_4 v'_3 - 3z^2 (\alpha'_2)^2 v'_2}{z\beta_1\beta_4 + 3(1+u^2)(\alpha'_2)^2},	
\end{align}
with $\alpha'_2 = w/(\hbar k_{\theta_{23}} v'_{2})$ and $\alpha'_3 = w/(\hbar k_{\theta_{23}} v'_{3})$. Both of these effective Fermi velocities vanish when
\begin{equation}
\frac{zw/\sqrt{\beta_1\beta_4}}{\hbar k_{\theta_{23}} \sqrt{v'_2 v'_3} } = \frac1{\sqrt{3}}. \label{eq: PT_z}
\end{equation}
The structure of this condition can be understood intuitively as follows. The factor $z$ is the ratio of the \emph{bare} tunneling amplitude between layers 2 and 3 with $w$, which is the tunneling amplitude in the bilayers 12 and 34. The effective tunneling between layers 2 and 3 is modified by the wavefunction renormalization factors $1/\sqrt{\beta_1}$ and $1/\sqrt{\beta_4}$, which generically reduce it due to the projection of the wavefunctions to layers 2 and 3, respectively. Because of the renormalizations, the final magic angle is dependent on all three twist angles. The dependence on $\theta_{23}$ is explicit, and by varying $\theta_{12}$ or $\theta_{34}$ one will change the $v'_{2}$ and $v'_{3}$, respectively. 
We note that Eq.~\eqref{eq: PT_z} may be rewritten as
\begin{equation}
    \sqrt{\alpha^2_1 + \alpha^2_4 + z^2\alpha^2_{23}} = \dfrac{1}{\sqrt{3}}, \label{eq: PT_noz}
\end{equation}
where $\alpha_{23}=w/(\hbar k_{\theta_{23}}v_F)$, with $v_F$ the Fermi velocity of a single graphene sheet. This magic-angle condition holds for both positive and negative twist angles.

We observe that the Fermi velocity drops to zero within the perturbative analysis for both Dirac point simultaneously, so that one does not end up with two closely spaced angles with approximately flat bands. Given that the Fermi velocities of the two uncoupled Dirac points are different, it is not obvious that this should happen, and as discussed in Appendix~\ref{app:incomm}, inclusion of incommensuration effects may change this result.

\begin{figure}
\includegraphics[scale=0.315]{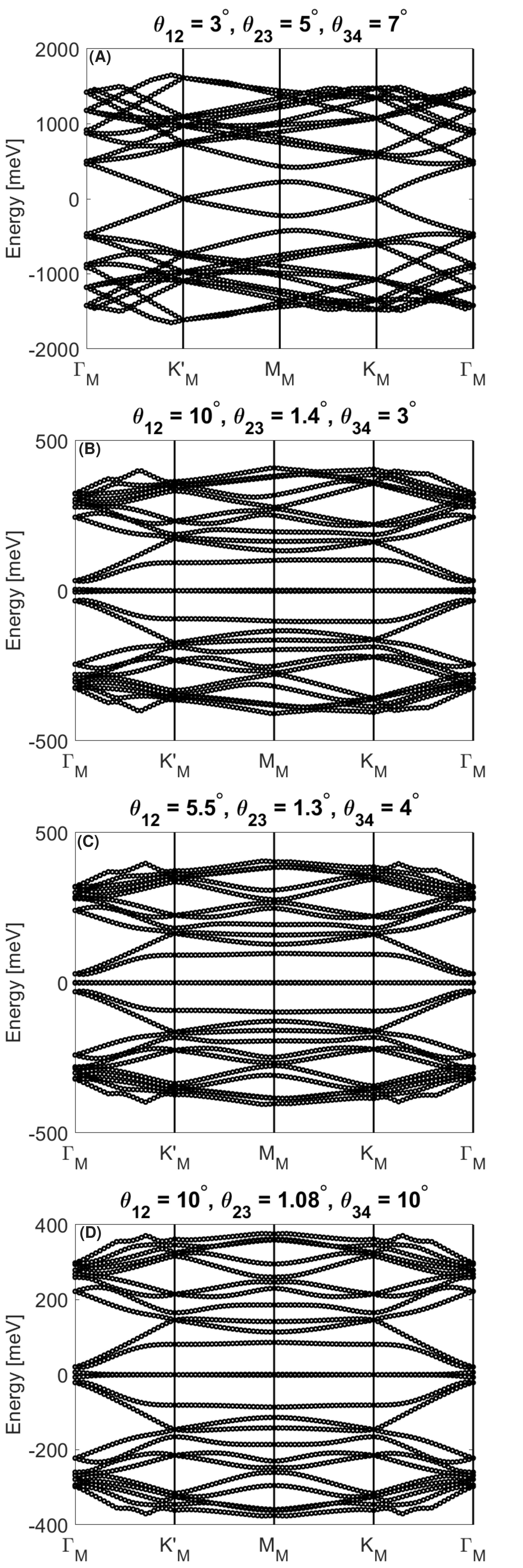}
\caption{Band spectra for twisted tetralayer graphene for combinations of angles that (a) do not yield flat bands; (b), (c), and (d) support flat bands. For sufficiently large $\theta_{12}$ and $\theta_{34}$ the magic $\theta_{23}$ approaches $1.08^\circ$, the magic angle of TBG.}
\label{fig:spectra}
\end{figure}

\subsection{Numerical Results}\label{sec:numerics}
We begin by showing numerical band structure results for a
representative triplet of twist angles $\theta_{12}$, $\theta_{23}$ and $\theta_{34}$ in Fig. \ref{fig:spectra}(a).
The calculations are performed by expanding Eq.~\eqref{HATBG} in plane waves, with $h_1$ and $h_2$ taken as the ${k}\cdot {p}$ approximations to the Hamiltonians near the relevant Dirac points of the 12 bilayer and 34 bilayer, respectively (obtained by numerically solving the BM model for each of these bilayers individually), and the off-diagonal tunneling operator is given by Eq.~\eqref{Tmumu}.
In all these calculations, the tunneling parameter $w$ is taken to be 110meV between each pair of neighboring layers, which is equivalent to $z=1$ in the perturbative analysis above. Notice that because $\theta_{12} \not = \theta_{34}$ there is asymmetry between the two valleys. Nevertheless, magic angles still occur in our model of the twisted tetralayer graphene system, and they manifest themselves in a qualitatively similar way to TBG, see Fig. \ref{fig:spectra}(b) and \ref{fig:spectra}(c).

\begin{figure}
  \includegraphics[width=\linewidth]{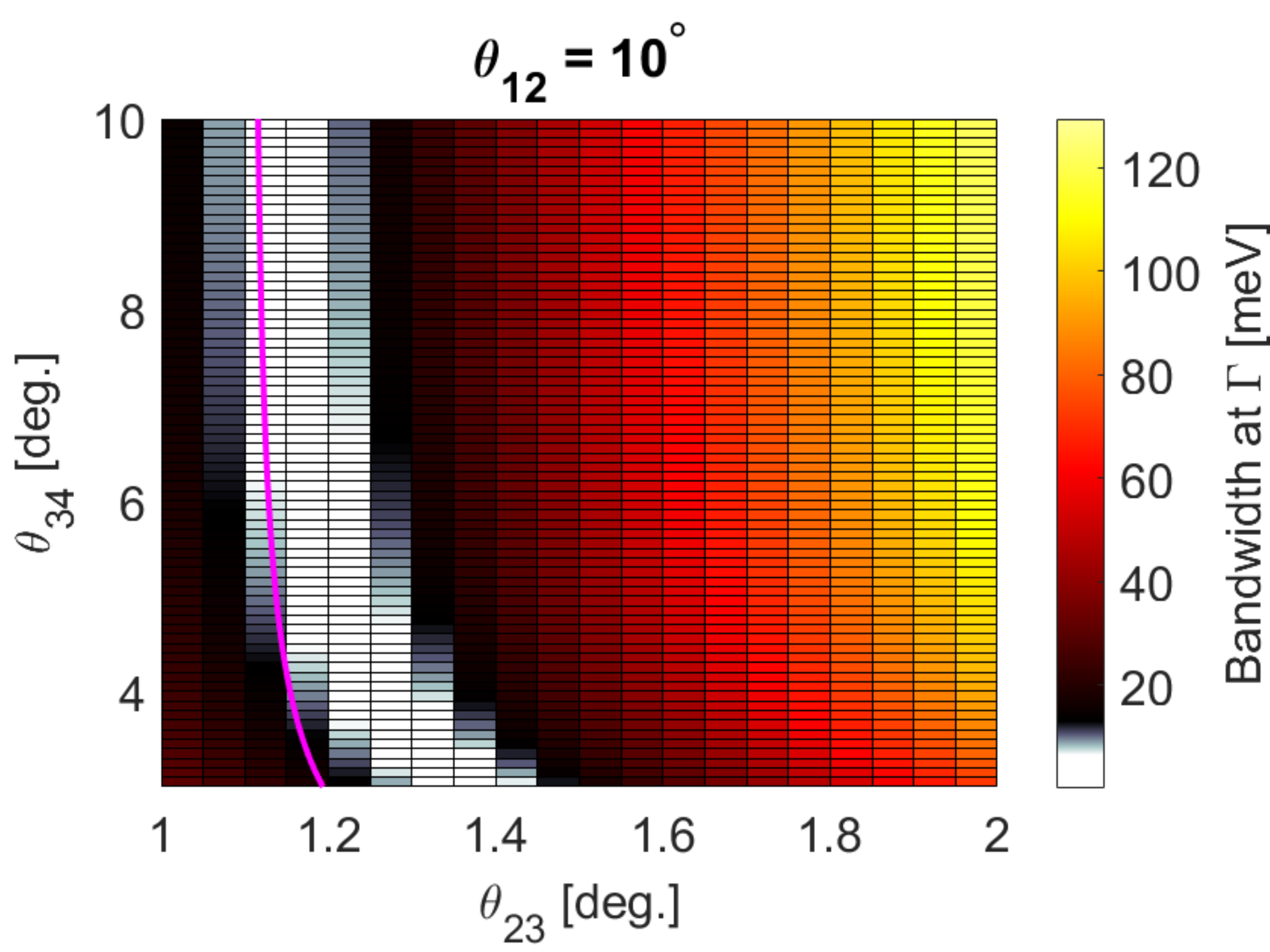}
\caption{Locations of the magic angles for twisted tetralayer graphene at fixed $\theta_{12} = 10^\circ$ and $\theta_{34}\in(3^\circ, 10^\circ), \theta_{23}\in(1^\circ, 2^\circ)$. Locations where the bandwidth is less than 10 meV are shown in white. The pink line shows the theoretical prediction given by Eq.~\eqref{eq: PT_noz}.}
\label{fig:bandwidth10}
\end{figure}

An interesting feature of this model is that, analogously to the unequal Fermi velocity system discussed above,  the system hosts flat bands for $\theta_{23}$ at different ``magic'' values, depending on the angles $\theta_{12}$ and $\theta_{34}$.  This is in contrast to
TBG, for which the twist angle for the primary magic angle is fixed at $\theta \approx 1.08^\circ$. Figures \ref{fig:spectra}(b) and \ref{fig:spectra}(c) show examples of this: the combinations of the twist angles are different for the pairs of figures, yet both sets of parameters produce flat bands. In general, magic angles will occur when $\theta_{23}$ is somewhat larger than the TBG magic angle, but for large $\theta_{12}$ and $\theta_{34}$ the first magic angle for $\theta_{23}$ converges to the TBG magic angle $1.08^{\circ}$. A bandstructure corresponding to this situation is shown in \ref{fig:spectra}(d).

Fig. \ref{fig:bandwidth10} shows a plot of the bandwidth of the lowest energy bands for the special case where $\theta_{12} = 10^\circ$.  Here we define the bandwidth as half the gap between the states of positive and negative energy closest to zero at the $\Gamma_M$ point ($\Gamma$ point of the mBZ), which typically has the widest separation between the two flat bands. As can be seen from the plot, the bandwidth is minimized for a continuum of twist angles. An important feature of this system in general, and in this example in particular, is the perfect swapping symmetry between $\theta_{12} \leftrightarrow \theta_{34}$:  Fig. \ref{fig:bandwidth10} appears identical when $\theta_{34}$ is fixed at $3^\circ$ and $\theta_{12}$ is varied over the same region of the parameter space.
More generally, we find that when $\theta_{12}$ and $\theta_{34}$ are not too small, the values of the angles at which we find flat bands adhere to Eq.~\eqref{eq: PT_noz} relatively well.  

An interesting observation about this behavior is that it is rather similar to that found in twisted trilayer systems, for example in Ref.~\onlinecite{Kaxiras}.  With a relatively large twist angle $\theta_{12}$, the Dirac point coming from this bilayer has little renormalization, so that it can be viewed as coming from an isolated graphene sheet.  The two relevant twist angles are then $\theta_{23}$ and $\theta_{34}$.  One can see in Fig.~\ref{fig:bandwidth10} that for large $\theta_{34}$ the flat band occurs when $\theta_{23}$ approaches the magic angle of a single twisted bilayer, while for smaller values of $\theta_{34}$, we find the flat band condition moves to larger values of $\theta_{23}$, precisely as found in Ref.~\cite{Kaxiras}.  Moreover, in the trilayer one loses the flat band behavior when both angles are smaller than $\sim 3^{\circ}$, which is precisely the situation in which we find results in our own approach to become unreliable.

\begin{figure}
  \includegraphics[width=\linewidth]{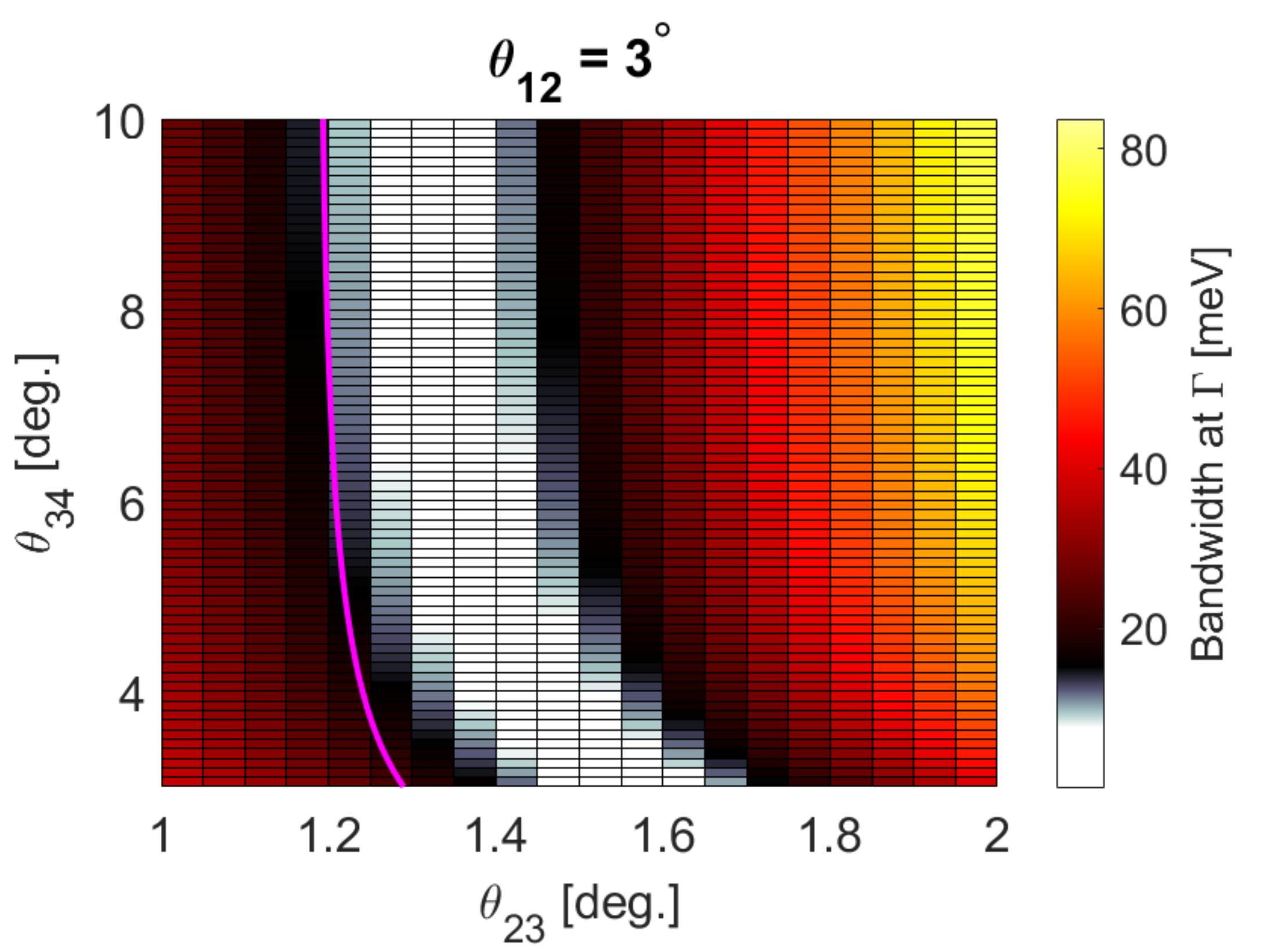}
\caption{Locations of the magic angles for twisted tetralayer graphene at fixed $\theta_{12} = 3^\circ$ and $\theta_{34}\in(3^\circ, 10^\circ), \theta_{23}\in(1^\circ, 2^\circ)$. Locations where the bandwidth is less than 10 meV are shown in white. The pink line shows the theoretical prediction given by Eq.~\eqref{eq: PT_noz}.}
\label{fig:bandwidth3}
\end{figure}

Fig.~\ref{fig:bandwidth3} shows correponding results for a situation in which the twist angle which is being held constant is much smaller than in Fig.~\ref{fig:bandwidth10}. The result is that the perturbative result is less faithful in matching the numerics. This is unsurprising since we expect our method to become increasingly unreliable as the two outer twist angles are made smaller and smaller.


\section{Summary and Discussion}\label{sec:sum}
In conclusion, we have introduced a model of twisted bilayer graphene in which the Fermi velocities of the Dirac points of each layer may be different. We have demonstrated that generically this asymmetry does not spoil the ``magic'' flat band phenomenon.   We argued that such models are relevant for systems with asymmetric screening, for which there are unequal interaction renormalizations of the Fermi velocities, and for tetralayer systems, when the main effect of the outermost layers is a slowing of the Fermi velocities of the Dirac points associated with the two inner layers.  This situation is realized when the outermost twist angles, $\theta_{12}$ and $\theta_{34}$, are not too small.  A perturbative analysis for the Fermi velocity of Dirac points of the fully coupled systems explains the locations of the flat bands under certain conditions, and interestingly shows that for both Dirac points this vanishes at the same twist angle ($\theta_{23}$ for the tetralayer).  Our numerical results also support the existence of a single minimum bandwidth as a function of twist angle for this system.

For the tetralayer system, open questions remain on the impact of the formal incommensuration between the moir{\'e} lattices of the outer pairs of layers relative to the moir{\'e} lattice associated with the inner pair. In Appendix~\ref{app:incomm} we study the impact of retaining a subset of the incommensurate reciprocal lattice vectors $\bf{g}$ and $\bf{g}'$ that define the outer moir{\'e} lattices.  Specifically we use degenerate perturbation theory to calculate the correction to the energy (accurate to first-order in the tunneling amplitude) at the $\Gamma_M$ and $M_M$ points of the lowest energy bands to obtain an estimate of their bandwidth.  The analysis indicates that the change in bandwidth is very small for most twist angles, but can become significant at the magic angles, perhaps not surprising as the degeneracy without the extra plane wave states coupled in is very nearly exact. Interestingly, we find within our estimation procedure that the magic angle breaks up into {\it two} closely spaced angles of maximal flatness, suggesting that our observation of a single magic angle found even with differing Dirac point Fermi velocities may not be precisely the case for the tetralayer realization of this system. Beyond this, we find that when the outer twist angles ($\theta_{12}$, $\theta_{34}$) are small enough, the change in bandwidth becomes sufficiently large as to indicate that $\mathbf{g}$ and $\mathbf{g}'$ with larger magnitudes should not be ignored (see Fig.\ref{fig:f} in Appendix~\ref{app:wave} and related discussion). For larger outer angles we believe our simpler treatment (in which incommensuration is ignored) correctly predicts that this system still hosts magic angles, and gives a good estimate of what these angles are.

One possible direction for future work is to treat the systems discussed in this work using a tight-binding model in order to investigate how well the continuum model approximation holds. For the ATBG system, this can be accomplished with a twisted bilayer graphene system where the nearest neighbor tunneling is different in the two layers. An application to the tetralayer system is less obvious, because one needs commensuration of all four lattices to define a unit cell.  Finding sets of such commensurate angles represents an interesting challenge.

Because of the change in Fermi velocities, the magic angles of the system acquires a certain level of tunability.  In principle this broadens the set of circumstances under which interaction effects can lead to collective phases such as Mott insulators and superconductivity, and possibly others with broken spin or valley symmetries.  In this sense the system we have studied in this work adds to the possible richness of physics in twisted graphene systems.

\section{Acknowledgements}
This work is supported in part by NSF Grant Nos. DMR-1350663, DMR-1914451, and ECCS-1936406, by the US-Israel Binational Foundation, and the Research
Corporation for Science Advancement through a Cottrell SEED grant. The authors thank the Aspen Center for Physics (NSF Grant No.
PHY-1607611) where part of this work was done.
\nocite{*}

\appendix
\section{Wavevector dependence of the overlap element $f_{\mu, \mu'}(\bf{g}, \bf{g}', \bf{k}, \bf{k}')$}
     \label{app:wave}
\begin{figure}[t]
        \centering
        \includegraphics[width=\linewidth]{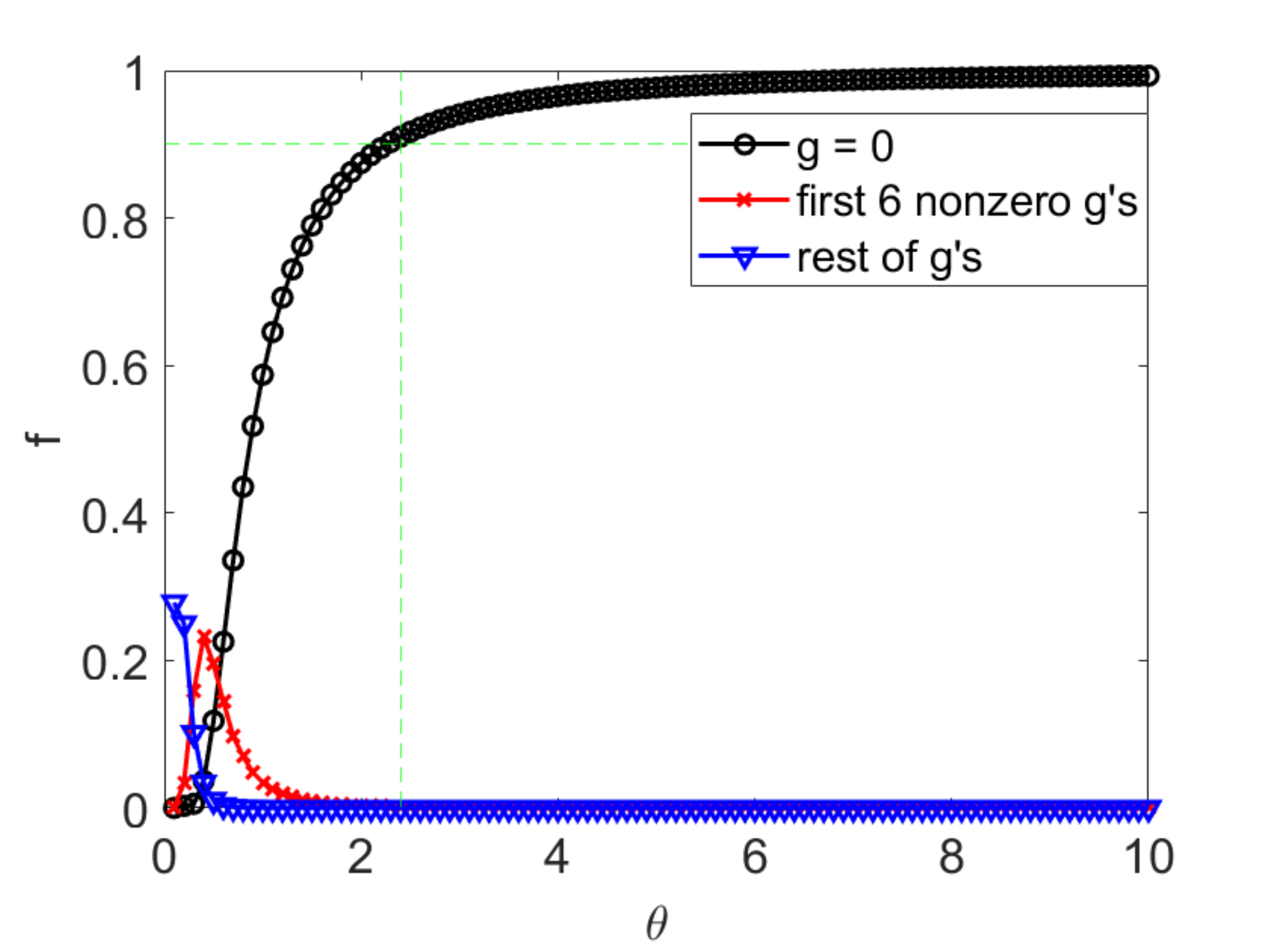}
        \caption{Overlap $f$ as a function of the twist angle $\theta_{12} = \theta_{34} = \theta$ for three different sets of momenta. The dashed lines indicate the 90\% ${\bf g} = 0$ cutoff for $f$ and its corresponding angle. In all three curves, ${\bf g}'=0$.}
        \label{fig:f}
    \end{figure}

     In Sec.~\ref{sec:4LHamiltonian}, we mentioned that the overlap element $f^{\mu\mu'}(\bf{g}, \bf{g}', \bf{k}, \bf{k}')$ is much larger for $\bf{g}= \bf{g}' = 0$ than for other values of $\bf{g},\bf{g}'$. To demonstrate this, we
     we define the quantity
     \begin{equation}
     f \equiv \dfrac{1}{12} \sum_{\mu \mu'} \sum_{j = 0}^2 \left| f_{j}^{\mu \mu'} \right|,
     \end{equation}
     with $f^{\mu\mu'}_j$ defined in Eq.~\eqref{eqn: fj} in the main text, and plot contributions to $f$ from different $\mathbf{g}$ as a function of twist angle $\theta$ in Fig. \ref{fig:f}.
      As can be seen in the figure, at large enough twist angles only the $\bf{g} = 0$ component is non-negligible. As the twist angle is made smaller, $f$ begins to find some support on the smallest magnitude nonzero reciprocal lattice vectors. There are six such vectors that all share the same magnitude; these six are summed together to generate the red curve marked with crosses in the figure. As $\theta$ is turned down still further, $f$ spreads out to larger magnitude wavevectors which are all summed together to give the blue triangle curve.

     Taken together, this figure shows that as long as the interbilayer twist angles $\theta_{12}$ and $\theta_{34}$ are not too small, then retaining only the $\bf{g} = 0$ wavevectors for the overlap element $f^{\mu\mu'}(\bf{g}, \bf{g}', \bf{k}, \bf{k}')$ is acceptable as a simplifying assumption. To make this concrete we demand that $f$ must contain at least 90\% of its weight on the $\bf{g} = 0$ lattice sites. This cutoff occurs at a twist angle of about $2.4^\circ$. Accordingly, none of the numerics discussed in Sec.~\ref{sec:numerics} involve a bilayer twist that is less than $2.4^\circ$.

     Discussion of the error associated with this approximation is discussed further in Appendix~\ref{app:incomm}.

\begin{figure}[t]
    \centering
    \includegraphics[width = \linewidth]{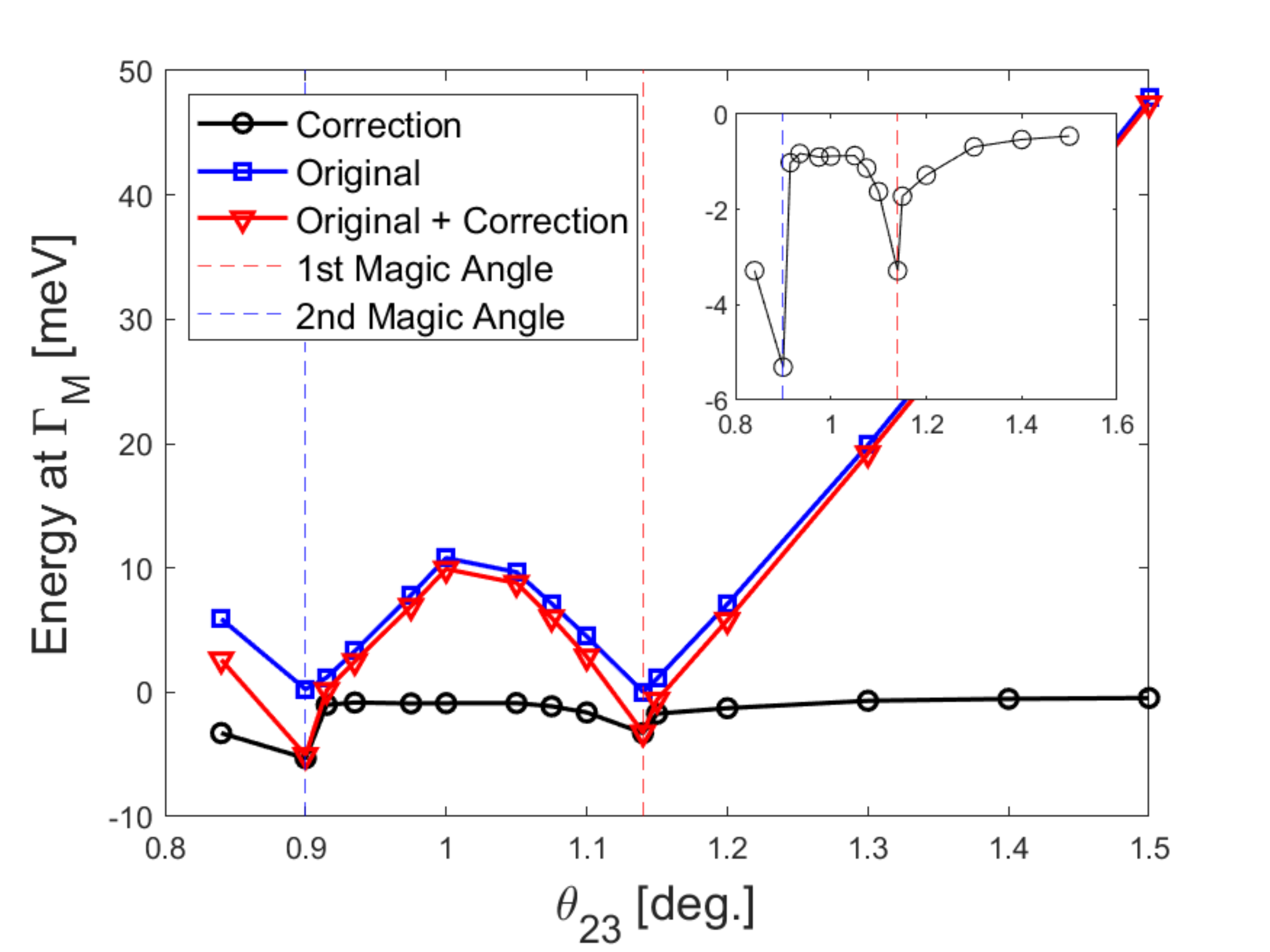}
    \caption{Estimate of correction to energy states nearest zero energy at $\Gamma_{M}$ point due to scattering by ${\bf g}_i^{(\lambda)} \ne 0$ in the moir\'e reciprocal lattice of bilayers $\lambda=12,34$ for $\theta_{12} = \theta_{34} = 10^\circ$, with $u=0.8$. Inset: Detail of the correction near the magic angles.}
    \label{fig:DPT1}
\end{figure}

\section{Effect of $\bf{g},\bf{g}' \ne 0$ on Tetralayer Bandwidth}\label{app:incomm}
In Section~\ref{sec:4LHamiltonian} we develop a simple model of a twisted four layer system in which the outer twist angles are well above magic angles, so that we can model the system as a pair of Dirac systems with different Fermi velocities that are coupled by an effective twisted bilayer tunneling term. In so doing we ignore the effective moir{\'e} periodicity of the two outer bilayers; including this formally renders the system aperiodic.  In this section we consider the impact of including the principle wavevectors that cause this aperiodicity.  In particular we develop an estimate of their impact on the flat-band phenomenon in the tetralayer system.

We begin by writing the total four-layer Hamiltonian $H$ as a sum of five individual operators,
\begin{align}
    H = H_0^{(12)} + H_{T}^{(12)} + H_0^{(34)} + H_{T}^{(34)} + H_{T}^{(23)}.
\end{align}
In this expression, $H_0^{(l\bar l)}$ represent Dirac Hamiltonians near the $K$ points of layers $l$ and $\bar l$ and $H_{T}^{(l\bar l)}$ is the tunnel coupling between them.  In the absence of $H^{(23)}_{T}$ the bilayer Hamiltonians for $\lambda=12,34$ can be diagonalized individually
\begin{align}
    H^{(\lambda)} &\equiv H_0^{(\lambda)} + H_{T}^{(\lambda)} \\
    &= \sum_n \sum_{{\bf k} \in \text{BZ}_{\lambda}} | w_n^{(\lambda)}({\bf k}) \rangle  \varepsilon^{(\lambda)}_n ({\bf k}) \langle w_n^{(\lambda)}({\bf k}) |.
    \end{align}
    where the ket $|w_n^{(\lambda)}(\bf{k})\rangle$ represents a state with crystal momentum $\bf{k}$ in band $n$.  In general, such a state contains wavevector content at all values of $\mathbf{k}+\mathbf{g}_i^{(\lambda)}$ where $\mathbf{g}_i^{(\lambda)}$ is moir{\'e} a reciprocal lattice vector of bilayer $\lambda$.

We now divide each of the TBG wavefunctions into two parts,
$| w_n^{\lambda}(\mathbf{k}) \rangle = | w_n^{\lambda,0}(\mathbf{k}) \rangle + | \delta w_n^{\lambda}(\mathbf{k}) \rangle $, where $| w_n^{\lambda,0}(\mathbf{k}) \rangle$ contains plane waves with wavevector $\bf{k}$, and $| \delta w_n^{\lambda}(\mathbf{k}) \rangle$ contains wavevectors $\mathbf{k}+\mathbf{g}_i$  with $\mathbf{g}_i \ne 0$.  We then write
Finally, we project to the two bands closest to zero energy, denoted by $n=\pm$.

The approximation scheme adopted in the main text involves
dropping the terms containing
$| \delta w_n^{\lambda}(\bf{k}) \rangle$ with $\lambda$ from $H$.  Denoting this as $H_0$,
we can also write
\begin{align}
    H_0 &= \sum_{\bf{k}} \sum_{m} | \varphi_m^{(0)}({\bf k}) \rangle E_m^{(0)}({\bf k}) \langle \varphi_m^{(0)} ({\bf k}) |.
\end{align}
Here, $E_m^{(0)}(\bf{k})$ is our approximation for the energy levels of the four-layer system, and $| \varphi_m^{(0)}(\bf{k}) \rangle$ are the corresponding wavefunctions.

We wish to estimate the error incurred by dropping the $\left| \delta w_n^{ij}(\bf{k}) \right\rangle$ terms from the Hamiltonian, particularly for $n=\pm$ bands whose states, as one approaches a flat band condition, become nearly degenerate.  Our approach is to re-introduce the largest of the Hamiltonian terms that were dropped, and diagonalize the resulting Hamiltonian within a relatively manageable subspace of the full Hilbert space.  Our analysis is essentially a form of degenerate perturbation theory, and so we expect results that are correct to linear order in the tunneling amplitude $w$.

\begin{widetext}
We thus write the Hamiltonian in the approximate form
\begin{align}
    H = H_0 + \sum_{n=\pm} \sum_{{\bf k}} \sum_{\lambda = 12, 34} \sum_{i=1}^6 \left\{ \left| \delta w^{(\lambda)}_n ({\bf k} - {\bf g}^{(\lambda)}_i) \right \rangle \varepsilon^{(\lambda)}_n ({\bf k}) \left \langle w^{(\lambda),(0)}_n ({\bf k}) \right| + \text{h.c.}\right\} \label{eqn: off-diag} + \mathcal O (w^2),
\end{align}
\end{widetext}
and diagonalize this Hamiltonian within a subspace of $\left\{|\phi_m^{(0)}(\bf{k})\rangle \right\}$, retaining $\mathbf{k}=\mathbf{k}_0$ and $\mathbf{k}=\mathbf{k}_0+\mathbf{g}_i^{(\lambda)}$ for the six shortest $\mathbf{g}_i^{(\lambda)}$ for each of $\lambda=12$ and $34$.
All the bands $m$ generated in our numerical diagonalization of $H_0$ are retained. Note that if one represents the band structure of $H_0$ in an extended zone scheme, this procedure for estimating the effects of scattering through the $\mathbf{g}_i^{(\lambda)}$ vectors amounts to retaining a small subset of states in each of the higher order Brillouin zones. Here the dimension of the Hilbert space is 244 corresponding to a cutoff radius of $\sqrt{61} k_\theta$. If more wavevectors are included in the calculation, the results do not noticeably change.

\begin{figure}[t]
    \centering
    \includegraphics[width = \linewidth]{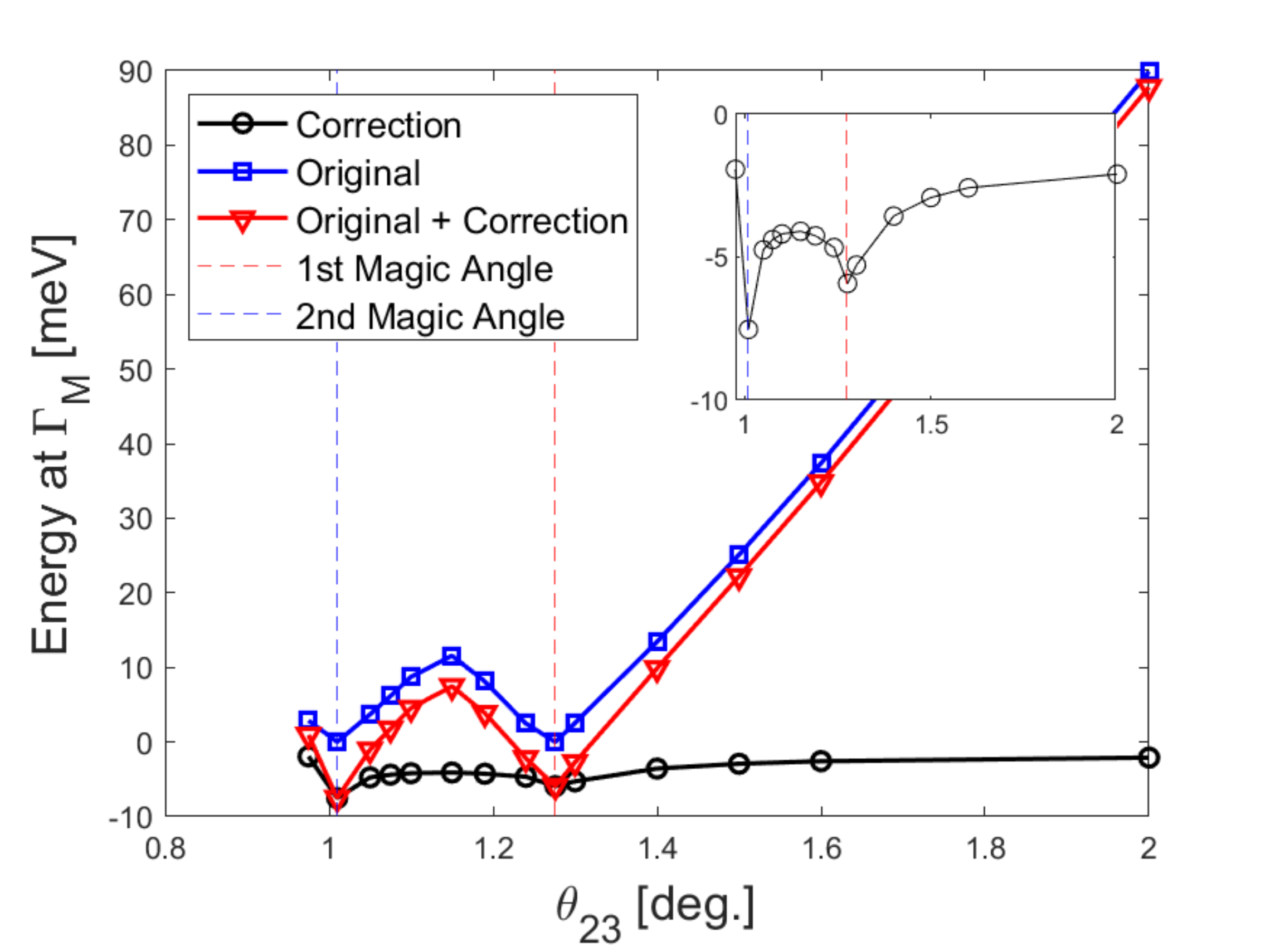}
    \caption{Estimate of correction to energy states nearest zero energy at $\Gamma_{M}$ point due to scattering by ${\bf g}_i^{(\lambda)} \ne 0$ in the moir\'e reciprocal lattice of bilayers $\lambda=12,34$ for $\theta_{12} = 6^\circ$ and  $\theta_{34} = 4^\circ$, with  $u=0.8$. Inset: Detail of the correction near the magic angles.}
    \label{fig:DPT2}
\end{figure}

\begin{figure}[b]
    \centering
    \includegraphics[width = \linewidth]{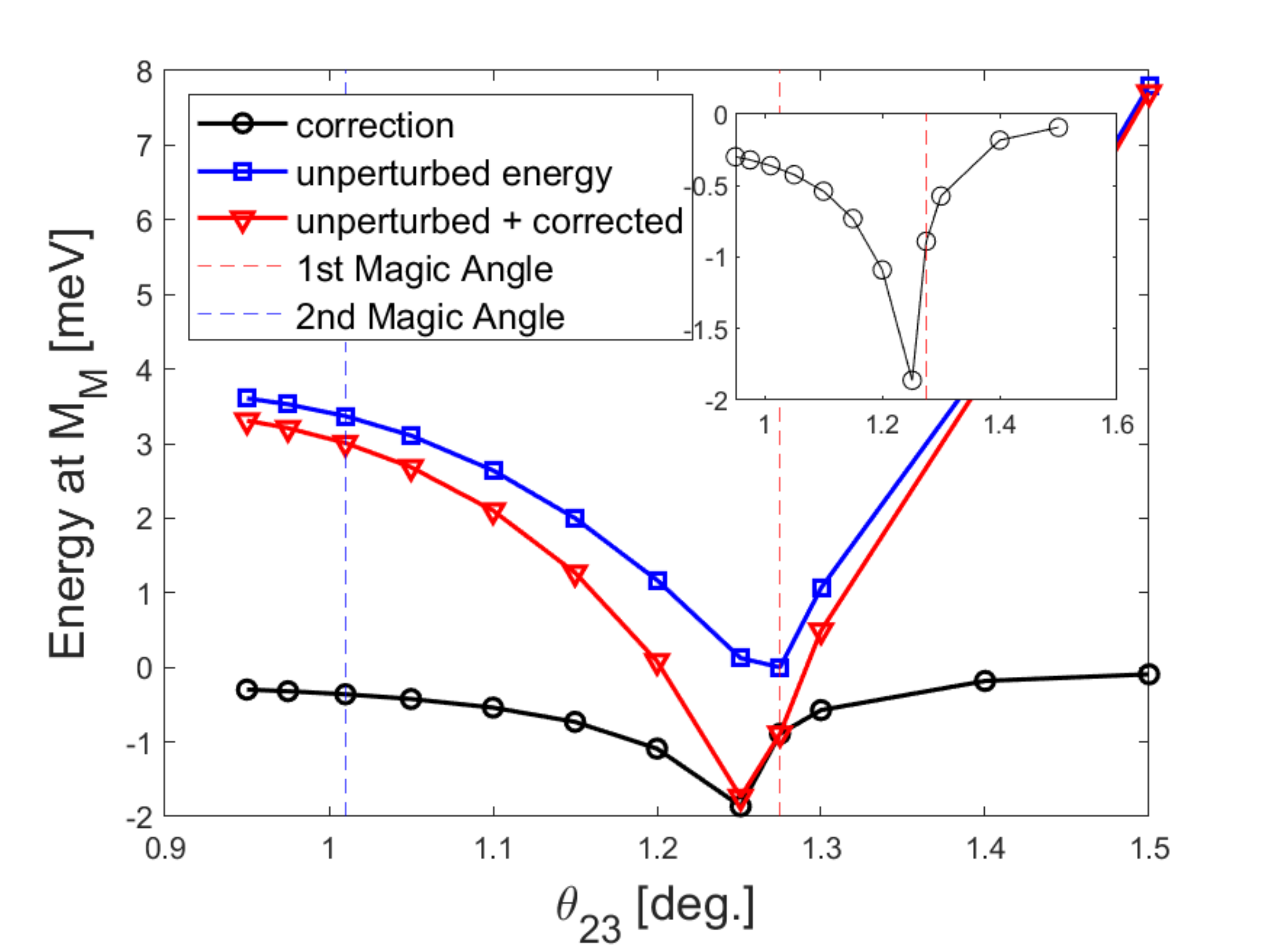}
    \caption{Estimate of correction to energy states nearest zero energy at $M_{M}$ point due to scattering by ${\bf g}_i^{(\lambda)} \ne 0$ in the moir\'e reciprocal lattice of bilayers $\lambda=12,34$ for $\theta_{12} = 6^\circ$ and  $\theta_{34} = 4^\circ$. Note the absence of a minimum in the separation of the two low energy bands at the second magic angle, which is also a feature of the unperturbed band structure.  The energy separation at that location in the Brillouin zone is quite small over a large range of angles.  Inset: Detail of the correction near the magic angles.}
    \label{fig:DPT3}
\end{figure}

Figures \ref{fig:DPT1}, \ref{fig:DPT2}, and \ref{fig:DPT3} illustrate representative results for the separation between the two bands nearest zero energy at $\Gamma_M$ and $M_M$, the $\Gamma$ and $M$ points of the (23) moir{\'e} Brillouin zone.  (Note that for the first magic angle, for a single twisted graphene bilayer the $\Gamma_M$ point is the location of greatest bandwidth.)  While in general the correction due to the coupling in of states by the $\bf{g}$ vectors is small, we see it becomes of order the bandwidth at the magic angles.  Nevertheless, we see that the band flattening survives their inclusion.  Interestingly, near but just away from the magic angles, the correction can actually cancel away the small bandwidth at the magic angle, in such a way that the angle of narrowest separation between the two low energy bands at $\Gamma_M$ splits into {\it two} closely spaced magic angles.  It is unclear if this small scale structure would survive the inclusion of further plane wave states coupled in by the $\bf{g}$ vectors.  However, the result is suggestive of the possibility that introducing asymmetry between the Dirac points coupled through the (23) interface using twisted outer layers could introduce extra structure not present in the single bilayer system.

\bibliography{thebibliography-new}

\begin{thebibliography}{77}%
\makeatletter
\providecommand \@ifxundefined [1]{%
 \@ifx{#1\undefined}
}%
\providecommand \@ifnum [1]{%
 \ifnum #1\expandafter \@firstoftwo
 \else \expandafter \@secondoftwo
 \fi
}%
\providecommand \@ifx [1]{%
 \ifx #1\expandafter \@firstoftwo
 \else \expandafter \@secondoftwo
 \fi
}%
\providecommand \natexlab [1]{#1}%
\providecommand \enquote  [1]{``#1''}%
\providecommand \bibnamefont  [1]{#1}%
\providecommand \bibfnamefont [1]{#1}%
\providecommand \citenamefont [1]{#1}%
\providecommand \href@noop [0]{\@secondoftwo}%
\providecommand \href [0]{\begingroup \@sanitize@url \@href}%
\providecommand \@href[1]{\@@startlink{#1}\@@href}%
\providecommand \@@href[1]{\endgroup#1\@@endlink}%
\providecommand \@sanitize@url [0]{\catcode `\\12\catcode `\$12\catcode
  `\&12\catcode `\#12\catcode `\^12\catcode `\_12\catcode `\%12\relax}%
\providecommand \@@startlink[1]{}%
\providecommand \@@endlink[0]{}%
\providecommand \url  [0]{\begingroup\@sanitize@url \@url }%
\providecommand \@url [1]{\endgroup\@href {#1}{\urlprefix }}%
\providecommand \urlprefix  [0]{URL }%
\providecommand \Eprint [0]{\href }%
\providecommand \doibase [0]{https://doi.org/}%
\providecommand \selectlanguage [0]{\@gobble}%
\providecommand \bibinfo  [0]{\@secondoftwo}%
\providecommand \bibfield  [0]{\@secondoftwo}%
\providecommand \translation [1]{[#1]}%
\providecommand \BibitemOpen [0]{}%
\providecommand \bibitemStop [0]{}%
\providecommand \bibitemNoStop [0]{.\EOS\space}%
\providecommand \EOS [0]{\spacefactor3000\relax}%
\providecommand \BibitemShut  [1]{\csname bibitem#1\endcsname}%
\let\auto@bib@innerbib\@empty
\bibitem [{\citenamefont {Lopes~dos Santos}\ \emph {et~al.}(2007)\citenamefont
  {Lopes~dos Santos}, \citenamefont {Peres},\ and\ \citenamefont
  {Castro~Neto}}]{Santos_2007}%
  \BibitemOpen
  \bibfield  {author} {\bibinfo {author} {\bibfnamefont {J.~M.~B.}\
  \bibnamefont {Lopes~dos Santos}}, \bibinfo {author} {\bibfnamefont
  {N.~M.~R.}\ \bibnamefont {Peres}},\ and\ \bibinfo {author} {\bibfnamefont
  {A.~H.}\ \bibnamefont {Castro~Neto}},\ }\bibfield  {title} {\bibinfo {title}
  {{Graphene Bilayer with a Twist: Electronic Structure}},\ }\href
  {https://doi.org/10.1103/PhysRevLett.99.256802} {\bibfield  {journal}
  {\bibinfo  {journal} {Phys. Rev. Lett.}\ }\textbf {\bibinfo {volume} {99}},\
  \bibinfo {pages} {256802} (\bibinfo {year} {2007})}\BibitemShut {NoStop}%
\bibitem [{\citenamefont {Su\'arez~Morell}\ \emph
  {et~al.}(2010{\natexlab{a}})\citenamefont {Su\'arez~Morell}, \citenamefont
  {Correa}, \citenamefont {Vargas}, \citenamefont {Pacheco},\ and\
  \citenamefont {Barticevic}}]{Morrell_2010}%
  \BibitemOpen
  \bibfield  {author} {\bibinfo {author} {\bibfnamefont {E.}~\bibnamefont
  {Su\'arez~Morell}}, \bibinfo {author} {\bibfnamefont {J.~D.}\ \bibnamefont
  {Correa}}, \bibinfo {author} {\bibfnamefont {P.}~\bibnamefont {Vargas}},
  \bibinfo {author} {\bibfnamefont {M.}~\bibnamefont {Pacheco}},\ and\ \bibinfo
  {author} {\bibfnamefont {Z.}~\bibnamefont {Barticevic}},\ }\bibfield  {title}
  {\bibinfo {title} {{Flat bands in slightly twisted bilayer graphene:
  Tight-binding calculations}},\ }\href
  {https://doi.org/10.1103/PhysRevB.82.121407} {\bibfield  {journal} {\bibinfo
  {journal} {Phys. Rev. B}\ }\textbf {\bibinfo {volume} {82}},\ \bibinfo
  {pages} {121407} (\bibinfo {year} {2010}{\natexlab{a}})}\BibitemShut
  {NoStop}%
\bibitem [{\citenamefont {Andrei}\ and\ \citenamefont
  {MacDonald}(2020)}]{Andrei_2020}%
  \BibitemOpen
  \bibfield  {author} {\bibinfo {author} {\bibfnamefont {E.~Y.}\ \bibnamefont
  {Andrei}}\ and\ \bibinfo {author} {\bibfnamefont {A.~H.}\ \bibnamefont
  {MacDonald}},\ }\bibfield  {title} {\bibinfo {title} {Graphene bilayers with
  a twist},\ }\href {https://doi.org/10.1038/s41563-020-00840-0} {\bibfield
  {journal} {\bibinfo  {journal} {Nature Materials}\ }\textbf {\bibinfo
  {volume} {19}},\ \bibinfo {pages} {1265} (\bibinfo {year}
  {2020})}\BibitemShut {NoStop}%
\bibitem [{\citenamefont {Carr}\ \emph {et~al.}(2017)\citenamefont {Carr},
  \citenamefont {Massatt}, \citenamefont {Fang}, \citenamefont {Cazeaux},
  \citenamefont {Luskin},\ and\ \citenamefont {Kaxiras}}]{Carr_2017}%
  \BibitemOpen
  \bibfield  {author} {\bibinfo {author} {\bibfnamefont {S.}~\bibnamefont
  {Carr}}, \bibinfo {author} {\bibfnamefont {D.}~\bibnamefont {Massatt}},
  \bibinfo {author} {\bibfnamefont {S.}~\bibnamefont {Fang}}, \bibinfo {author}
  {\bibfnamefont {P.}~\bibnamefont {Cazeaux}}, \bibinfo {author} {\bibfnamefont
  {M.}~\bibnamefont {Luskin}},\ and\ \bibinfo {author} {\bibfnamefont
  {E.}~\bibnamefont {Kaxiras}},\ }\bibfield  {title} {\bibinfo {title}
  {{Twistronics: Manipulating the electronic properties of two-dimensional
  layered structures through their twist angle}},\ }\href
  {https://doi.org/10.1103/PhysRevB.95.075420} {\bibfield  {journal} {\bibinfo
  {journal} {Phys. Rev. B}\ }\textbf {\bibinfo {volume} {95}},\ \bibinfo
  {pages} {075420} (\bibinfo {year} {2017})}\BibitemShut {NoStop}%
\bibitem [{\citenamefont {Ren}\ \emph {et~al.}(2020)\citenamefont {Ren},
  \citenamefont {Zhang}, \citenamefont {Liu},\ and\ \citenamefont
  {He}}]{Ren_2020}%
  \BibitemOpen
  \bibfield  {author} {\bibinfo {author} {\bibfnamefont {Y.-N.}\ \bibnamefont
  {Ren}}, \bibinfo {author} {\bibfnamefont {Y.}~\bibnamefont {Zhang}}, \bibinfo
  {author} {\bibfnamefont {Y.-W.}\ \bibnamefont {Liu}},\ and\ \bibinfo {author}
  {\bibfnamefont {L.}~\bibnamefont {He}},\ }\bibfield  {title} {\bibinfo
  {title} {Twistronics in graphene-based van der waals structures},\ }\href
  {https://doi.org/10.1088/1674-1056/abbbe2} {\bibfield  {journal} {\bibinfo
  {journal} {Chinese Physics B}\ }\textbf {\bibinfo {volume} {29}},\ \bibinfo
  {pages} {117303} (\bibinfo {year} {2020})}\BibitemShut {NoStop}%
\bibitem [{\citenamefont {Bistritzer}\ and\ \citenamefont
  {MacDonald}(2011)}]{BM}%
  \BibitemOpen
  \bibfield  {author} {\bibinfo {author} {\bibfnamefont {R.}~\bibnamefont
  {Bistritzer}}\ and\ \bibinfo {author} {\bibfnamefont {A.}~\bibnamefont
  {MacDonald}},\ }\bibfield  {title} {\bibinfo {title} {Moiré bands in twisted
  double-layer graphene},\ }\href
  {https://doi.org/https://doi.org/10.1073/pnas.1108174108} {\bibfield
  {journal} {\bibinfo  {journal} {PNAS}\ }\textbf {\bibinfo {volume} {108}},\
  \bibinfo {pages} {12233} (\bibinfo {year} {2011})}\BibitemShut {NoStop}%
\bibitem [{\citenamefont {Cao}\ \emph {et~al.}(2018{\natexlab{a}})\citenamefont
  {Cao}, \citenamefont {Fatemi}, \citenamefont {Demir}, \citenamefont {Fang},
  \citenamefont {Tomarken}, \citenamefont {Luo}, \citenamefont
  {Sanchez-Yamagishi}, \citenamefont {Watanabe}, \citenamefont {Taniguchi},
  \citenamefont {Kaxiras}, \citenamefont {Ashoori},\ and\ \citenamefont
  {Jarillo-Herrero}}]{Cao_2018a}%
  \BibitemOpen
  \bibfield  {author} {\bibinfo {author} {\bibfnamefont {Y.}~\bibnamefont
  {Cao}}, \bibinfo {author} {\bibfnamefont {V.}~\bibnamefont {Fatemi}},
  \bibinfo {author} {\bibfnamefont {A.}~\bibnamefont {Demir}}, \bibinfo
  {author} {\bibfnamefont {S.}~\bibnamefont {Fang}}, \bibinfo {author}
  {\bibfnamefont {S.~L.}\ \bibnamefont {Tomarken}}, \bibinfo {author}
  {\bibfnamefont {J.~Y.}\ \bibnamefont {Luo}}, \bibinfo {author} {\bibfnamefont
  {J.~D.}\ \bibnamefont {Sanchez-Yamagishi}}, \bibinfo {author} {\bibfnamefont
  {K.}~\bibnamefont {Watanabe}}, \bibinfo {author} {\bibfnamefont
  {T.}~\bibnamefont {Taniguchi}}, \bibinfo {author} {\bibfnamefont
  {E.}~\bibnamefont {Kaxiras}}, \bibinfo {author} {\bibfnamefont {R.~C.}\
  \bibnamefont {Ashoori}},\ and\ \bibinfo {author} {\bibfnamefont
  {P.}~\bibnamefont {Jarillo-Herrero}},\ }\bibfield  {title} {\bibinfo {title}
  {Correlated insulator behaviour at half-filling in magic-angle graphene
  superlattices},\ }\href {https://doi.org/10.1038/nature26154} {\bibfield
  {journal} {\bibinfo  {journal} {Nature}\ }\textbf {\bibinfo {volume} {556}},\
  \bibinfo {pages} {80} (\bibinfo {year} {2018}{\natexlab{a}})}\BibitemShut
  {NoStop}%
\bibitem [{\citenamefont {Cao}\ \emph {et~al.}(2018{\natexlab{b}})\citenamefont
  {Cao}, \citenamefont {Fatemi}, \citenamefont {Fang}, \citenamefont
  {Watanabe}, \citenamefont {Taniguchi}, \citenamefont {Kaxiras},\ and\
  \citenamefont {Jarillo-Herrero}}]{Cao_2018b}%
  \BibitemOpen
  \bibfield  {author} {\bibinfo {author} {\bibfnamefont {Y.}~\bibnamefont
  {Cao}}, \bibinfo {author} {\bibfnamefont {V.}~\bibnamefont {Fatemi}},
  \bibinfo {author} {\bibfnamefont {S.}~\bibnamefont {Fang}}, \bibinfo {author}
  {\bibfnamefont {K.}~\bibnamefont {Watanabe}}, \bibinfo {author}
  {\bibfnamefont {T.}~\bibnamefont {Taniguchi}}, \bibinfo {author}
  {\bibfnamefont {E.}~\bibnamefont {Kaxiras}},\ and\ \bibinfo {author}
  {\bibfnamefont {P.}~\bibnamefont {Jarillo-Herrero}},\ }\bibfield  {title}
  {\bibinfo {title} {Unconventional superconductivity in magic-angle graphene
  superlattices},\ }\href {https://doi.org/10.1038/nature26160} {\bibfield
  {journal} {\bibinfo  {journal} {Nature}\ }\textbf {\bibinfo {volume} {556}},\
  \bibinfo {pages} {43} (\bibinfo {year} {2018}{\natexlab{b}})}\BibitemShut
  {NoStop}%
\bibitem [{\citenamefont {Yankowitz}\ \emph {et~al.}(2019)\citenamefont
  {Yankowitz}, \citenamefont {Chen}, \citenamefont {Polshyn}, \citenamefont
  {Zhang}, \citenamefont {Watanabe}, \citenamefont {Taniguchi}, \citenamefont
  {Graf}, \citenamefont {Young},\ and\ \citenamefont {Dean}}]{Yankowitz_2019}%
  \BibitemOpen
  \bibfield  {author} {\bibinfo {author} {\bibfnamefont {M.}~\bibnamefont
  {Yankowitz}}, \bibinfo {author} {\bibfnamefont {S.}~\bibnamefont {Chen}},
  \bibinfo {author} {\bibfnamefont {H.}~\bibnamefont {Polshyn}}, \bibinfo
  {author} {\bibfnamefont {Y.}~\bibnamefont {Zhang}}, \bibinfo {author}
  {\bibfnamefont {K.}~\bibnamefont {Watanabe}}, \bibinfo {author}
  {\bibfnamefont {T.}~\bibnamefont {Taniguchi}}, \bibinfo {author}
  {\bibfnamefont {D.}~\bibnamefont {Graf}}, \bibinfo {author} {\bibfnamefont
  {A.~F.}\ \bibnamefont {Young}},\ and\ \bibinfo {author} {\bibfnamefont
  {C.~R.}\ \bibnamefont {Dean}},\ }\bibfield  {title} {\bibinfo {title}
  {{Tuning superconductivity in twisted bilayer graphene}},\ }\href
  {https://doi.org/10.1126/science.aav1910} {\bibfield  {journal} {\bibinfo
  {journal} {Science}\ }\textbf {\bibinfo {volume} {363}},\ \bibinfo {pages}
  {1059} (\bibinfo {year} {2019})}\BibitemShut {NoStop}%
\bibitem [{\citenamefont {Wong}\ \emph {et~al.}(2020)\citenamefont {Wong},
  \citenamefont {Nuckolls}, \citenamefont {Oh}, \citenamefont {Lian},
  \citenamefont {Xie}, \citenamefont {Jeon}, \citenamefont {Watanabe},
  \citenamefont {Taniguchi}, \citenamefont {Bernevig},\ and\ \citenamefont
  {Yazdani}}]{Wong_2020}%
  \BibitemOpen
  \bibfield  {author} {\bibinfo {author} {\bibfnamefont {D.}~\bibnamefont
  {Wong}}, \bibinfo {author} {\bibfnamefont {K.~P.}\ \bibnamefont {Nuckolls}},
  \bibinfo {author} {\bibfnamefont {M.}~\bibnamefont {Oh}}, \bibinfo {author}
  {\bibfnamefont {B.}~\bibnamefont {Lian}}, \bibinfo {author} {\bibfnamefont
  {Y.}~\bibnamefont {Xie}}, \bibinfo {author} {\bibfnamefont {S.}~\bibnamefont
  {Jeon}}, \bibinfo {author} {\bibfnamefont {K.}~\bibnamefont {Watanabe}},
  \bibinfo {author} {\bibfnamefont {T.}~\bibnamefont {Taniguchi}}, \bibinfo
  {author} {\bibfnamefont {B.~A.}\ \bibnamefont {Bernevig}},\ and\ \bibinfo
  {author} {\bibfnamefont {A.}~\bibnamefont {Yazdani}},\ }\bibfield  {title}
  {\bibinfo {title} {Cascade of electronic transitions in magic-angle twisted
  bilayer graphene},\ }\href {https://doi.org/10.1038/s41586-020-2339-0}
  {\bibfield  {journal} {\bibinfo  {journal} {Nature}\ }\textbf {\bibinfo
  {volume} {582}},\ \bibinfo {pages} {198} (\bibinfo {year}
  {2020})}\BibitemShut {NoStop}%
\bibitem [{\citenamefont {Cea}\ \emph {et~al.}(2020)\citenamefont {Cea},
  \citenamefont {Pantaleón},\ and\ \citenamefont {Guinea}}]{hBN1}%
  \BibitemOpen
  \bibfield  {author} {\bibinfo {author} {\bibfnamefont {T.}~\bibnamefont
  {Cea}}, \bibinfo {author} {\bibfnamefont {P.~A.}\ \bibnamefont
  {Pantaleón}},\ and\ \bibinfo {author} {\bibfnamefont {F.}~\bibnamefont
  {Guinea}},\ }\bibfield  {title} {\bibinfo {title} {Band structure of twisted
  bilayer graphene on hexagonal boron nitride},\ }\href
  {https://doi.org/https://doi.org/10.1103/PhysRevB.102.155136} {\bibfield
  {journal} {\bibinfo  {journal} {Phys. Rev. B}\ }\textbf {\bibinfo {volume}
  {102}},\ \bibinfo {pages} {155136} (\bibinfo {year} {2020})}\BibitemShut
  {NoStop}%
\bibitem [{\citenamefont {Zhang}\ \emph {et~al.}(2019)\citenamefont {Zhang},
  \citenamefont {Mao},\ and\ \citenamefont {Senthil}}]{hBN2}%
  \BibitemOpen
  \bibfield  {author} {\bibinfo {author} {\bibfnamefont {Y.-H.}\ \bibnamefont
  {Zhang}}, \bibinfo {author} {\bibfnamefont {D.}~\bibnamefont {Mao}},\ and\
  \bibinfo {author} {\bibfnamefont {T.}~\bibnamefont {Senthil}},\ }\bibfield
  {title} {\bibinfo {title} {{Twisted bilayer graphene aligned with hexagonal
  boron nitride: Anomalous Hall effect and a lattice model}},\ }\href
  {https://doi.org/10.1103/PhysRevResearch.1.033126} {\bibfield  {journal}
  {\bibinfo  {journal} {Phys. Rev. Research}\ }\textbf {\bibinfo {volume}
  {1}},\ \bibinfo {pages} {033126} (\bibinfo {year} {2019})}\BibitemShut
  {NoStop}%
\bibitem [{\citenamefont {Lin}\ and\ \citenamefont {Ni}(2020)}]{hBN3}%
  \BibitemOpen
  \bibfield  {author} {\bibinfo {author} {\bibfnamefont {X.}~\bibnamefont
  {Lin}}\ and\ \bibinfo {author} {\bibfnamefont {J.}~\bibnamefont {Ni}},\
  }\bibfield  {title} {\bibinfo {title} {{Symmetry breaking in the double
  moir\'e superlattices of relaxed twisted bilayer graphene on hexagonal boron
  nitride}},\ }\href {https://doi.org/10.1103/PhysRevB.102.035441} {\bibfield
  {journal} {\bibinfo  {journal} {Phys. Rev. B}\ }\textbf {\bibinfo {volume}
  {102}},\ \bibinfo {pages} {035441} (\bibinfo {year} {2020})}\BibitemShut
  {NoStop}%
\bibitem [{\citenamefont {Yang}\ \emph {et~al.}(2020)\citenamefont {Yang},
  \citenamefont {Li}, \citenamefont {Yin}, \citenamefont {Xu}, \citenamefont
  {Mullan}, \citenamefont {Taniguchi}, \citenamefont {Watanabe}, \citenamefont
  {Geim}, \citenamefont {Novoselov},\ and\ \citenamefont
  {Mishchenko}}]{Yang_2020}%
  \BibitemOpen
  \bibfield  {author} {\bibinfo {author} {\bibfnamefont {Y.}~\bibnamefont
  {Yang}}, \bibinfo {author} {\bibfnamefont {J.}~\bibnamefont {Li}}, \bibinfo
  {author} {\bibfnamefont {J.}~\bibnamefont {Yin}}, \bibinfo {author}
  {\bibfnamefont {S.}~\bibnamefont {Xu}}, \bibinfo {author} {\bibfnamefont
  {C.}~\bibnamefont {Mullan}}, \bibinfo {author} {\bibfnamefont
  {T.}~\bibnamefont {Taniguchi}}, \bibinfo {author} {\bibfnamefont
  {K.}~\bibnamefont {Watanabe}}, \bibinfo {author} {\bibfnamefont {A.~K.}\
  \bibnamefont {Geim}}, \bibinfo {author} {\bibfnamefont {K.~S.}\ \bibnamefont
  {Novoselov}},\ and\ \bibinfo {author} {\bibfnamefont {A.}~\bibnamefont
  {Mishchenko}},\ }\bibfield  {title} {\bibinfo {title} {In situ manipulation
  of van der waals heterostructures for twistronics},\ }\href
  {https://doi.org/10.1126/sciadv.abd3655} {\bibfield  {journal} {\bibinfo
  {journal} {Science Advances}\ }\textbf {\bibinfo {volume} {6}},\ \bibinfo
  {pages} {eabd3655} (\bibinfo {year} {2020})}\BibitemShut {NoStop}%
\bibitem [{\citenamefont {An{\dj}elkovi{\'{c}}}\ \emph
  {et~al.}(2020)\citenamefont {An{\dj}elkovi{\'{c}}}, \citenamefont
  {Milovanovi{\'{c}}}, \citenamefont {Covaci},\ and\ \citenamefont
  {Peeters}}]{Andelkovic2020}%
  \BibitemOpen
  \bibfield  {author} {\bibinfo {author} {\bibfnamefont {M.}~\bibnamefont
  {An{\dj}elkovi{\'{c}}}}, \bibinfo {author} {\bibfnamefont {S.~P.}\
  \bibnamefont {Milovanovi{\'{c}}}}, \bibinfo {author} {\bibfnamefont
  {L.}~\bibnamefont {Covaci}},\ and\ \bibinfo {author} {\bibfnamefont {F.~M.}\
  \bibnamefont {Peeters}},\ }\bibfield  {title} {\bibinfo {title} {Double
  moir{\'e} with a twist: Supermoir{\'e} in encapsulated graphene},\ }\href
  {https://doi.org/10.1021/acs.nanolett.9b04058} {\bibfield  {journal}
  {\bibinfo  {journal} {Nano Letters}\ }\textbf {\bibinfo {volume} {20}},\
  \bibinfo {pages} {979} (\bibinfo {year} {2020})}\BibitemShut {NoStop}%
\bibitem [{\citenamefont {Wang}\ \emph {et~al.}(2019)\citenamefont {Wang},
  \citenamefont {Wang}, \citenamefont {Yin}, \citenamefont {Tóvári},
  \citenamefont {Yang}, \citenamefont {Lin}, \citenamefont {Holwill},
  \citenamefont {Birkbeck}, \citenamefont {Perello}, \citenamefont {Xu},
  \citenamefont {Zultak}, \citenamefont {Gorbachev}, \citenamefont {Kretinin},
  \citenamefont {Taniguchi}, \citenamefont {Watanabe}, \citenamefont {Morozov},
  \citenamefont {Anđelković}, \citenamefont {Milovanović}, \citenamefont
  {Covaci}, \citenamefont {Peeters}, \citenamefont {Mishchenko}, \citenamefont
  {Geim}, \citenamefont {Novoselov}, \citenamefont {Fal’ko}, \citenamefont
  {Knothe},\ and\ \citenamefont {Woods}}]{Wang2019}%
  \BibitemOpen
  \bibfield  {author} {\bibinfo {author} {\bibfnamefont {Z.}~\bibnamefont
  {Wang}}, \bibinfo {author} {\bibfnamefont {Y.~B.}\ \bibnamefont {Wang}},
  \bibinfo {author} {\bibfnamefont {J.}~\bibnamefont {Yin}}, \bibinfo {author}
  {\bibfnamefont {E.}~\bibnamefont {Tóvári}}, \bibinfo {author}
  {\bibfnamefont {Y.}~\bibnamefont {Yang}}, \bibinfo {author} {\bibfnamefont
  {L.}~\bibnamefont {Lin}}, \bibinfo {author} {\bibfnamefont {M.}~\bibnamefont
  {Holwill}}, \bibinfo {author} {\bibfnamefont {J.}~\bibnamefont {Birkbeck}},
  \bibinfo {author} {\bibfnamefont {D.~J.}\ \bibnamefont {Perello}}, \bibinfo
  {author} {\bibfnamefont {S.}~\bibnamefont {Xu}}, \bibinfo {author}
  {\bibfnamefont {J.}~\bibnamefont {Zultak}}, \bibinfo {author} {\bibfnamefont
  {R.~V.}\ \bibnamefont {Gorbachev}}, \bibinfo {author} {\bibfnamefont {A.~V.}\
  \bibnamefont {Kretinin}}, \bibinfo {author} {\bibfnamefont {T.}~\bibnamefont
  {Taniguchi}}, \bibinfo {author} {\bibfnamefont {K.}~\bibnamefont {Watanabe}},
  \bibinfo {author} {\bibfnamefont {S.~V.}\ \bibnamefont {Morozov}}, \bibinfo
  {author} {\bibfnamefont {M.}~\bibnamefont {Anđelković}}, \bibinfo {author}
  {\bibfnamefont {S.~P.}\ \bibnamefont {Milovanović}}, \bibinfo {author}
  {\bibfnamefont {L.}~\bibnamefont {Covaci}}, \bibinfo {author} {\bibfnamefont
  {F.~M.}\ \bibnamefont {Peeters}}, \bibinfo {author} {\bibfnamefont
  {A.}~\bibnamefont {Mishchenko}}, \bibinfo {author} {\bibfnamefont {A.~K.}\
  \bibnamefont {Geim}}, \bibinfo {author} {\bibfnamefont {K.~S.}\ \bibnamefont
  {Novoselov}}, \bibinfo {author} {\bibfnamefont {V.~I.}\ \bibnamefont
  {Fal’ko}}, \bibinfo {author} {\bibfnamefont {A.}~\bibnamefont {Knothe}},\
  and\ \bibinfo {author} {\bibfnamefont {C.~R.}\ \bibnamefont {Woods}},\
  }\bibfield  {title} {\bibinfo {title} {{Composite super-moir\'e; lattices in
  double-aligned graphene heterostructures}},\ }\href
  {https://doi.org/10.1126/sciadv.aay8897} {\bibfield  {journal} {\bibinfo
  {journal} {Science Advances}\ }\textbf {\bibinfo {volume} {5}},\ \bibinfo
  {pages} {8897} (\bibinfo {year} {2019})}\BibitemShut {NoStop}%
\bibitem [{\citenamefont {Wang}\ \emph {et~al.}(2020)\citenamefont {Wang},
  \citenamefont {Shih}, \citenamefont {Ghiotto}, \citenamefont {Xian},
  \citenamefont {Rhodes}, \citenamefont {Tan}, \citenamefont {Claassen},
  \citenamefont {Kennes}, \citenamefont {Bai}, \citenamefont {Kim},\ and\
  \citenamefont {et~al.}}]{Wang_2020}%
  \BibitemOpen
  \bibfield  {author} {\bibinfo {author} {\bibfnamefont {L.}~\bibnamefont
  {Wang}}, \bibinfo {author} {\bibfnamefont {E.-M.}\ \bibnamefont {Shih}},
  \bibinfo {author} {\bibfnamefont {A.}~\bibnamefont {Ghiotto}}, \bibinfo
  {author} {\bibfnamefont {L.}~\bibnamefont {Xian}}, \bibinfo {author}
  {\bibfnamefont {D.~A.}\ \bibnamefont {Rhodes}}, \bibinfo {author}
  {\bibfnamefont {C.}~\bibnamefont {Tan}}, \bibinfo {author} {\bibfnamefont
  {M.}~\bibnamefont {Claassen}}, \bibinfo {author} {\bibfnamefont {D.~M.}\
  \bibnamefont {Kennes}}, \bibinfo {author} {\bibfnamefont {Y.}~\bibnamefont
  {Bai}}, \bibinfo {author} {\bibfnamefont {B.}~\bibnamefont {Kim}},\ and\
  \bibinfo {author} {\bibnamefont {et~al.}},\ }\bibfield  {title} {\bibinfo
  {title} {Correlated electronic phases in twisted bilayer transition metal
  dichalcogenides},\ }\href {https://doi.org/10.1038/s41563-020-0708-6}
  {\bibfield  {journal} {\bibinfo  {journal} {Nature Materials}\ }\textbf
  {\bibinfo {volume} {19}},\ \bibinfo {pages} {861–866} (\bibinfo {year}
  {2020})}\BibitemShut {NoStop}%
\bibitem [{\citenamefont {Zhang}(2019)}]{Zhang_2019}%
  \BibitemOpen
  \bibfield  {author} {\bibinfo {author} {\bibfnamefont {L.}~\bibnamefont
  {Zhang}},\ }\bibfield  {title} {\bibinfo {title} {{Lowest-energy moiré band
  formed by Dirac zero modes in twisted bilayer graphene}},\ }\href
  {https://doi.org/https://doi.org/10.1016/j.scib.2019.03.010} {\bibfield
  {journal} {\bibinfo  {journal} {Science Bulletin}\ }\textbf {\bibinfo
  {volume} {64}},\ \bibinfo {pages} {495} (\bibinfo {year} {2019})}\BibitemShut
  {NoStop}%
\bibitem [{\citenamefont {Li}\ \emph {et~al.}(2021)\citenamefont {Li},
  \citenamefont {Hu}, \citenamefont {Feng}, \citenamefont {Zhou}, \citenamefont
  {An}, \citenamefont {Law}, \citenamefont {Wang},\ and\ \citenamefont
  {Lin}}]{Li_2021}%
  \BibitemOpen
  \bibfield  {author} {\bibinfo {author} {\bibfnamefont {E.}~\bibnamefont
  {Li}}, \bibinfo {author} {\bibfnamefont {J.-X.}\ \bibnamefont {Hu}}, \bibinfo
  {author} {\bibfnamefont {X.}~\bibnamefont {Feng}}, \bibinfo {author}
  {\bibfnamefont {Z.}~\bibnamefont {Zhou}}, \bibinfo {author} {\bibfnamefont
  {L.}~\bibnamefont {An}}, \bibinfo {author} {\bibfnamefont {K.~T.}\
  \bibnamefont {Law}}, \bibinfo {author} {\bibfnamefont {N.}~\bibnamefont
  {Wang}},\ and\ \bibinfo {author} {\bibfnamefont {N.}~\bibnamefont {Lin}},\
  }\href@noop {} {\bibinfo {title} {Lattice reconstruction induced multiple
  ultra-flat bands in twisted bilayer wse2}} (\bibinfo {year} {2021}),\ \Eprint
  {https://arxiv.org/abs/2103.06479} {arXiv:2103.06479 [cond-mat.mes-hall]}
  \BibitemShut {NoStop}%
\bibitem [{\citenamefont {Naik}\ and\ \citenamefont {Jain}(2018)}]{Naik_2018}%
  \BibitemOpen
  \bibfield  {author} {\bibinfo {author} {\bibfnamefont {M.~H.}\ \bibnamefont
  {Naik}}\ and\ \bibinfo {author} {\bibfnamefont {M.}~\bibnamefont {Jain}},\
  }\bibfield  {title} {\bibinfo {title} {{Ultraflatbands and Shear Solitons in
  Moir\'e Patterns of Twisted Bilayer Transition Metal Dichalcogenides}},\
  }\href {https://doi.org/10.1103/PhysRevLett.121.266401} {\bibfield  {journal}
  {\bibinfo  {journal} {Phys. Rev. Lett.}\ }\textbf {\bibinfo {volume} {121}},\
  \bibinfo {pages} {266401} (\bibinfo {year} {2018})}\BibitemShut {NoStop}%
\bibitem [{\citenamefont {Naik}\ \emph {et~al.}(2020)\citenamefont {Naik},
  \citenamefont {Kundu}, \citenamefont {Maity},\ and\ \citenamefont
  {Jain}}]{Naik_2020}%
  \BibitemOpen
  \bibfield  {author} {\bibinfo {author} {\bibfnamefont {M.~H.}\ \bibnamefont
  {Naik}}, \bibinfo {author} {\bibfnamefont {S.}~\bibnamefont {Kundu}},
  \bibinfo {author} {\bibfnamefont {I.}~\bibnamefont {Maity}},\ and\ \bibinfo
  {author} {\bibfnamefont {M.}~\bibnamefont {Jain}},\ }\bibfield  {title}
  {\bibinfo {title} {{Origin and evolution of ultraflat bands in twisted
  bilayer transition metal dichalcogenides: Realization of triangular quantum
  dots}},\ }\href {https://doi.org/10.1103/PhysRevB.102.075413} {\bibfield
  {journal} {\bibinfo  {journal} {Phys. Rev. B}\ }\textbf {\bibinfo {volume}
  {102}},\ \bibinfo {pages} {075413} (\bibinfo {year} {2020})}\BibitemShut
  {NoStop}%
\bibitem [{\citenamefont {Zhan}\ \emph {et~al.}(2020)\citenamefont {Zhan},
  \citenamefont {Zhang}, \citenamefont {Lv}, \citenamefont {Zhong},
  \citenamefont {Yu}, \citenamefont {Guinea}, \citenamefont {Silva-Guill\'en},\
  and\ \citenamefont {Yuan}}]{Zhan_2020}%
  \BibitemOpen
  \bibfield  {author} {\bibinfo {author} {\bibfnamefont {Z.}~\bibnamefont
  {Zhan}}, \bibinfo {author} {\bibfnamefont {Y.}~\bibnamefont {Zhang}},
  \bibinfo {author} {\bibfnamefont {P.}~\bibnamefont {Lv}}, \bibinfo {author}
  {\bibfnamefont {H.}~\bibnamefont {Zhong}}, \bibinfo {author} {\bibfnamefont
  {G.}~\bibnamefont {Yu}}, \bibinfo {author} {\bibfnamefont {F.}~\bibnamefont
  {Guinea}}, \bibinfo {author} {\bibfnamefont {J.~A.}\ \bibnamefont
  {Silva-Guill\'en}},\ and\ \bibinfo {author} {\bibfnamefont {S.}~\bibnamefont
  {Yuan}},\ }\bibfield  {title} {\bibinfo {title} {{Tunability of multiple
  ultraflat bands and effect of spin-orbit coupling in twisted bilayer
  transition metal dichalcogenides}},\ }\href
  {https://doi.org/10.1103/PhysRevB.102.241106} {\bibfield  {journal} {\bibinfo
   {journal} {Phys. Rev. B}\ }\textbf {\bibinfo {volume} {102}},\ \bibinfo
  {pages} {241106} (\bibinfo {year} {2020})}\BibitemShut {NoStop}%
\bibitem [{\citenamefont {Haddadi}\ \emph {et~al.}(2020)\citenamefont
  {Haddadi}, \citenamefont {Wu}, \citenamefont {Kruchkov},\ and\ \citenamefont
  {Yazyev}}]{Haddadi}%
  \BibitemOpen
  \bibfield  {author} {\bibinfo {author} {\bibfnamefont {F.}~\bibnamefont
  {Haddadi}}, \bibinfo {author} {\bibfnamefont {Q.}~\bibnamefont {Wu}},
  \bibinfo {author} {\bibfnamefont {A.~J.}\ \bibnamefont {Kruchkov}},\ and\
  \bibinfo {author} {\bibfnamefont {O.~V.}\ \bibnamefont {Yazyev}},\ }\bibfield
   {title} {\bibinfo {title} {Moir\'e flat bands in twisted double bilayer
  graphene},\ }\href
  {https://doi.org/https://doi.org/10.1021/acs.nanolett.9b05117} {\bibfield
  {journal} {\bibinfo  {journal} {Nano. Lett.}\ }\textbf {\bibinfo {volume}
  {20}},\ \bibinfo {pages} {2410–} (\bibinfo {year} {2020})}\BibitemShut
  {NoStop}%
\bibitem [{\citenamefont {Chebrolu}\ \emph {et~al.}(2019)\citenamefont
  {Chebrolu}, \citenamefont {Chittari},\ and\ \citenamefont {Jung}}]{Cheroblu}%
  \BibitemOpen
  \bibfield  {author} {\bibinfo {author} {\bibfnamefont {N.~R.}\ \bibnamefont
  {Chebrolu}}, \bibinfo {author} {\bibfnamefont {B.~L.}\ \bibnamefont
  {Chittari}},\ and\ \bibinfo {author} {\bibfnamefont {J.}~\bibnamefont
  {Jung}},\ }\bibfield  {title} {\bibinfo {title} {Flatbands in twisted double
  bilayer graphene},\ }\href
  {https://doi.org/https://doi.org/10.1103/PhysRevB.99.235417} {\bibfield
  {journal} {\bibinfo  {journal} {Phys. Rev. B}\ }\textbf {\bibinfo {volume}
  {99}},\ \bibinfo {pages} {235417} (\bibinfo {year} {2019})}\BibitemShut
  {NoStop}%
\bibitem [{\citenamefont {Burg}\ \emph {et~al.}(2019)\citenamefont {Burg},
  \citenamefont {Zhu}, \citenamefont {Taniguchi}, \citenamefont {Watanabe},
  \citenamefont {MacDonald},\ and\ \citenamefont {Tutuc}}]{Burg2019}%
  \BibitemOpen
  \bibfield  {author} {\bibinfo {author} {\bibfnamefont {G.~W.}\ \bibnamefont
  {Burg}}, \bibinfo {author} {\bibfnamefont {J.}~\bibnamefont {Zhu}}, \bibinfo
  {author} {\bibfnamefont {T.}~\bibnamefont {Taniguchi}}, \bibinfo {author}
  {\bibfnamefont {K.}~\bibnamefont {Watanabe}}, \bibinfo {author}
  {\bibfnamefont {A.~H.}\ \bibnamefont {MacDonald}},\ and\ \bibinfo {author}
  {\bibfnamefont {E.}~\bibnamefont {Tutuc}},\ }\bibfield  {title} {\bibinfo
  {title} {{Correlated Insulating States in Twisted Double Bilayer Graphene}},\
  }\href {https://doi.org/10.1103/PhysRevLett.123.197702} {\bibfield  {journal}
  {\bibinfo  {journal} {Phys. Rev. Lett.}\ }\textbf {\bibinfo {volume} {123}},\
  \bibinfo {pages} {197702} (\bibinfo {year} {2019})}\BibitemShut {NoStop}%
\bibitem [{\citenamefont {Koshino}(2019)}]{Koshino2019}%
  \BibitemOpen
  \bibfield  {author} {\bibinfo {author} {\bibfnamefont {M.}~\bibnamefont
  {Koshino}},\ }\bibfield  {title} {\bibinfo {title} {{Band structure and
  topological properties of twisted double bilayer graphene}},\ }\href
  {https://doi.org/10.1103/PhysRevB.99.235406} {\bibfield  {journal} {\bibinfo
  {journal} {Phys. Rev. B}\ }\textbf {\bibinfo {volume} {99}},\ \bibinfo
  {pages} {235406} (\bibinfo {year} {2019})}\BibitemShut {NoStop}%
\bibitem [{\citenamefont {He}\ \emph {et~al.}(2021)\citenamefont {He},
  \citenamefont {Li}, \citenamefont {Cai}, \citenamefont {Liu}, \citenamefont
  {Watanabe}, \citenamefont {Taniguchi}, \citenamefont {Xu},\ and\
  \citenamefont {Yankowitz}}]{He_2021}%
  \BibitemOpen
  \bibfield  {author} {\bibinfo {author} {\bibfnamefont {M.}~\bibnamefont
  {He}}, \bibinfo {author} {\bibfnamefont {Y.}~\bibnamefont {Li}}, \bibinfo
  {author} {\bibfnamefont {J.}~\bibnamefont {Cai}}, \bibinfo {author}
  {\bibfnamefont {Y.}~\bibnamefont {Liu}}, \bibinfo {author} {\bibfnamefont
  {K.}~\bibnamefont {Watanabe}}, \bibinfo {author} {\bibfnamefont
  {T.}~\bibnamefont {Taniguchi}}, \bibinfo {author} {\bibfnamefont
  {X.}~\bibnamefont {Xu}},\ and\ \bibinfo {author} {\bibfnamefont
  {M.}~\bibnamefont {Yankowitz}},\ }\bibfield  {title} {\bibinfo {title}
  {{Symmetry breaking in twisted double bilayer graphene}},\ }\href@noop {}
  {\bibfield  {journal} {\bibinfo  {journal} {Nat. Phys. 17}\ ,\ \bibinfo
  {pages} {26–30}} (\bibinfo {year} {2021})}\BibitemShut {NoStop}%
\bibitem [{\citenamefont {Zhang}\ \emph {et~al.}(2021)\citenamefont {Zhang},
  \citenamefont {Zhu}, \citenamefont {Kahn}, \citenamefont {Li}, \citenamefont
  {Yang}, \citenamefont {Herbig}, \citenamefont {Wu}, \citenamefont {Li},
  \citenamefont {Watanabe}, \citenamefont {Taniguchi}, \citenamefont {Cabrini},
  \citenamefont {Zettl}, \citenamefont {Zaletel}, \citenamefont {Wang},\ and\
  \citenamefont {Crommie}}]{Zhang_2021}%
  \BibitemOpen
  \bibfield  {author} {\bibinfo {author} {\bibfnamefont {C.}~\bibnamefont
  {Zhang}}, \bibinfo {author} {\bibfnamefont {T.}~\bibnamefont {Zhu}}, \bibinfo
  {author} {\bibfnamefont {S.}~\bibnamefont {Kahn}}, \bibinfo {author}
  {\bibfnamefont {S.}~\bibnamefont {Li}}, \bibinfo {author} {\bibfnamefont
  {B.}~\bibnamefont {Yang}}, \bibinfo {author} {\bibfnamefont {C.}~\bibnamefont
  {Herbig}}, \bibinfo {author} {\bibfnamefont {X.}~\bibnamefont {Wu}}, \bibinfo
  {author} {\bibfnamefont {H.}~\bibnamefont {Li}}, \bibinfo {author}
  {\bibfnamefont {K.}~\bibnamefont {Watanabe}}, \bibinfo {author}
  {\bibfnamefont {T.}~\bibnamefont {Taniguchi}}, \bibinfo {author}
  {\bibfnamefont {S.}~\bibnamefont {Cabrini}}, \bibinfo {author} {\bibfnamefont
  {A.}~\bibnamefont {Zettl}}, \bibinfo {author} {\bibfnamefont {M.~P.}\
  \bibnamefont {Zaletel}}, \bibinfo {author} {\bibfnamefont {F.}~\bibnamefont
  {Wang}},\ and\ \bibinfo {author} {\bibfnamefont {M.~F.}\ \bibnamefont
  {Crommie}},\ }\bibfield  {title} {\bibinfo {title} {{Visualizing delocalized
  correlated electronic states in twisted double bilayer graphene}},\ }\href
  {https://doi.org/10.1038/s41467-021-22711-1} {\bibfield  {journal} {\bibinfo
  {journal} {Nat. Commun. 12}\ } (\bibinfo {year} {2021})}\BibitemShut
  {NoStop}%
\bibitem [{\citenamefont {Cao}\ \emph {et~al.}(2020)\citenamefont {Cao},
  \citenamefont {Rodan-Legrain}, \citenamefont {Rubies-Bigorda}, \citenamefont
  {Min~Park}, \citenamefont {Watanabe}, \citenamefont {Taniguchi},\ and\
  \citenamefont {Jarillo-Herrero}}]{Cao_2020}%
  \BibitemOpen
  \bibfield  {author} {\bibinfo {author} {\bibfnamefont {Y.}~\bibnamefont
  {Cao}}, \bibinfo {author} {\bibfnamefont {D.}~\bibnamefont {Rodan-Legrain}},
  \bibinfo {author} {\bibfnamefont {O.}~\bibnamefont {Rubies-Bigorda}},
  \bibinfo {author} {\bibfnamefont {J.}~\bibnamefont {Min~Park}}, \bibinfo
  {author} {\bibfnamefont {K.}~\bibnamefont {Watanabe}}, \bibinfo {author}
  {\bibfnamefont {T.}~\bibnamefont {Taniguchi}},\ and\ \bibinfo {author}
  {\bibfnamefont {P.}~\bibnamefont {Jarillo-Herrero}},\ }\bibfield  {title}
  {\bibinfo {title} {{Tunable correlated states and spin-polarized phases in
  twisted bilayer–bilayer graphene}},\ }\href
  {https://doi.org/10.1038/s41586-020-2260-6} {\bibfield  {journal} {\bibinfo
  {journal} {Nature 583}\ ,\ \bibinfo {pages} {215–220}} (\bibinfo {year}
  {2020})}\BibitemShut {NoStop}%
\bibitem [{\citenamefont {Fang}\ and\ \citenamefont
  {Kaxiras}(2016)}]{Fang_2016}%
  \BibitemOpen
  \bibfield  {author} {\bibinfo {author} {\bibfnamefont {S.}~\bibnamefont
  {Fang}}\ and\ \bibinfo {author} {\bibfnamefont {E.}~\bibnamefont {Kaxiras}},\
  }\bibfield  {title} {\bibinfo {title} {{Electronic structure theory of weakly
  interacting bilayers}},\ }\href {https://doi.org/10.1103/PhysRevB.93.235153}
  {\bibfield  {journal} {\bibinfo  {journal} {Phys. Rev. B}\ }\textbf {\bibinfo
  {volume} {93}},\ \bibinfo {pages} {235153} (\bibinfo {year}
  {2016})}\BibitemShut {NoStop}%
\bibitem [{\citenamefont {Culchac}\ \emph {et~al.}(2020)\citenamefont
  {Culchac}, \citenamefont {Del~Grande}, \citenamefont {Rodrigo}, \citenamefont
  {Chico},\ and\ \citenamefont {Morell}}]{Culchac_2020}%
  \BibitemOpen
  \bibfield  {author} {\bibinfo {author} {\bibfnamefont {F.}~\bibnamefont
  {Culchac}}, \bibinfo {author} {\bibfnamefont {R.~R.}\ \bibnamefont
  {Del~Grande}}, \bibinfo {author} {\bibfnamefont {B.~C.}\ \bibnamefont
  {Rodrigo}}, \bibinfo {author} {\bibfnamefont {L.}~\bibnamefont {Chico}},\
  and\ \bibinfo {author} {\bibfnamefont {E.~S.}\ \bibnamefont {Morell}},\
  }\bibfield  {title} {\bibinfo {title} {Flat bands and gaps in twisted double
  bilayer graphene},\ }\href {https://doi.org/10.1039/C9NR10830K} {\bibfield
  {journal} {\bibinfo  {journal} {Nanoscale 12}\ ,\ \bibinfo {pages} {5014}}
  (\bibinfo {year} {2020})}\BibitemShut {NoStop}%
\bibitem [{\citenamefont {Choi}\ and\ \citenamefont {Choi}(2019)}]{Choi_2019}%
  \BibitemOpen
  \bibfield  {author} {\bibinfo {author} {\bibfnamefont {Y.~W.}\ \bibnamefont
  {Choi}}\ and\ \bibinfo {author} {\bibfnamefont {H.~J.}\ \bibnamefont
  {Choi}},\ }\bibfield  {title} {\bibinfo {title} {{Intrinsic band gap and
  electrically tunable flat bands in twisted double bilayer graphene}},\ }\href
  {https://doi.org/10.1103/PhysRevB.100.201402} {\bibfield  {journal} {\bibinfo
   {journal} {Phys. Rev. B}\ }\textbf {\bibinfo {volume} {100}},\ \bibinfo
  {pages} {201402} (\bibinfo {year} {2019})}\BibitemShut {NoStop}%
\bibitem [{\citenamefont {Liu}\ \emph {et~al.}(2020)\citenamefont {Liu},
  \citenamefont {Hao}, \citenamefont {Khalaf}, \citenamefont {Lee},
  \citenamefont {Ronen}, \citenamefont {Yoo}, \citenamefont {Haei~Najafabadi},
  \citenamefont {Watanabe}, \citenamefont {Taniguchi}, \citenamefont
  {Vishwanath},\ and\ \citenamefont {et~al.}}]{Liu_2020}%
  \BibitemOpen
  \bibfield  {author} {\bibinfo {author} {\bibfnamefont {X.}~\bibnamefont
  {Liu}}, \bibinfo {author} {\bibfnamefont {Z.}~\bibnamefont {Hao}}, \bibinfo
  {author} {\bibfnamefont {E.}~\bibnamefont {Khalaf}}, \bibinfo {author}
  {\bibfnamefont {J.~Y.}\ \bibnamefont {Lee}}, \bibinfo {author} {\bibfnamefont
  {Y.}~\bibnamefont {Ronen}}, \bibinfo {author} {\bibfnamefont
  {H.}~\bibnamefont {Yoo}}, \bibinfo {author} {\bibfnamefont {D.}~\bibnamefont
  {Haei~Najafabadi}}, \bibinfo {author} {\bibfnamefont {K.}~\bibnamefont
  {Watanabe}}, \bibinfo {author} {\bibfnamefont {T.}~\bibnamefont {Taniguchi}},
  \bibinfo {author} {\bibfnamefont {A.}~\bibnamefont {Vishwanath}},\ and\
  \bibinfo {author} {\bibnamefont {et~al.}},\ }\bibfield  {title} {\bibinfo
  {title} {Tunable spin-polarized correlated states in twisted double bilayer
  graphene},\ }\href {https://doi.org/10.1038/s41586-020-2458-7} {\bibfield
  {journal} {\bibinfo  {journal} {Nature}\ }\textbf {\bibinfo {volume} {583}},\
  \bibinfo {pages} {221–225} (\bibinfo {year} {2020})}\BibitemShut {NoStop}%
\bibitem [{\citenamefont {Lee}\ \emph {et~al.}(2019)\citenamefont {Lee},
  \citenamefont {Khalaf}, \citenamefont {Liu}, \citenamefont {Liu},
  \citenamefont {Hao}, \citenamefont {Kim},\ and\ \citenamefont
  {Vishwanath}}]{Lee_2019}%
  \BibitemOpen
  \bibfield  {author} {\bibinfo {author} {\bibfnamefont {J.~Y.}\ \bibnamefont
  {Lee}}, \bibinfo {author} {\bibfnamefont {E.}~\bibnamefont {Khalaf}},
  \bibinfo {author} {\bibfnamefont {S.}~\bibnamefont {Liu}}, \bibinfo {author}
  {\bibfnamefont {X.}~\bibnamefont {Liu}}, \bibinfo {author} {\bibfnamefont
  {Z.}~\bibnamefont {Hao}}, \bibinfo {author} {\bibfnamefont {P.}~\bibnamefont
  {Kim}},\ and\ \bibinfo {author} {\bibfnamefont {A.}~\bibnamefont
  {Vishwanath}},\ }\bibfield  {title} {\bibinfo {title} {Theory of correlated
  insulating behaviour and spin-triplet superconductivity in twisted double
  bilayer graphene},\ }\href {https://doi.org/10.1038/s41467-019-12981-1}
  {\bibfield  {journal} {\bibinfo  {journal} {Nature Communications}\ }\textbf
  {\bibinfo {volume} {10}},\ \bibinfo {pages} {5333} (\bibinfo {year}
  {2019})}\BibitemShut {NoStop}%
\bibitem [{\citenamefont {Adak}\ \emph {et~al.}(2020)\citenamefont {Adak},
  \citenamefont {Sinha}, \citenamefont {Ghorai}, \citenamefont {Sangani},
  \citenamefont {Watanabe}, \citenamefont {Taniguchi}, \citenamefont
  {Sensarma},\ and\ \citenamefont {Deshmukh}}]{Adak_2020}%
  \BibitemOpen
  \bibfield  {author} {\bibinfo {author} {\bibfnamefont {P.~C.}\ \bibnamefont
  {Adak}}, \bibinfo {author} {\bibfnamefont {S.}~\bibnamefont {Sinha}},
  \bibinfo {author} {\bibfnamefont {U.}~\bibnamefont {Ghorai}}, \bibinfo
  {author} {\bibfnamefont {L.~D.~V.}\ \bibnamefont {Sangani}}, \bibinfo
  {author} {\bibfnamefont {K.}~\bibnamefont {Watanabe}}, \bibinfo {author}
  {\bibfnamefont {T.}~\bibnamefont {Taniguchi}}, \bibinfo {author}
  {\bibfnamefont {R.}~\bibnamefont {Sensarma}},\ and\ \bibinfo {author}
  {\bibfnamefont {M.~M.}\ \bibnamefont {Deshmukh}},\ }\bibfield  {title}
  {\bibinfo {title} {{Tunable bandwidths and gaps in twisted double bilayer
  graphene on the verge of correlations}},\ }\href
  {https://doi.org/10.1103/PhysRevB.101.125428} {\bibfield  {journal} {\bibinfo
   {journal} {Phys. Rev. B}\ }\textbf {\bibinfo {volume} {101}},\ \bibinfo
  {pages} {125428} (\bibinfo {year} {2020})}\BibitemShut {NoStop}%
\bibitem [{\citenamefont {Zhu}\ \emph {et~al.}(2020)\citenamefont {Zhu},
  \citenamefont {Carr}, \citenamefont {Massatt}, \citenamefont {Luskin},\ and\
  \citenamefont {Kaxiras}}]{Kaxiras}%
  \BibitemOpen
  \bibfield  {author} {\bibinfo {author} {\bibfnamefont {Z.}~\bibnamefont
  {Zhu}}, \bibinfo {author} {\bibfnamefont {S.}~\bibnamefont {Carr}}, \bibinfo
  {author} {\bibfnamefont {D.}~\bibnamefont {Massatt}}, \bibinfo {author}
  {\bibfnamefont {M.}~\bibnamefont {Luskin}},\ and\ \bibinfo {author}
  {\bibfnamefont {E.}~\bibnamefont {Kaxiras}},\ }\bibfield  {title} {\bibinfo
  {title} {Twisted trilayer graphene: a precisely tunable platform for
  correlated electrons},\ }\href
  {https://doi.org/https://doi.org/10.1103/PhysRevLett.125.11640} {\bibfield
  {journal} {\bibinfo  {journal} {Phys. Rev. Lett.}\ }\textbf {\bibinfo
  {volume} {125}},\ \bibinfo {pages} {116404} (\bibinfo {year}
  {2020})}\BibitemShut {NoStop}%
\bibitem [{\citenamefont {Park}\ \emph {et~al.}(2021)\citenamefont {Park},
  \citenamefont {Cao}, \citenamefont {Watanabe}, \citenamefont {Taniguchi},\
  and\ \citenamefont {Jarillo-Herrero}}]{Park_2021}%
  \BibitemOpen
  \bibfield  {author} {\bibinfo {author} {\bibfnamefont {J.~M.}\ \bibnamefont
  {Park}}, \bibinfo {author} {\bibfnamefont {Y.}~\bibnamefont {Cao}}, \bibinfo
  {author} {\bibfnamefont {K.}~\bibnamefont {Watanabe}}, \bibinfo {author}
  {\bibfnamefont {T.}~\bibnamefont {Taniguchi}},\ and\ \bibinfo {author}
  {\bibfnamefont {P.}~\bibnamefont {Jarillo-Herrero}},\ }\bibfield  {title}
  {\bibinfo {title} {Tunable strongly coupled superconductivity in magic-angle
  twisted trilayer graphene},\ }\href
  {https://doi.org/10.1038/s41586-021-03192-0} {\bibfield  {journal} {\bibinfo
  {journal} {Nature}\ }\textbf {\bibinfo {volume} {590}},\ \bibinfo {pages}
  {249} (\bibinfo {year} {2021})}\BibitemShut {NoStop}%
\bibitem [{\citenamefont {Hao}\ \emph {et~al.}(2021)\citenamefont {Hao},
  \citenamefont {Zimmerman}, \citenamefont {Ledwith}, \citenamefont {Khalaf},
  \citenamefont {Najafabadi}, \citenamefont {Watanabe}, \citenamefont
  {Taniguchi}, \citenamefont {Vishwanath},\ and\ \citenamefont
  {Kim}}]{Hao_2021}%
  \BibitemOpen
  \bibfield  {author} {\bibinfo {author} {\bibfnamefont {Z.}~\bibnamefont
  {Hao}}, \bibinfo {author} {\bibfnamefont {A.~M.}\ \bibnamefont {Zimmerman}},
  \bibinfo {author} {\bibfnamefont {P.}~\bibnamefont {Ledwith}}, \bibinfo
  {author} {\bibfnamefont {E.}~\bibnamefont {Khalaf}}, \bibinfo {author}
  {\bibfnamefont {D.~H.}\ \bibnamefont {Najafabadi}}, \bibinfo {author}
  {\bibfnamefont {K.}~\bibnamefont {Watanabe}}, \bibinfo {author}
  {\bibfnamefont {T.}~\bibnamefont {Taniguchi}}, \bibinfo {author}
  {\bibfnamefont {A.}~\bibnamefont {Vishwanath}},\ and\ \bibinfo {author}
  {\bibfnamefont {P.}~\bibnamefont {Kim}},\ }\bibfield  {title} {\bibinfo
  {title} {{Electric field{\textendash}tunable superconductivity in
  alternating-twist magic-angle trilayer graphene}},\ }\href
  {https://doi.org/10.1126/science.abg0399} {\bibfield  {journal} {\bibinfo
  {journal} {Science}\ }\textbf {\bibinfo {volume} {371}},\ \bibinfo {pages}
  {1133} (\bibinfo {year} {2021})}\BibitemShut {NoStop}%
\bibitem [{\citenamefont {Su\'arez~Morell}\ \emph {et~al.}(2013)\citenamefont
  {Su\'arez~Morell}, \citenamefont {Pacheco}, \citenamefont {Chico},\ and\
  \citenamefont {Brey}}]{Suarez_2013}%
  \BibitemOpen
  \bibfield  {author} {\bibinfo {author} {\bibfnamefont {E.}~\bibnamefont
  {Su\'arez~Morell}}, \bibinfo {author} {\bibfnamefont {M.}~\bibnamefont
  {Pacheco}}, \bibinfo {author} {\bibfnamefont {L.}~\bibnamefont {Chico}},\
  and\ \bibinfo {author} {\bibfnamefont {L.}~\bibnamefont {Brey}},\ }\bibfield
  {title} {\bibinfo {title} {{Electronic properties of twisted trilayer
  graphene}},\ }\href {https://doi.org/10.1103/PhysRevB.87.125414} {\bibfield
  {journal} {\bibinfo  {journal} {Phys. Rev. B}\ }\textbf {\bibinfo {volume}
  {87}},\ \bibinfo {pages} {125414} (\bibinfo {year} {2013})}\BibitemShut
  {NoStop}%
\bibitem [{\citenamefont {Chen}\ \emph {et~al.}(2016)\citenamefont {Chen},
  \citenamefont {Xin}, \citenamefont {Jiang}, \citenamefont {Liu},
  \citenamefont {Chen},\ and\ \citenamefont {Tian}}]{Chen_2016}%
  \BibitemOpen
  \bibfield  {author} {\bibinfo {author} {\bibfnamefont {X.-D.}\ \bibnamefont
  {Chen}}, \bibinfo {author} {\bibfnamefont {W.}~\bibnamefont {Xin}}, \bibinfo
  {author} {\bibfnamefont {W.-S.}\ \bibnamefont {Jiang}}, \bibinfo {author}
  {\bibfnamefont {Z.-B.}\ \bibnamefont {Liu}}, \bibinfo {author} {\bibfnamefont
  {Y.}~\bibnamefont {Chen}},\ and\ \bibinfo {author} {\bibfnamefont {J.-G.}\
  \bibnamefont {Tian}},\ }\bibfield  {title} {\bibinfo {title} {{High-Precision
  Twist-Controlled Bilayer and Trilayer Graphene}},\ }\href
  {https://doi.org/https://doi.org/10.1002/adma.201505129} {\bibfield
  {journal} {\bibinfo  {journal} {Advanced Materials}\ }\textbf {\bibinfo
  {volume} {28}},\ \bibinfo {pages} {2563} (\bibinfo {year}
  {2016})}\BibitemShut {NoStop}%
\bibitem [{\citenamefont {Zuo}\ \emph {et~al.}(2018)\citenamefont {Zuo},
  \citenamefont {Qiao}, \citenamefont {Ma}, \citenamefont {Yin}, \citenamefont
  {Sun}, \citenamefont {Zhang}, \citenamefont {Guan},\ and\ \citenamefont
  {He}}]{Zuo_2018}%
  \BibitemOpen
  \bibfield  {author} {\bibinfo {author} {\bibfnamefont {W.-J.}\ \bibnamefont
  {Zuo}}, \bibinfo {author} {\bibfnamefont {J.-B.}\ \bibnamefont {Qiao}},
  \bibinfo {author} {\bibfnamefont {D.-L.}\ \bibnamefont {Ma}}, \bibinfo
  {author} {\bibfnamefont {L.-J.}\ \bibnamefont {Yin}}, \bibinfo {author}
  {\bibfnamefont {G.}~\bibnamefont {Sun}}, \bibinfo {author} {\bibfnamefont
  {J.-Y.}\ \bibnamefont {Zhang}}, \bibinfo {author} {\bibfnamefont {L.-Y.}\
  \bibnamefont {Guan}},\ and\ \bibinfo {author} {\bibfnamefont
  {L.}~\bibnamefont {He}},\ }\bibfield  {title} {\bibinfo {title} {{Scanning
  tunneling microscopy and spectroscopy of twisted trilayer graphene}},\ }\href
  {https://doi.org/10.1103/PhysRevB.97.035440} {\bibfield  {journal} {\bibinfo
  {journal} {Phys. Rev. B}\ }\textbf {\bibinfo {volume} {97}},\ \bibinfo
  {pages} {035440} (\bibinfo {year} {2018})}\BibitemShut {NoStop}%
\bibitem [{\citenamefont {Ma}\ \emph {et~al.}(2021)\citenamefont {Ma},
  \citenamefont {Li}, \citenamefont {Zheng}, \citenamefont {Xiao},
  \citenamefont {Jiang}, \citenamefont {Gao},\ and\ \citenamefont
  {Xie}}]{Ma_2021}%
  \BibitemOpen
  \bibfield  {author} {\bibinfo {author} {\bibfnamefont {Z.}~\bibnamefont
  {Ma}}, \bibinfo {author} {\bibfnamefont {S.}~\bibnamefont {Li}}, \bibinfo
  {author} {\bibfnamefont {Y.-W.}\ \bibnamefont {Zheng}}, \bibinfo {author}
  {\bibfnamefont {M.-M.}\ \bibnamefont {Xiao}}, \bibinfo {author}
  {\bibfnamefont {H.}~\bibnamefont {Jiang}}, \bibinfo {author} {\bibfnamefont
  {J.-H.}\ \bibnamefont {Gao}},\ and\ \bibinfo {author} {\bibfnamefont
  {X.}~\bibnamefont {Xie}},\ }\bibfield  {title} {\bibinfo {title}
  {{Topological flat bands in twisted trilayer graphene}},\ }\href
  {https://doi.org/https://doi.org/10.1016/j.scib.2020.10.004} {\bibfield
  {journal} {\bibinfo  {journal} {Science Bulletin}\ }\textbf {\bibinfo
  {volume} {66}},\ \bibinfo {pages} {18} (\bibinfo {year} {2021})}\BibitemShut
  {NoStop}%
\bibitem [{\citenamefont {Xu}\ \emph {et~al.}(2021)\citenamefont {Xu},
  \citenamefont {Al~Ezzi}, \citenamefont {Balakrishnan}, \citenamefont
  {Garcia-Ruiz}, \citenamefont {Tsim}, \citenamefont {Mullan}, \citenamefont
  {Barrier}, \citenamefont {Xin}, \citenamefont {Piot}, \citenamefont
  {Taniguchi},\ and\ \citenamefont {et~al.}}]{Xu_2021}%
  \BibitemOpen
  \bibfield  {author} {\bibinfo {author} {\bibfnamefont {S.}~\bibnamefont
  {Xu}}, \bibinfo {author} {\bibfnamefont {M.~M.}\ \bibnamefont {Al~Ezzi}},
  \bibinfo {author} {\bibfnamefont {N.}~\bibnamefont {Balakrishnan}}, \bibinfo
  {author} {\bibfnamefont {A.}~\bibnamefont {Garcia-Ruiz}}, \bibinfo {author}
  {\bibfnamefont {B.}~\bibnamefont {Tsim}}, \bibinfo {author} {\bibfnamefont
  {C.}~\bibnamefont {Mullan}}, \bibinfo {author} {\bibfnamefont
  {J.}~\bibnamefont {Barrier}}, \bibinfo {author} {\bibfnamefont
  {N.}~\bibnamefont {Xin}}, \bibinfo {author} {\bibfnamefont {B.~A.}\
  \bibnamefont {Piot}}, \bibinfo {author} {\bibfnamefont {T.}~\bibnamefont
  {Taniguchi}},\ and\ \bibinfo {author} {\bibnamefont {et~al.}},\ }\bibfield
  {title} {\bibinfo {title} {Tunable van hove singularities and correlated
  states in twisted monolayer–bilayer graphene},\ }\href
  {https://doi.org/10.1038/s41567-021-01172-9} {\bibfield  {journal} {\bibinfo
  {journal} {Nature Physics}\ }\textbf {\bibinfo {volume} {17}},\ \bibinfo
  {pages} {619–626} (\bibinfo {year} {2021})}\BibitemShut {NoStop}%
\bibitem [{\citenamefont {Li}\ \emph {et~al.}(2019)\citenamefont {Li},
  \citenamefont {Wu},\ and\ \citenamefont {MacDonald}}]{Li_2019}%
  \BibitemOpen
  \bibfield  {author} {\bibinfo {author} {\bibfnamefont {X.}~\bibnamefont
  {Li}}, \bibinfo {author} {\bibfnamefont {F.}~\bibnamefont {Wu}},\ and\
  \bibinfo {author} {\bibfnamefont {A.~H.}\ \bibnamefont {MacDonald}},\
  }\href@noop {} {\bibinfo {title} {Electronic structure of single-twist
  trilayer graphene}} (\bibinfo {year} {2019}),\ \Eprint
  {https://arxiv.org/abs/1907.12338} {arXiv:1907.12338 [cond-mat.mtrl-sci]}
  \BibitemShut {NoStop}%
\bibitem [{\citenamefont {Khalaf}\ \emph {et~al.}(2019)\citenamefont {Khalaf},
  \citenamefont {Kruchkov}, \citenamefont {Tarnopolsky},\ and\ \citenamefont
  {Vishwanath}}]{Hierarchy}%
  \BibitemOpen
  \bibfield  {author} {\bibinfo {author} {\bibfnamefont {E.}~\bibnamefont
  {Khalaf}}, \bibinfo {author} {\bibfnamefont {A.~J.}\ \bibnamefont
  {Kruchkov}}, \bibinfo {author} {\bibfnamefont {G.}~\bibnamefont
  {Tarnopolsky}},\ and\ \bibinfo {author} {\bibfnamefont {A.}~\bibnamefont
  {Vishwanath}},\ }\bibfield  {title} {\bibinfo {title} {Magic angle hierarchy
  in twisted graphene multilayers},\ }\href
  {https://doi.org/https://doi.org/10.1103/PhysRevB.100.085109} {\bibfield
  {journal} {\bibinfo  {journal} {Phys. Rev. B}\ }\textbf {\bibinfo {volume}
  {100}},\ \bibinfo {pages} {085109} (\bibinfo {year} {2019})}\BibitemShut
  {NoStop}%
\bibitem [{\citenamefont {Denner}\ \emph {et~al.}(2020)\citenamefont {Denner},
  \citenamefont {Lado},\ and\ \citenamefont {Zilberberg}}]{Denner2020}%
  \BibitemOpen
  \bibfield  {author} {\bibinfo {author} {\bibfnamefont {M.~M.}\ \bibnamefont
  {Denner}}, \bibinfo {author} {\bibfnamefont {J.~L.}\ \bibnamefont {Lado}},\
  and\ \bibinfo {author} {\bibfnamefont {O.}~\bibnamefont {Zilberberg}},\
  }\bibfield  {title} {\bibinfo {title} {{Antichiral states in twisted graphene
  multilayers}},\ }\href {https://doi.org/10.1103/PhysRevResearch.2.043190}
  {\bibfield  {journal} {\bibinfo  {journal} {Phys. Rev. Research}\ }\textbf
  {\bibinfo {volume} {2}},\ \bibinfo {pages} {043190} (\bibinfo {year}
  {2020})}\BibitemShut {NoStop}%
\bibitem [{\citenamefont {Tritsaris}\ \emph {et~al.}(2020)\citenamefont
  {Tritsaris}, \citenamefont {Carr}, \citenamefont {Zhu}, \citenamefont {Xie},
  \citenamefont {Torrisi}, \citenamefont {Tang}, \citenamefont {Mattheakis},
  \citenamefont {Larson},\ and\ \citenamefont {Kaxiras}}]{Tritsaris_2020}%
  \BibitemOpen
  \bibfield  {author} {\bibinfo {author} {\bibfnamefont {G.~A.}\ \bibnamefont
  {Tritsaris}}, \bibinfo {author} {\bibfnamefont {S.}~\bibnamefont {Carr}},
  \bibinfo {author} {\bibfnamefont {Z.}~\bibnamefont {Zhu}}, \bibinfo {author}
  {\bibfnamefont {Y.}~\bibnamefont {Xie}}, \bibinfo {author} {\bibfnamefont
  {S.~B.}\ \bibnamefont {Torrisi}}, \bibinfo {author} {\bibfnamefont
  {J.}~\bibnamefont {Tang}}, \bibinfo {author} {\bibfnamefont {M.}~\bibnamefont
  {Mattheakis}}, \bibinfo {author} {\bibfnamefont {D.~T.}\ \bibnamefont
  {Larson}},\ and\ \bibinfo {author} {\bibfnamefont {E.}~\bibnamefont
  {Kaxiras}},\ }\bibfield  {title} {\bibinfo {title} {{Electronic structure
  calculations of twisted multi-layer graphene superlattices}},\ }\href
  {https://doi.org/10.1088/2053-1583/ab8f62} {\bibfield  {journal} {\bibinfo
  {journal} {2D Materials}\ }\textbf {\bibinfo {volume} {7}},\ \bibinfo {pages}
  {035028} (\bibinfo {year} {2020})}\BibitemShut {NoStop}%
\bibitem [{\citenamefont {Gupta}\ \emph {et~al.}(2020)\citenamefont {Gupta},
  \citenamefont {Walia}, \citenamefont {Mogera},\ and\ \citenamefont
  {Kulkarni}}]{Gupta2020}%
  \BibitemOpen
  \bibfield  {author} {\bibinfo {author} {\bibfnamefont {N.}~\bibnamefont
  {Gupta}}, \bibinfo {author} {\bibfnamefont {S.}~\bibnamefont {Walia}},
  \bibinfo {author} {\bibfnamefont {U.}~\bibnamefont {Mogera}},\ and\ \bibinfo
  {author} {\bibfnamefont {G.~U.}\ \bibnamefont {Kulkarni}},\ }\bibfield
  {title} {\bibinfo {title} {{Twist-Dependent Raman and Electron Diffraction
  Correlations in Twisted Multilayer Graphene}},\ }\href
  {https://doi.org/10.1021/acs.jpclett.0c00582} {\bibfield  {journal} {\bibinfo
   {journal} {The Journal of Physical Chemistry Letters}\ }\textbf {\bibinfo
  {volume} {11}},\ \bibinfo {pages} {2797} (\bibinfo {year}
  {2020})}\BibitemShut {NoStop}%
\bibitem [{\citenamefont {Kerelsky}\ \emph {et~al.}(2021)\citenamefont
  {Kerelsky}, \citenamefont {Rubio-Verd{\'u}}, \citenamefont {Xian},
  \citenamefont {Kennes}, \citenamefont {Halbertal}, \citenamefont {Finney},
  \citenamefont {Song}, \citenamefont {Turkel}, \citenamefont {Wang},
  \citenamefont {Watanabe}, \citenamefont {Taniguchi}, \citenamefont {Hone},
  \citenamefont {Dean}, \citenamefont {Basov}, \citenamefont {Rubio},\ and\
  \citenamefont {Pasupathy}}]{Kerelskye_2021}%
  \BibitemOpen
  \bibfield  {author} {\bibinfo {author} {\bibfnamefont {A.}~\bibnamefont
  {Kerelsky}}, \bibinfo {author} {\bibfnamefont {C.}~\bibnamefont
  {Rubio-Verd{\'u}}}, \bibinfo {author} {\bibfnamefont {L.}~\bibnamefont
  {Xian}}, \bibinfo {author} {\bibfnamefont {D.~M.}\ \bibnamefont {Kennes}},
  \bibinfo {author} {\bibfnamefont {D.}~\bibnamefont {Halbertal}}, \bibinfo
  {author} {\bibfnamefont {N.}~\bibnamefont {Finney}}, \bibinfo {author}
  {\bibfnamefont {L.}~\bibnamefont {Song}}, \bibinfo {author} {\bibfnamefont
  {S.}~\bibnamefont {Turkel}}, \bibinfo {author} {\bibfnamefont
  {L.}~\bibnamefont {Wang}}, \bibinfo {author} {\bibfnamefont {K.}~\bibnamefont
  {Watanabe}}, \bibinfo {author} {\bibfnamefont {T.}~\bibnamefont {Taniguchi}},
  \bibinfo {author} {\bibfnamefont {J.}~\bibnamefont {Hone}}, \bibinfo {author}
  {\bibfnamefont {C.}~\bibnamefont {Dean}}, \bibinfo {author} {\bibfnamefont
  {D.~N.}\ \bibnamefont {Basov}}, \bibinfo {author} {\bibfnamefont
  {A.}~\bibnamefont {Rubio}},\ and\ \bibinfo {author} {\bibfnamefont {A.~N.}\
  \bibnamefont {Pasupathy}},\ }\bibfield  {title} {\bibinfo {title}
  {{Moir{\'e}less correlations in ABCA graphene}},\ }\href
  {https://doi.org/10.1073/pnas.2017366118} {\bibfield  {journal} {\bibinfo
  {journal} {Proceedings of the National Academy of Sciences}\ }\textbf
  {\bibinfo {volume} {118}},\ \bibinfo {pages} {e2017366118} (\bibinfo {year}
  {2021})}\BibitemShut {NoStop}%
\bibitem [{\citenamefont {Zhou}\ \emph
  {et~al.}(2021{\natexlab{a}})\citenamefont {Zhou}, \citenamefont {Xie},
  \citenamefont {Taniguchi}, \citenamefont {Watanabe},\ and\ \citenamefont
  {Young}}]{2021}%
  \BibitemOpen
  \bibfield  {author} {\bibinfo {author} {\bibfnamefont {H.}~\bibnamefont
  {Zhou}}, \bibinfo {author} {\bibfnamefont {T.}~\bibnamefont {Xie}}, \bibinfo
  {author} {\bibfnamefont {T.}~\bibnamefont {Taniguchi}}, \bibinfo {author}
  {\bibfnamefont {K.}~\bibnamefont {Watanabe}},\ and\ \bibinfo {author}
  {\bibfnamefont {A.~F.}\ \bibnamefont {Young}},\ }\bibfield  {title} {\bibinfo
  {title} {Superconductivity in rhombohedral trilayer graphene},\ }\href
  {https://doi.org/10.1038/s41586-021-03926-0} {\bibfield  {journal} {\bibinfo
  {journal} {Nature}\ }\textbf {\bibinfo {volume} {598}},\ \bibinfo {pages}
  {434–438} (\bibinfo {year} {2021}{\natexlab{a}})}\BibitemShut {NoStop}%
\bibitem [{\citenamefont {Zhou}\ \emph
  {et~al.}(2021{\natexlab{b}})\citenamefont {Zhou}, \citenamefont {Holleis},
  \citenamefont {Saito}, \citenamefont {Cohen}, \citenamefont {Huynh},
  \citenamefont {Patterson}, \citenamefont {Yang}, \citenamefont {Taniguchi},
  \citenamefont {Watanabe},\ and\ \citenamefont {Young}}]{zhou2021isospin}%
  \BibitemOpen
  \bibfield  {author} {\bibinfo {author} {\bibfnamefont {H.}~\bibnamefont
  {Zhou}}, \bibinfo {author} {\bibfnamefont {L.}~\bibnamefont {Holleis}},
  \bibinfo {author} {\bibfnamefont {Y.}~\bibnamefont {Saito}}, \bibinfo
  {author} {\bibfnamefont {L.}~\bibnamefont {Cohen}}, \bibinfo {author}
  {\bibfnamefont {W.}~\bibnamefont {Huynh}}, \bibinfo {author} {\bibfnamefont
  {C.~L.}\ \bibnamefont {Patterson}}, \bibinfo {author} {\bibfnamefont
  {F.}~\bibnamefont {Yang}}, \bibinfo {author} {\bibfnamefont {T.}~\bibnamefont
  {Taniguchi}}, \bibinfo {author} {\bibfnamefont {K.}~\bibnamefont
  {Watanabe}},\ and\ \bibinfo {author} {\bibfnamefont {A.~F.}\ \bibnamefont
  {Young}},\ }\href@noop {} {\bibinfo {title} {Isospin magnetism and
  spin-triplet superconductivity in bernal bilayer graphene}} (\bibinfo {year}
  {2021}{\natexlab{b}}),\ \Eprint {https://arxiv.org/abs/2110.11317}
  {arXiv:2110.11317 [cond-mat.mes-hall]} \BibitemShut {NoStop}%
\bibitem [{\citenamefont {de~la Barrera}\ \emph {et~al.}(2021)\citenamefont
  {de~la Barrera}, \citenamefont {Aronson}, \citenamefont {Zheng},
  \citenamefont {Watanabe}, \citenamefont {Taniguchi}, \citenamefont {Ma},
  \citenamefont {Jarillo-Herrero},\ and\ \citenamefont
  {Ashoori}}]{delabarrera2021cascade}%
  \BibitemOpen
  \bibfield  {author} {\bibinfo {author} {\bibfnamefont {S.~C.}\ \bibnamefont
  {de~la Barrera}}, \bibinfo {author} {\bibfnamefont {S.}~\bibnamefont
  {Aronson}}, \bibinfo {author} {\bibfnamefont {Z.}~\bibnamefont {Zheng}},
  \bibinfo {author} {\bibfnamefont {K.}~\bibnamefont {Watanabe}}, \bibinfo
  {author} {\bibfnamefont {T.}~\bibnamefont {Taniguchi}}, \bibinfo {author}
  {\bibfnamefont {Q.}~\bibnamefont {Ma}}, \bibinfo {author} {\bibfnamefont
  {P.}~\bibnamefont {Jarillo-Herrero}},\ and\ \bibinfo {author} {\bibfnamefont
  {R.}~\bibnamefont {Ashoori}},\ }\href@noop {} {\bibinfo {title} {Cascade of
  isospin phase transitions in bernal bilayer graphene at zero magnetic field}}
  (\bibinfo {year} {2021}),\ \Eprint {https://arxiv.org/abs/2110.13907}
  {arXiv:2110.13907 [cond-mat.mes-hall]} \BibitemShut {NoStop}%
\bibitem [{\citenamefont {Seiler}\ \emph {et~al.}(2021)\citenamefont {Seiler},
  \citenamefont {Geisenhof}, \citenamefont {Winterer}, \citenamefont
  {Watanabe}, \citenamefont {Taniguchi}, \citenamefont {Xu}, \citenamefont
  {Zhang},\ and\ \citenamefont {Weitz}}]{seiler2021quantum}%
  \BibitemOpen
  \bibfield  {author} {\bibinfo {author} {\bibfnamefont {A.~M.}\ \bibnamefont
  {Seiler}}, \bibinfo {author} {\bibfnamefont {F.~R.}\ \bibnamefont
  {Geisenhof}}, \bibinfo {author} {\bibfnamefont {F.}~\bibnamefont {Winterer}},
  \bibinfo {author} {\bibfnamefont {K.}~\bibnamefont {Watanabe}}, \bibinfo
  {author} {\bibfnamefont {T.}~\bibnamefont {Taniguchi}}, \bibinfo {author}
  {\bibfnamefont {T.}~\bibnamefont {Xu}}, \bibinfo {author} {\bibfnamefont
  {F.}~\bibnamefont {Zhang}},\ and\ \bibinfo {author} {\bibfnamefont {R.~T.}\
  \bibnamefont {Weitz}},\ }\href@noop {} {\bibinfo {title} {Quantum cascade of
  new correlated phases in trigonally warped bilayer graphene}} (\bibinfo
  {year} {2021}),\ \Eprint {https://arxiv.org/abs/2111.06413} {arXiv:2111.06413
  [cond-mat.mes-hall]} \BibitemShut {NoStop}%
\bibitem [{\citenamefont {Li}\ \emph {et~al.}(2022)\citenamefont {Li},
  \citenamefont {Eaton}, \citenamefont {Fertig},\ and\ \citenamefont
  {Seradjeh}}]{Li_2022}%
  \BibitemOpen
  \bibfield  {author} {\bibinfo {author} {\bibfnamefont {Y.}~\bibnamefont
  {Li}}, \bibinfo {author} {\bibfnamefont {A.}~\bibnamefont {Eaton}}, \bibinfo
  {author} {\bibfnamefont {H.~A.}\ \bibnamefont {Fertig}},\ and\ \bibinfo
  {author} {\bibfnamefont {B.}~\bibnamefont {Seradjeh}},\ }\bibfield  {title}
  {\bibinfo {title} {{Dirac Magic and Lifshitz Transitions in AA-Stacked
  Twisted Multilayer Graphene}},\ }\href
  {https://doi.org/10.1103/PhysRevLett.128.026404} {\bibfield  {journal}
  {\bibinfo  {journal} {Phys. Rev. Lett.}\ }\textbf {\bibinfo {volume} {128}},\
  \bibinfo {pages} {026404} (\bibinfo {year} {2022})}\BibitemShut {NoStop}%
\bibitem [{\citenamefont {Tarnopolsky}\ \emph
  {et~al.}(2019{\natexlab{a}})\citenamefont {Tarnopolsky}, \citenamefont
  {Kruchkov},\ and\ \citenamefont {Vishwanath}}]{Tarnopolsky_2019}%
  \BibitemOpen
  \bibfield  {author} {\bibinfo {author} {\bibfnamefont {G.}~\bibnamefont
  {Tarnopolsky}}, \bibinfo {author} {\bibfnamefont {A.~J.}\ \bibnamefont
  {Kruchkov}},\ and\ \bibinfo {author} {\bibfnamefont {A.}~\bibnamefont
  {Vishwanath}},\ }\bibfield  {title} {\bibinfo {title} {{Origin of Magic
  Angles in Twisted Bilayer Graphene}},\ }\href
  {https://doi.org/10.1103/PhysRevLett.122.106405} {\bibfield  {journal}
  {\bibinfo  {journal} {Phys. Rev. Lett.}\ }\textbf {\bibinfo {volume} {122}},\
  \bibinfo {pages} {106405} (\bibinfo {year} {2019}{\natexlab{a}})}\BibitemShut
  {NoStop}%
\bibitem [{\citenamefont {Hejazi}\ \emph {et~al.}(2019)\citenamefont {Hejazi},
  \citenamefont {Liu}, \citenamefont {Shapourian}, \citenamefont {Chen},\ and\
  \citenamefont {Balents}}]{Balents}%
  \BibitemOpen
  \bibfield  {author} {\bibinfo {author} {\bibfnamefont {K.}~\bibnamefont
  {Hejazi}}, \bibinfo {author} {\bibfnamefont {C.}~\bibnamefont {Liu}},
  \bibinfo {author} {\bibfnamefont {H.}~\bibnamefont {Shapourian}}, \bibinfo
  {author} {\bibfnamefont {X.}~\bibnamefont {Chen}},\ and\ \bibinfo {author}
  {\bibfnamefont {L.}~\bibnamefont {Balents}},\ }\bibfield  {title} {\bibinfo
  {title} {Multiple topological transitions in twisted bilayer graphene near
  the first magic angle},\ }\href
  {https://doi.org/https://doi.org/10.1103/PhysRevB.99.035111} {\bibfield
  {journal} {\bibinfo  {journal} {Phys. Rev. B}\ }\textbf {\bibinfo {volume}
  {99}},\ \bibinfo {pages} {035111} (\bibinfo {year} {2019})}\BibitemShut
  {NoStop}%
\bibitem [{\citenamefont {Nam}\ and\ \citenamefont {Koshino}(2017)}]{Nam_2017}%
  \BibitemOpen
  \bibfield  {author} {\bibinfo {author} {\bibfnamefont {N.~N.~T.}\
  \bibnamefont {Nam}}\ and\ \bibinfo {author} {\bibfnamefont {M.}~\bibnamefont
  {Koshino}},\ }\bibfield  {title} {\bibinfo {title} {{Lattice relaxation and
  energy band modulation in twisted bilayer graphene}},\ }\href
  {https://doi.org/10.1103/PhysRevB.96.075311} {\bibfield  {journal} {\bibinfo
  {journal} {Phys. Rev. B}\ }\textbf {\bibinfo {volume} {96}},\ \bibinfo
  {pages} {075311} (\bibinfo {year} {2017})}\BibitemShut {NoStop}%
\bibitem [{\citenamefont {Zou}\ \emph {et~al.}(2018)\citenamefont {Zou},
  \citenamefont {Po}, \citenamefont {Vishwanath},\ and\ \citenamefont
  {Senthil}}]{Zou_2018}%
  \BibitemOpen
  \bibfield  {author} {\bibinfo {author} {\bibfnamefont {L.}~\bibnamefont
  {Zou}}, \bibinfo {author} {\bibfnamefont {H.~C.}\ \bibnamefont {Po}},
  \bibinfo {author} {\bibfnamefont {A.}~\bibnamefont {Vishwanath}},\ and\
  \bibinfo {author} {\bibfnamefont {T.}~\bibnamefont {Senthil}},\ }\bibfield
  {title} {\bibinfo {title} {{Band structure of twisted bilayer graphene:
  Emergent symmetries, commensurate approximants, and Wannier obstructions}},\
  }\href {https://doi.org/10.1103/PhysRevB.98.085435} {\bibfield  {journal}
  {\bibinfo  {journal} {Phys. Rev. B}\ }\textbf {\bibinfo {volume} {98}},\
  \bibinfo {pages} {085435} (\bibinfo {year} {2018})}\BibitemShut {NoStop}%
\bibitem [{\citenamefont {Yuan}\ and\ \citenamefont {Fu}(2018)}]{Yuan_2018}%
  \BibitemOpen
  \bibfield  {author} {\bibinfo {author} {\bibfnamefont {N.~F.~Q.}\
  \bibnamefont {Yuan}}\ and\ \bibinfo {author} {\bibfnamefont {L.}~\bibnamefont
  {Fu}},\ }\bibfield  {title} {\bibinfo {title} {{Model for the metal-insulator
  transition in graphene superlattices and beyond}},\ }\href
  {https://doi.org/10.1103/PhysRevB.98.045103} {\bibfield  {journal} {\bibinfo
  {journal} {Phys. Rev. B}\ }\textbf {\bibinfo {volume} {98}},\ \bibinfo
  {pages} {045103} (\bibinfo {year} {2018})}\BibitemShut {NoStop}%
\bibitem [{\citenamefont {Lin}\ and\ \citenamefont
  {Tom\'anek}(2018)}]{Lin_2018}%
  \BibitemOpen
  \bibfield  {author} {\bibinfo {author} {\bibfnamefont {X.}~\bibnamefont
  {Lin}}\ and\ \bibinfo {author} {\bibfnamefont {D.}~\bibnamefont
  {Tom\'anek}},\ }\bibfield  {title} {\bibinfo {title} {{Minimum model for the
  electronic structure of twisted bilayer graphene and related structures}},\
  }\href {https://doi.org/10.1103/PhysRevB.98.081410} {\bibfield  {journal}
  {\bibinfo  {journal} {Phys. Rev. B}\ }\textbf {\bibinfo {volume} {98}},\
  \bibinfo {pages} {081410} (\bibinfo {year} {2018})}\BibitemShut {NoStop}%
\bibitem [{\citenamefont {Kang}\ and\ \citenamefont {Vafek}(2018)}]{Kang_2018}%
  \BibitemOpen
  \bibfield  {author} {\bibinfo {author} {\bibfnamefont {J.}~\bibnamefont
  {Kang}}\ and\ \bibinfo {author} {\bibfnamefont {O.}~\bibnamefont {Vafek}},\
  }\bibfield  {title} {\bibinfo {title} {{Symmetry, Maximally Localized Wannier
  States, and a Low-Energy Model for Twisted Bilayer Graphene Narrow Bands}},\
  }\href {https://doi.org/10.1103/PhysRevX.8.031088} {\bibfield  {journal}
  {\bibinfo  {journal} {Phys. Rev. X}\ }\textbf {\bibinfo {volume} {8}},\
  \bibinfo {pages} {031088} (\bibinfo {year} {2018})}\BibitemShut {NoStop}%
\bibitem [{\citenamefont {Rademaker}\ and\ \citenamefont
  {Mellado}(2018)}]{Rademaker_2018}%
  \BibitemOpen
  \bibfield  {author} {\bibinfo {author} {\bibfnamefont {L.}~\bibnamefont
  {Rademaker}}\ and\ \bibinfo {author} {\bibfnamefont {P.}~\bibnamefont
  {Mellado}},\ }\bibfield  {title} {\bibinfo {title} {{Charge-transfer
  insulation in twisted bilayer graphene}},\ }\href
  {https://doi.org/10.1103/PhysRevB.98.235158} {\bibfield  {journal} {\bibinfo
  {journal} {Phys. Rev. B}\ }\textbf {\bibinfo {volume} {98}},\ \bibinfo
  {pages} {235158} (\bibinfo {year} {2018})}\BibitemShut {NoStop}%
\bibitem [{\citenamefont {Po}\ \emph {et~al.}(2019)\citenamefont {Po},
  \citenamefont {Zou}, \citenamefont {Senthil},\ and\ \citenamefont
  {Vishwanath}}]{Po_2019}%
  \BibitemOpen
  \bibfield  {author} {\bibinfo {author} {\bibfnamefont {H.~C.}\ \bibnamefont
  {Po}}, \bibinfo {author} {\bibfnamefont {L.}~\bibnamefont {Zou}}, \bibinfo
  {author} {\bibfnamefont {T.}~\bibnamefont {Senthil}},\ and\ \bibinfo {author}
  {\bibfnamefont {A.}~\bibnamefont {Vishwanath}},\ }\bibfield  {title}
  {\bibinfo {title} {{Faithful tight-binding models and fragile topology of
  magic-angle bilayer graphene}},\ }\href
  {https://doi.org/10.1103/PhysRevB.99.195455} {\bibfield  {journal} {\bibinfo
  {journal} {Phys. Rev. B}\ }\textbf {\bibinfo {volume} {99}},\ \bibinfo
  {pages} {195455} (\bibinfo {year} {2019})}\BibitemShut {NoStop}%
\bibitem [{\citenamefont {Qiao}\ \emph {et~al.}(2018)\citenamefont {Qiao},
  \citenamefont {Yin},\ and\ \citenamefont {He}}]{Qiao_2018}%
  \BibitemOpen
  \bibfield  {author} {\bibinfo {author} {\bibfnamefont {J.-B.}\ \bibnamefont
  {Qiao}}, \bibinfo {author} {\bibfnamefont {L.-J.}\ \bibnamefont {Yin}},\ and\
  \bibinfo {author} {\bibfnamefont {L.}~\bibnamefont {He}},\ }\bibfield
  {title} {\bibinfo {title} {{Twisted graphene bilayer around the first magic
  angle engineered by heterostrain}},\ }\href
  {https://doi.org/10.1103/PhysRevB.98.235402} {\bibfield  {journal} {\bibinfo
  {journal} {Phys. Rev. B}\ }\textbf {\bibinfo {volume} {98}},\ \bibinfo
  {pages} {235402} (\bibinfo {year} {2018})}\BibitemShut {NoStop}%
\bibitem [{\citenamefont {Carr}\ \emph {et~al.}(2019)\citenamefont {Carr},
  \citenamefont {Fang}, \citenamefont {Zhu},\ and\ \citenamefont
  {Kaxiras}}]{Carr_2019}%
  \BibitemOpen
  \bibfield  {author} {\bibinfo {author} {\bibfnamefont {S.}~\bibnamefont
  {Carr}}, \bibinfo {author} {\bibfnamefont {S.}~\bibnamefont {Fang}}, \bibinfo
  {author} {\bibfnamefont {Z.}~\bibnamefont {Zhu}},\ and\ \bibinfo {author}
  {\bibfnamefont {E.}~\bibnamefont {Kaxiras}},\ }\bibfield  {title} {\bibinfo
  {title} {{Exact continuum model for low-energy electronic states of twisted
  bilayer graphene}},\ }\href
  {https://doi.org/10.1103/PhysRevResearch.1.013001} {\bibfield  {journal}
  {\bibinfo  {journal} {Phys. Rev. Research}\ }\textbf {\bibinfo {volume}
  {1}},\ \bibinfo {pages} {013001} (\bibinfo {year} {2019})}\BibitemShut
  {NoStop}%
\bibitem [{\citenamefont {Guinea}\ and\ \citenamefont
  {Walet}(2019)}]{Guinea_2019}%
  \BibitemOpen
  \bibfield  {author} {\bibinfo {author} {\bibfnamefont {F.}~\bibnamefont
  {Guinea}}\ and\ \bibinfo {author} {\bibfnamefont {N.~R.}\ \bibnamefont
  {Walet}},\ }\bibfield  {title} {\bibinfo {title} {{Continuum models for
  twisted bilayer graphene: Effect of lattice deformation and hopping
  parameters}},\ }\href {https://doi.org/10.1103/PhysRevB.99.205134} {\bibfield
   {journal} {\bibinfo  {journal} {Phys. Rev. B}\ }\textbf {\bibinfo {volume}
  {99}},\ \bibinfo {pages} {205134} (\bibinfo {year} {2019})}\BibitemShut
  {NoStop}%
\bibitem [{\citenamefont {Po}\ \emph {et~al.}(2018)\citenamefont {Po},
  \citenamefont {Zou}, \citenamefont {Vishwanath},\ and\ \citenamefont
  {Senthil}}]{Po_2018}%
  \BibitemOpen
  \bibfield  {author} {\bibinfo {author} {\bibfnamefont {H.~C.}\ \bibnamefont
  {Po}}, \bibinfo {author} {\bibfnamefont {L.}~\bibnamefont {Zou}}, \bibinfo
  {author} {\bibfnamefont {A.}~\bibnamefont {Vishwanath}},\ and\ \bibinfo
  {author} {\bibfnamefont {T.}~\bibnamefont {Senthil}},\ }\bibfield  {title}
  {\bibinfo {title} {{Origin of Mott Insulating Behavior and Superconductivity
  in Twisted Bilayer Graphene}},\ }\href
  {https://doi.org/10.1103/PhysRevX.8.031089} {\bibfield  {journal} {\bibinfo
  {journal} {Phys. Rev. X}\ }\textbf {\bibinfo {volume} {8}},\ \bibinfo {pages}
  {031089} (\bibinfo {year} {2018})}\BibitemShut {NoStop}%
\bibitem [{\citenamefont {González}\ \emph {et~al.}(1994)\citenamefont
  {González}, \citenamefont {Guinea},\ and\ \citenamefont
  {Vozmediano}}]{Gonzalez_1994}%
  \BibitemOpen
  \bibfield  {author} {\bibinfo {author} {\bibfnamefont {J.}~\bibnamefont
  {González}}, \bibinfo {author} {\bibfnamefont {F.}~\bibnamefont {Guinea}},\
  and\ \bibinfo {author} {\bibfnamefont {M.}~\bibnamefont {Vozmediano}},\
  }\bibfield  {title} {\bibinfo {title} {{Non-Fermi liquid behavior of
  electrons in the half-filled honeycomb lattice (A renormalization group
  approach)}},\ }\href
  {https://doi.org/https://doi.org/10.1016/0550-3213(94)90410-3} {\bibfield
  {journal} {\bibinfo  {journal} {Nuclear Physics B}\ }\textbf {\bibinfo
  {volume} {424}},\ \bibinfo {pages} {595} (\bibinfo {year}
  {1994})}\BibitemShut {NoStop}%
\bibitem [{\citenamefont {Veyrat}\ \emph {et~al.}(2020)\citenamefont {Veyrat},
  \citenamefont {D{\'e}prez}, \citenamefont {Coissard}, \citenamefont {Li},
  \citenamefont {Gay}, \citenamefont {Watanabe}, \citenamefont {Taniguchi},
  \citenamefont {Han}, \citenamefont {Piot}, \citenamefont {Sellier},\ and\
  \citenamefont {Sac{\'e}p{\'e}}}]{Veyrat_2020}%
  \BibitemOpen
  \bibfield  {author} {\bibinfo {author} {\bibfnamefont {L.}~\bibnamefont
  {Veyrat}}, \bibinfo {author} {\bibfnamefont {C.}~\bibnamefont {D{\'e}prez}},
  \bibinfo {author} {\bibfnamefont {A.}~\bibnamefont {Coissard}}, \bibinfo
  {author} {\bibfnamefont {X.}~\bibnamefont {Li}}, \bibinfo {author}
  {\bibfnamefont {F.}~\bibnamefont {Gay}}, \bibinfo {author} {\bibfnamefont
  {K.}~\bibnamefont {Watanabe}}, \bibinfo {author} {\bibfnamefont
  {T.}~\bibnamefont {Taniguchi}}, \bibinfo {author} {\bibfnamefont
  {Z.}~\bibnamefont {Han}}, \bibinfo {author} {\bibfnamefont {B.~A.}\
  \bibnamefont {Piot}}, \bibinfo {author} {\bibfnamefont {H.}~\bibnamefont
  {Sellier}},\ and\ \bibinfo {author} {\bibfnamefont {B.}~\bibnamefont
  {Sac{\'e}p{\'e}}},\ }\bibfield  {title} {\bibinfo {title} {{Helical quantum
  Hall phase in graphene on SrTiO3}},\ }\href
  {https://doi.org/10.1126/science.aax8201} {\bibfield  {journal} {\bibinfo
  {journal} {Science}\ }\textbf {\bibinfo {volume} {367}},\ \bibinfo {pages}
  {781} (\bibinfo {year} {2020})}\BibitemShut {NoStop}%
\bibitem [{\citenamefont {Koshino}\ \emph {et~al.}(2018)\citenamefont
  {Koshino}, \citenamefont {Yuan}, \citenamefont {Koretsune}, \citenamefont
  {Ochi}, \citenamefont {Kuroki},\ and\ \citenamefont {Fu}}]{Koshino_2018}%
  \BibitemOpen
  \bibfield  {author} {\bibinfo {author} {\bibfnamefont {M.}~\bibnamefont
  {Koshino}}, \bibinfo {author} {\bibfnamefont {N.~F.~Q.}\ \bibnamefont
  {Yuan}}, \bibinfo {author} {\bibfnamefont {T.}~\bibnamefont {Koretsune}},
  \bibinfo {author} {\bibfnamefont {M.}~\bibnamefont {Ochi}}, \bibinfo {author}
  {\bibfnamefont {K.}~\bibnamefont {Kuroki}},\ and\ \bibinfo {author}
  {\bibfnamefont {L.}~\bibnamefont {Fu}},\ }\bibfield  {title} {\bibinfo
  {title} {{Maximally Localized Wannier Orbitals and the Extended Hubbard Model
  for Twisted Bilayer Graphene}},\ }\href
  {https://doi.org/10.1103/PhysRevX.8.031087} {\bibfield  {journal} {\bibinfo
  {journal} {Phys. Rev. X}\ }\textbf {\bibinfo {volume} {8}},\ \bibinfo {pages}
  {031087} (\bibinfo {year} {2018})}\BibitemShut {NoStop}%
\bibitem [{\citenamefont {Tang}\ \emph {et~al.}(2018)\citenamefont {Tang},
  \citenamefont {Leaw}, \citenamefont {Rodrigues}, \citenamefont {Herbut},
  \citenamefont {Sengupta}, \citenamefont {Assaad},\ and\ \citenamefont
  {Adam}}]{Tang_2018}%
  \BibitemOpen
  \bibfield  {author} {\bibinfo {author} {\bibfnamefont {H.-K.}\ \bibnamefont
  {Tang}}, \bibinfo {author} {\bibfnamefont {J.~N.}\ \bibnamefont {Leaw}},
  \bibinfo {author} {\bibfnamefont {J.~N.~B.}\ \bibnamefont {Rodrigues}},
  \bibinfo {author} {\bibfnamefont {I.~F.}\ \bibnamefont {Herbut}}, \bibinfo
  {author} {\bibfnamefont {P.}~\bibnamefont {Sengupta}}, \bibinfo {author}
  {\bibfnamefont {F.~F.}\ \bibnamefont {Assaad}},\ and\ \bibinfo {author}
  {\bibfnamefont {S.}~\bibnamefont {Adam}},\ }\bibfield  {title} {\bibinfo
  {title} {{The role of electron-electron interactions in two-dimensional Dirac
  fermions}},\ }\href {https://doi.org/10.1126/science.aao2934} {\bibfield
  {journal} {\bibinfo  {journal} {Science}\ }\textbf {\bibinfo {volume}
  {361}},\ \bibinfo {pages} {570} (\bibinfo {year} {2018})}\BibitemShut
  {NoStop}%
\bibitem [{\citenamefont {Kotov}\ \emph {et~al.}(2012)\citenamefont {Kotov},
  \citenamefont {Uchoa}, \citenamefont {Pereira}, \citenamefont {Guinea},\ and\
  \citenamefont {Castro~Neto}}]{Kotov_12}%
  \BibitemOpen
  \bibfield  {author} {\bibinfo {author} {\bibfnamefont {V.~N.}\ \bibnamefont
  {Kotov}}, \bibinfo {author} {\bibfnamefont {B.}~\bibnamefont {Uchoa}},
  \bibinfo {author} {\bibfnamefont {V.~M.}\ \bibnamefont {Pereira}}, \bibinfo
  {author} {\bibfnamefont {F.}~\bibnamefont {Guinea}},\ and\ \bibinfo {author}
  {\bibfnamefont {A.~H.}\ \bibnamefont {Castro~Neto}},\ }\bibfield  {title}
  {\bibinfo {title} {{Electron-Electron Interactions in Graphene: Current
  Status and Perspectives}},\ }\href
  {https://doi.org/10.1103/RevModPhys.84.1067} {\bibfield  {journal} {\bibinfo
  {journal} {Rev. Mod. Phys.}\ }\textbf {\bibinfo {volume} {84}},\ \bibinfo
  {pages} {1067} (\bibinfo {year} {2012})}\BibitemShut {NoStop}%
\bibitem [{\citenamefont {Tarnopolsky}\ \emph
  {et~al.}(2019{\natexlab{b}})\citenamefont {Tarnopolsky}, \citenamefont
  {Kruchkov},\ and\ \citenamefont {Vishwanath}}]{Origin}%
  \BibitemOpen
  \bibfield  {author} {\bibinfo {author} {\bibfnamefont {G.}~\bibnamefont
  {Tarnopolsky}}, \bibinfo {author} {\bibfnamefont {A.~J.}\ \bibnamefont
  {Kruchkov}},\ and\ \bibinfo {author} {\bibfnamefont {A.}~\bibnamefont
  {Vishwanath}},\ }\bibfield  {title} {\bibinfo {title} {Magic angle hierarchy
  in twisted graphene multilayers},\ }\href
  {https://doi.org/https://doi.org/10.1103/PhysRevLett.122.10640} {\bibfield
  {journal} {\bibinfo  {journal} {Phys. Rev. Lett.}\ }\textbf {\bibinfo
  {volume} {122}},\ \bibinfo {pages} {106405} (\bibinfo {year}
  {2019}{\natexlab{b}})}\BibitemShut {NoStop}%
\bibitem [{\citenamefont {Stajic}(2019)}]{Ash2019}%
  \BibitemOpen
  \bibfield  {author} {\bibinfo {author} {\bibfnamefont {J.}~\bibnamefont
  {Stajic}},\ }\bibfield  {title} {\bibinfo {title} {{Twisted multilayer
  graphene}},\ }\href {https://doi.org/10.1126/science.365.6456.879-b}
  {\bibfield  {journal} {\bibinfo  {journal} {Science}\ }\textbf {\bibinfo
  {volume} {365}},\ \bibinfo {pages} {879} (\bibinfo {year}
  {2019})}\BibitemShut {NoStop}%
\bibitem [{\citenamefont {Shen}\ \emph {et~al.}(2020)\citenamefont {Shen},
  \citenamefont {Chu}, \citenamefont {Wu}, \citenamefont {Li}, \citenamefont
  {Wang}, \citenamefont {Zhao}, \citenamefont {Tang}, \citenamefont {Liu},
  \citenamefont {Tian}, \citenamefont {Watanabe}, \citenamefont {Taniguchi},
  \citenamefont {Yang}, \citenamefont {Meng}, \citenamefont {Shi},
  \citenamefont {Yazyev},\ and\ \citenamefont {Zhang}}]{Shen_2020}%
  \BibitemOpen
  \bibfield  {author} {\bibinfo {author} {\bibfnamefont {C.}~\bibnamefont
  {Shen}}, \bibinfo {author} {\bibfnamefont {Y.}~\bibnamefont {Chu}}, \bibinfo
  {author} {\bibfnamefont {Q.}~\bibnamefont {Wu}}, \bibinfo {author}
  {\bibfnamefont {N.}~\bibnamefont {Li}}, \bibinfo {author} {\bibfnamefont
  {S.}~\bibnamefont {Wang}}, \bibinfo {author} {\bibfnamefont {Y.}~\bibnamefont
  {Zhao}}, \bibinfo {author} {\bibfnamefont {J.}~\bibnamefont {Tang}}, \bibinfo
  {author} {\bibfnamefont {J.}~\bibnamefont {Liu}}, \bibinfo {author}
  {\bibfnamefont {J.}~\bibnamefont {Tian}}, \bibinfo {author} {\bibfnamefont
  {K.}~\bibnamefont {Watanabe}}, \bibinfo {author} {\bibfnamefont
  {T.}~\bibnamefont {Taniguchi}}, \bibinfo {author} {\bibfnamefont
  {R.}~\bibnamefont {Yang}}, \bibinfo {author} {\bibfnamefont {Z.~Y.}\
  \bibnamefont {Meng}}, \bibinfo {author} {\bibfnamefont {D.}~\bibnamefont
  {Shi}}, \bibinfo {author} {\bibfnamefont {O.~V.}\ \bibnamefont {Yazyev}},\
  and\ \bibinfo {author} {\bibfnamefont {G.}~\bibnamefont {Zhang}},\ }\bibfield
   {title} {\bibinfo {title} {{Correlated states in twisted double bilayer
  graphene}},\ }\href {https://doi.org/10.1038/s41567-020-0825-9} {\bibfield
  {journal} {\bibinfo  {journal} {Nat. Phys. 16}\ ,\ \bibinfo {pages}
  {520–525}} (\bibinfo {year} {2020})}\BibitemShut {NoStop}%
\bibitem [{\citenamefont {Su\'arez~Morell}\ \emph
  {et~al.}(2010{\natexlab{b}})\citenamefont {Su\'arez~Morell}, \citenamefont
  {Correa}, \citenamefont {Vargas}, \citenamefont {Pacheco},\ and\
  \citenamefont {Barticevic}}]{Morell_2010}%
  \BibitemOpen
  \bibfield  {author} {\bibinfo {author} {\bibfnamefont {E.}~\bibnamefont
  {Su\'arez~Morell}}, \bibinfo {author} {\bibfnamefont {J.~D.}\ \bibnamefont
  {Correa}}, \bibinfo {author} {\bibfnamefont {P.}~\bibnamefont {Vargas}},
  \bibinfo {author} {\bibfnamefont {M.}~\bibnamefont {Pacheco}},\ and\ \bibinfo
  {author} {\bibfnamefont {Z.}~\bibnamefont {Barticevic}},\ }\bibfield  {title}
  {\bibinfo {title} {{Flat bands in slightly twisted bilayer graphene:
  Tight-binding calculations}},\ }\href
  {https://doi.org/10.1103/PhysRevB.82.121407} {\bibfield  {journal} {\bibinfo
  {journal} {Phys. Rev. B}\ }\textbf {\bibinfo {volume} {82}},\ \bibinfo
  {pages} {121407} (\bibinfo {year} {2010}{\natexlab{b}})}\BibitemShut
  {NoStop}%
\bibitem [{\citenamefont {Zhang}\ \emph {et~al.}(2020)\citenamefont {Zhang},
  \citenamefont {Wang}, \citenamefont {Watanabe}, \citenamefont {Taniguchi},
  \citenamefont {Ueno}, \citenamefont {Tutuc},\ and\ \citenamefont
  {LeRoy}}]{Zhang2020}%
  \BibitemOpen
  \bibfield  {author} {\bibinfo {author} {\bibfnamefont {Z.}~\bibnamefont
  {Zhang}}, \bibinfo {author} {\bibfnamefont {Y.}~\bibnamefont {Wang}},
  \bibinfo {author} {\bibfnamefont {K.}~\bibnamefont {Watanabe}}, \bibinfo
  {author} {\bibfnamefont {T.}~\bibnamefont {Taniguchi}}, \bibinfo {author}
  {\bibfnamefont {K.}~\bibnamefont {Ueno}}, \bibinfo {author} {\bibfnamefont
  {E.}~\bibnamefont {Tutuc}},\ and\ \bibinfo {author} {\bibfnamefont {B.~J.}\
  \bibnamefont {LeRoy}},\ }\bibfield  {title} {\bibinfo {title} {Flat bands in
  twisted bilayer transition metal dichalcogenides},\ }\href
  {https://doi.org/10.1038/s41567-020-0958-x} {\bibfield  {journal} {\bibinfo
  {journal} {Nature Physics}\ }\textbf {\bibinfo {volume} {16}},\ \bibinfo
  {pages} {1093} (\bibinfo {year} {2020})}\BibitemShut {NoStop}%
\end{thebibliography}%


%

\end{document}